\documentclass[aps,prd, twocolumn,superscriptaddress,nofootinbib,preprintnumbers]{revtex4-1}

\usepackage{amsmath}
\usepackage{graphicx}
\usepackage{amssymb}
\usepackage{xcolor}
\usepackage{enumitem}

\usepackage[colorlinks,citecolor=blue]{hyperref}

\newcommand{\be}{\begin{equation}}
\newcommand{\ee}{\end{equation}}
\newcommand{\bea}{\begin{eqnarray}}
\newcommand{\eea}{\end{eqnarray}}
\newcommand{\met}{\not{\!\! \rm E}_{T}}
\newcommand{\nn}{\nonumber}

\allowdisplaybreaks[4]
\newcommand{\tabincell}[2]{\begin{tabular}{@{}#1@{}}#2\end{tabular}}

\begin{document}

\title{Double Parton Scattering of Weak Gauge Boson Productions \\at the 13 TeV and 100 TeV Proton-Proton Colliders}

\author{Qing-Hong Cao}
\email{qinghongcao@pku.edu.cn}
\affiliation{Department of Physics and State Key Laboratory of Nuclear Physics and Technology, Peking University, Beijing 100871, China}
\affiliation{Collaborative Innovation Center of Quantum Matter, Beijing 100871, China}
\affiliation{Center for High Energy Physics, Peking University, Beijing 100871, China}

\author{Yandong Liu}
\email{ydliu@pku.edu.cn}
\affiliation{Department of Physics and State Key Laboratory of Nuclear Physics and Technology, Peking University, Beijing 100871, China}

\author{Ke-Pan Xie}
\email{kpxie@pku.edu.cn}
\affiliation{Department of Physics and State Key Laboratory of Nuclear Physics and Technology, Peking University, Beijing 100871, China}

\author{Bin Yan}
\email{yanbin1@msu.edu}
\affiliation{Department of Physics and State Key Laboratory of Nuclear Physics and Technology, Peking University, Beijing 100871, China}
\affiliation{Department of Physics and Astronomy, Michigan State University, East Lansing, MI 48824 U.S.A.}

\begin{abstract}

We study double parton scattering (DPS) processes involving electroweak gauge bosons at the 13 TeV and 100 TeV proton-proton colliders. Specifically, we focus on three DPS channels: $W$-boson plus two jets ($W\otimes jj$), $Z$-boson plus two jets ($Z\otimes jj$), and same-sign $W$ pair production ($W^\pm\otimes W^\pm$). We demonstrate that the $Z\otimes jj$ process, which has not been paid too much attentions, is the best channel for measuring effective cross section $\sigma_{\rm eff}$. The accuracy of $\sigma_{\rm eff}$ measurement in the three DPS channels, especially the $W^\pm\otimes W^\pm$ production, is significantly improved at the 100~TeV colliders. We advocate that combined analysis of the three DPS channels could test the universality of  effective cross section $\sigma_{\rm eff}$. 

\end{abstract}

\maketitle

\section{Introduction}\label{section:introduction}

\begin{figure}
  \centering
    \includegraphics[scale=0.35]{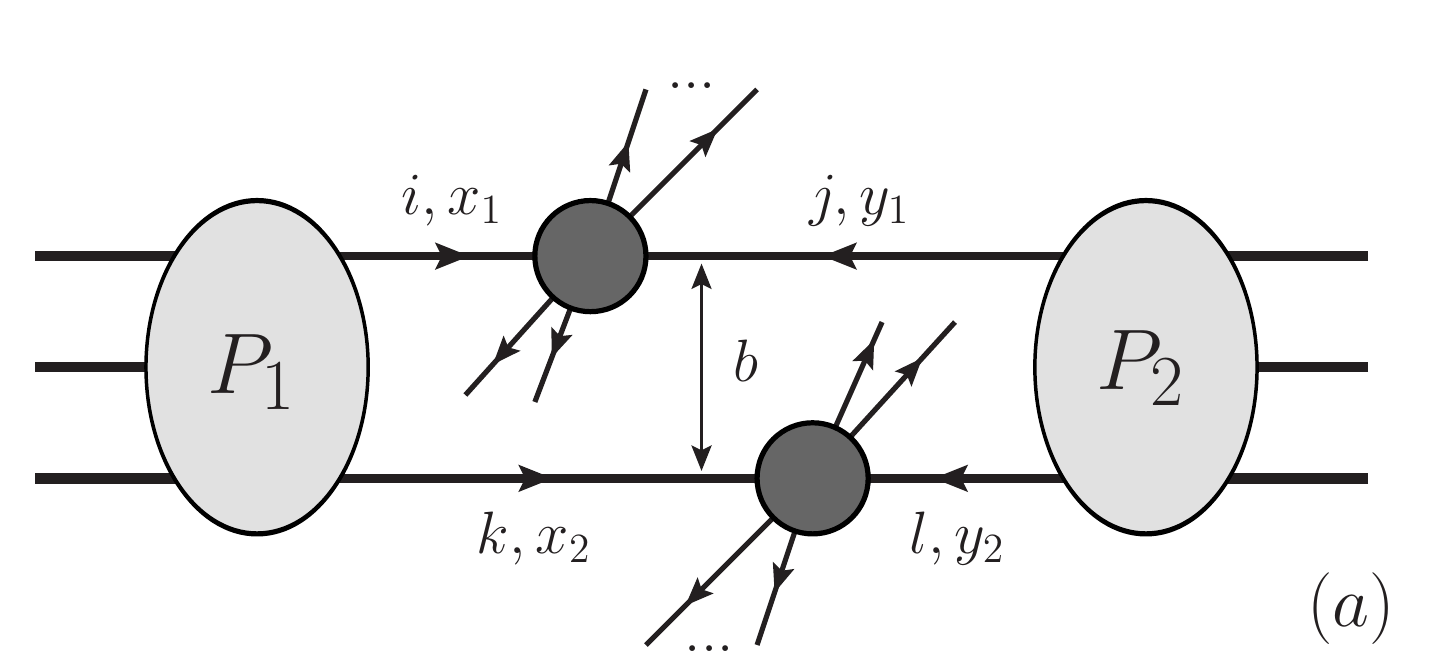}
    \includegraphics[scale=0.35]{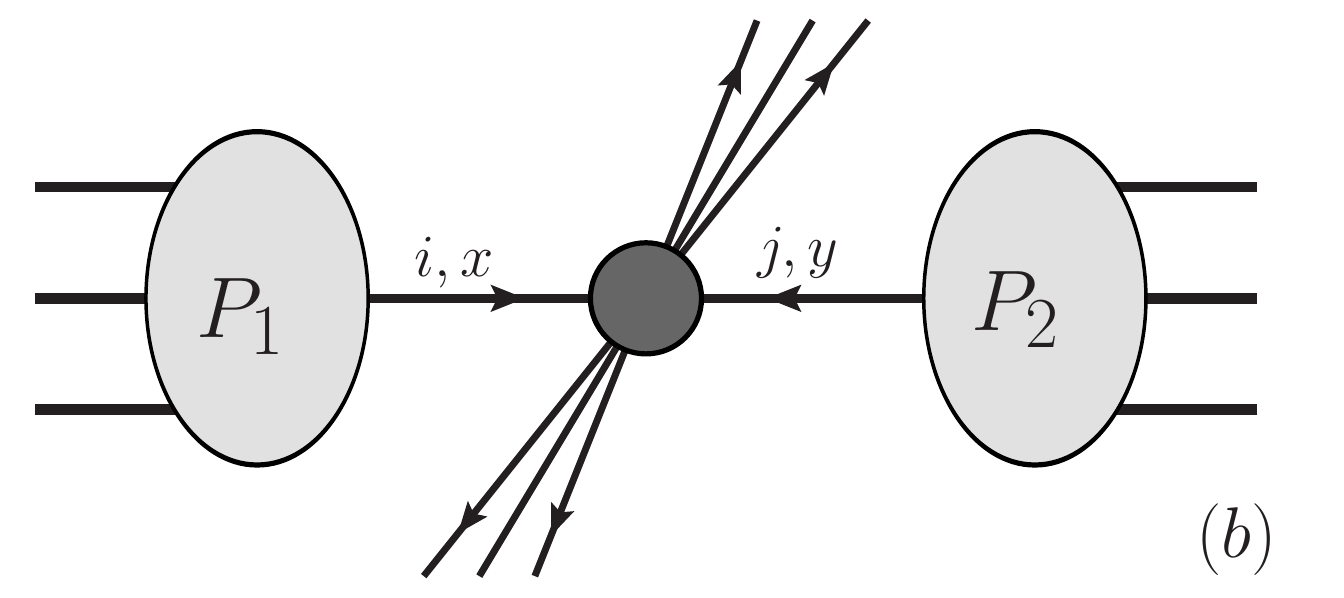}
  \caption{Pictorial illustration of double parton scattering (a) and single parton scattering (b).}
  \label{SPSandDPS} 
\end{figure}

The precise measurement of multiple parton interactions (MPI) is very important to improve our understanding of proton. In such process, two or more short distance subprocesses occur in one given hadronic interaction. The correlations and distributions of multiple partons within a proton relate directly to the transverse spatial structure of the proton, but those effects are highly suppressed by the momentum transfer of hard scattering. A typical MPI at low scales is the double parton scattering (DPS), in which two pairs of partons participant in hard interactions in a single proton-proton collision, as illustrated in Fig.~\ref{SPSandDPS}(a). As the simplest MPI process, DPS is different from the standard picture of hadron-hadron collision in which one parton from each proton partakes in the hard scattering named as single parton scattering (SPS); see Fig.~\ref{SPSandDPS}(b).

The cross-section of a DPS process that contains two subprocesses $A$ and $B$ (denoted as $A\otimes B$) can be estimated as following
\be\label{DPSapprox}
\sigma^{\rm DPS}_{A\otimes B}\approx \frac{1}{1+\delta_{AB}}\frac{\sigma^{\rm SPS}_A\otimes \sigma^{\rm SPS}_B}{\sigma_{\rm eff}},
\ee
where $\sigma_{\rm eff}$ is an effective cross-section ($\sim15$ mb) that reflecting the structure of the proton, and the symmetry factor $\delta_{AB}$ is introduced to avoid double counting, which is 1 for $A=B$ and 0 otherwise. $\sigma^{\rm SPS}_{A(B)}$ is the SPS cross section of subprocess $A(B)$, respectively. Given the large value of $\sigma_{\rm eff}$, it is usually expected that the effects of DPS are negligible or described in the parametrization of underlying events. However, the cross-section of DPS can be sizably enhanced with increasing collider energy $\sqrt{s}$ if the subprocesses involve the sea quark or gluon in the initial state as the parton distribution function (PDF) of both sea quarks and gluons grows dramatically in small $x$ region. Therefore, at high energy colliders, some of the DPS processes could yield enough signal events to be discovered. For the search of new physics beyond the Standard Model (SM), the DPS processes can also be considerable backgrounds.

Owing to the unprecedented energy of the LHC, one expects to measure the model parameter $\sigma_{\rm eff}$ precisely through various DPS processes. That would shed lights on MPI in hadron collisions; for example, there are a few open questions concerning DPS: 
\begin{enumerate}
\addtolength{\itemsep}{-0.5 em} 
\item how well can one measure $\sigma_{\rm eff}$ in hadron collisions?
\item does $\sigma_{\rm eff}$ vary with colliding energies? 
\item is $\sigma_{\rm eff}$ universal for different DPS processes? 
\end{enumerate}
In this paper, we investigate then these problems at the 13 TeV LHC and also at a future hadron collider with a center of mass energy of 100 TeV, e.g.  SppC~\cite{CEPC-SppCStudyGroup:2015csa} and FCC-hh~\cite{Mangano:2017tke}. In order to overcome the huge suppression of $\sigma_{\rm eff}$, one should consider those DPS processes involving two sizable SPS subprocesses. Table~\ref{table:sps} displays the cross sections of three SPS processes of interest to us. The jets in the dijet ($jj$) production are required to satisfy the kinematic cuts of $p_T^j>25$ GeV, $|\eta^j|<$ 5 and $\Delta R_{jj}\geq 0.4$, where $p_T$ and $\eta$ denotes the transverse momentum and rapidity, respectively, and $\Delta R_{mn}\equiv \sqrt{(\eta^m-\eta^n)^2 + (\phi^m-\phi^n)^2}$ represents the angular distance between the object $m$ and $n$ with $\phi$ being the azimuthal angle. Combining any two SPS processes in the list might yield a sizable DPS process. 

Among the possibilities, $jj\otimes jj$ provides the largest cross-section of DPS. Indeed the 4 jets final state is a good channel to measure the DPS~\cite{Aaboud:2016dea, Abe:1993rv, ATLAS-CONF-2015-058, Akesson:1986iv, Alitti:1991rd}, but triggering the jets is challenging in high-energy hadron collisions. In contrast, the $W\otimes jj$ and $Z\otimes jj$ processes exhibit charged leptons in the final state and can be easily detected~\cite{Kumar:2016oyn, Chatrchyan:2013xxa, Aad:2013bjm}. 
The pure electroweak processes, $W^\pm\otimes W^\mp$ or $W^\pm\otimes Z$, have sizable production rates, but it is challenging to extract them from the enormous SPS diboson backgrounds. However, the same-sign $W^\pm\otimes W^\pm$ channel has rather low SM backgrounds and is promising~\cite{CMS-PAS-FSQ-13-001, Myska:2012dj, Myska:2013duq, CMS-PAS-FSQ-16-009}. In addition, the production rate of quarkoniums, e.g. $J/\psi$, also has the potential to be a subprocess of DPS, and it has been studied both theoretically~\cite{Kom:2011bd, Baranov:2011ch, Novoselov:2011ff, Luszczak:2011zp, Maciula:2017egq, Lansberg:2017chq, Borschensky:2016nkv} and experimentally~\cite{Aaij:2012dz, Abazov:2014qba, Aaij:2016bqq}. But the precision calculation of SPS $J/\psi$ associated production processes is still an ongoing problem~\cite{Li:2013csa, Sun:2014gca}, which limits the accuracy of experimental measurement. Table~\ref{table:alldata} shows the effective cross section $\sigma_{\rm eff}$ measured by different experiments and energies. The results do not converge into a single value and have large errors. The average $\sigma_{\rm eff}$ is approximately 15~mb. In this work we will study the $W\otimes jj$, $Z\otimes jj$ and $W^\pm \otimes W^\pm$ channels and explore the potential of measuring $\sigma_{\rm eff}\sim 10-20~{\rm mb}$ at the 13~TeV and 100~TeV colliders. 

\begin{table}
\caption{The cross section of the SPS processes of interest to us at the 13 TeV LHC and at the 100~TeV  SppC/FCC-hh. }
\label{table:sps}
\begin{tabular}{c|c|c|c} \hline
SPS Process & $pp\to jj$ & $pp\to W$ & $pp\to Z$ \\ \hline
13~TeV& $\sim10^8$ pb & $\sim10^5$ pb & $\sim10^4$ pb \\ \hline
100 TeV& $\sim10^9$ pb & $\sim10^6$ pb & $\sim10^5$ pb \\ \hline
\end{tabular}
\end{table}

The paper is organized as follows. We introduce the framework and various double parton models in Sec.~\ref{section:framework}. A comparison of two double parton models is presented in Sec.~\ref{section:parton_lumi}. We use the simply factorized model to investigate the phenomenologies of the three DPS processes in Sec.~\ref{section:wjj}-Sec.~\ref{section:ww}.
Finally, we present a combined analysis of the three DPS channels and conclude in Sec.~\ref{section:discussion}. 

\begin{table}
\footnotesize
\caption{Recent $\sigma_{\rm eff}$ measurements by different experiments and energies.}
\label{table:alldata}
\begin{tabular}{l|c|c|c|r} 
\hline
DPS channel & $\sigma_{\rm eff} ({\rm mb})$& Collaboration & Collider & Luminosity \\ \hline
$jj\otimes jj$ & $12.1^{+10.7}_{-5.4}$ & CDF~\cite{Abe:1993rv} & \tabincell{c}{1.8~TeV\\ Tevatron} & $325~{\rm nb}^{-1}$ \\ \hline
$J/\Psi\otimes D$ & 
\tabincell{c}{$14.9^{+2.6}_{-3.1}$\\ $17.6^{+3.1}_{-4.0}$\\ $12.8^{+2.6}_{-3.2}$\\ $18.0^{+4.8}_{-5.5}$}
& LHCb~\cite{Aaij:2012dz} & \tabincell{c}{7~TeV\\LHC} & $355~{\rm pb}^{-1}$ \\ \hline
$W\otimes jj$ & $15.0_{- 4.2}^{+ 5.8}$ & ATLAS~\cite{Aad:2013bjm} & \tabincell{c}{7~TeV\\ LHC} & $36~{\rm pb}^{-1}$ \\ \hline
$W^\pm\otimes W^\pm$ & $>5.91$ & CMS~\cite{CMS-PAS-FSQ-13-001} & \tabincell{c}{8~TeV\\ LHC} & $19.7~{\rm fb}^{-1}$ \\ \hline
$W\otimes jj $ & $ 20.7_{ - 6.6}^{+ 6.6}$ & CMS~\cite{Chatrchyan:2013xxa} & \tabincell{c}{7~TeV\\ LHC} & $5~{\rm fb}^{-1}$ \\ \hline
$\gamma j\otimes jj$  & \tabincell{c}{ 
$12.7^{+1.3}_{-1.3}$ \\ $14.5^{+3.3}_{-5.3}$} & D0~\cite{Abazov:2014fha} & \tabincell{c}{1.96~TeV\\ Tevatron} & $8.1~{\rm fb}^{-1}$ \\ \hline
$jj\otimes jj$ & $16.1^{+6.4}_{-7.0}$ & ATLAS~\cite{ATLAS-CONF-2015-058} & \tabincell{c}{7~TeV\\LHC} & $37.3~{\rm pb}^{-1}$ \\ \hline
$\gamma\gamma\otimes jj$ & $19.3^{+7.9}_{-7.9}$ & D0~\cite{Abazov:2015nnn} & \tabincell{c}{1.96~TeV\\ Tevatron} & $8.7~{\rm fb}^{-1}$ \\ \hline
$J/\Psi\otimes J/\Psi$ & $4.80^{+2.55}_{-2.55}$ & D0~\cite{Abazov:2014qba} & \tabincell{c}{1.96~TeV\\ Tevatron}& $8.1~{\rm fb}^{-1}$ \\ \hline
$J/\Psi\otimes J/\Psi$ & 
\tabincell{c}{$14.4_{ - 4.9}^{ + 4.9}$ \\ $9.2_{ - 3.9}^{+ 3.9}$ \\$11.3_{ - 1.5}^{ + 1.5}$}
 & LHCb~\cite{Aaij:2016bqq} & \tabincell{c}{13~TeV\\ LHC}& $5~{\rm fb}^{-1}$ \\ \hline
\end{tabular}
\end{table}

\section{Framework}\label{section:framework}

According to factorization theorem~\cite{Collins:1989gx}, the inclusive cross section of SPS is expressed as
\be
\sigma^{\rm SPS}_Y=\sum_{ij}\int dxdyf_i(x,\mu_F)f_j(y,\mu_F)\hat\sigma_{ij}^Y(x,y),
\ee 
where $\hat\sigma_{ij}^Y$ is the inclusive cross section of parton scattering $ij\to Y$, and the parton distribution functions (PDF) $f_i(x,\mu_F)$ represents the probability of finding a parton $i$ with a momentum fraction $x$ and scale $\mu_F$ in a proton. The physical meaning of this equation can be read clearly in Fig.~\ref{SPSandDPS}(b). Unlike SPS, however, the cross section of DPS doesn't have a well-proved mathematica expression yet. In general, DPS cross section can be written down as~\cite{Gaunt:2010pi, Gaunt:2009re}
\bea\label{DPSfac}
\sigma^{\rm DPS}_{A\otimes B}&=&\frac{1}{1+\delta_{AB}}\sum_{ijkl}\int dx_1dy_1dx_2dy_2d^2b\nn\\
&\times&\Gamma_{ik}(x_1,x_2,\mu_F,\mu'_F;b)\Gamma_{jl}(y_1,y_2,\mu_F,\mu'_F;b)\nn\\
&\times&\hat\sigma_{ij}^A(x_1,y_1)\hat\sigma_{kl}^B(x_2,y_2),
\eea
where $\Gamma_{ik}(x_1,x_2,\mu_F,\mu'_F;b)$ represents the probability of finding two partons $i$ (with momentum fraction $x_1$ and scale $\mu_F$) and $k$ (with momentum fraction $x_2$ and scale $\mu'_F$) with a transverse distance separation $b$. And $\Gamma_{jl}(y_1,y_2,\mu_F,\mu'_F;b)$ has a similar meaning. In addition, $\hat\sigma_{ij}^A$ and $\hat\sigma_{kl}^B$ are the subprocess cross sections for inclusive $ij\to A$ and $kl\to B$, respectively. See Fig.~\ref{SPSandDPS}(a) for a pictorial illustration. Ignoring the transverse correlation of partons, $\Gamma_{ik}$ can be factorized as~\cite{Gaunt:2010pi, Gaunt:2009re}
\be
\Gamma_{ik}(x_1,x_2,\mu_F,\mu'_F;b)=D_{ik}(x_1,x_2,\mu_F,\mu'_F)F(b),
\ee
where the double PDF (dPDF) $D_{ik}$ describes the longitudinal structure of double partons while $F(b)$ represents the effective transverse overlap area of partonic interactions that produces the characteristic phenomena of the DPS process. The $F(b)$ is usually assumed to be the same for all parton pairs involved in the DPS process of interest. Integrating over the distance $b$ yields the {\it master formula} in our study, 
\begin{widetext}
\be
\label{DPSfactorize}
\sigma^{\rm DPS}_{A\otimes B}=\frac{1}{1+\delta_{AB}}\frac{1}{\sigma_{\text{eff}}}\sum_{ijkl}\int dx_1dy_1dx_2dy_2 D_{ik}(x_1,x_2,\mu_F,\mu'_F) D_{jl}(y_1,y_2,\mu_F,\mu'_F)\hat\sigma_{ij}^A(x_1,y_1)\hat\sigma_{kl}^B(x_2,y_2),
\ee
\end{widetext}
where the effective cross section,
\be
\sigma_{\text{eff}}^{-1}\equiv\int d^2b\big(F(b)\big)^2,
\ee
is sensitive to the transverse size of incoming protons. Its value is difficult to derive from the parton model assumptions and has to be determined from experiments. 

Although the dPDF should be measured in experiments, one often assume it can be built up from the single parton PDFs. Various construction approaches have been proposed~\cite{Snigirev:2003cq, Korotkikh:2004bz, Gaunt:2009re, Rinaldi:2016mlk, Golec-Biernat:2016vbt}. In general, the dPDF can be written as 
\be
D_{ik}(x_1,x_2,\mu_F,\mu'_F)=f_i(x_1,\mu_F)f_k(x_2,\mu'_F)\rho_{ik}(x_1,x_2),\nn
\ee
where $\rho_{ik}$ describes the correlation between the two partons. A simple model is to ignore longitudinal momentum correlations of the two parton and only demands their momentum sum less than the momentum of their mother proton, i.e.  
\be\label{SF_model}
\rho_{ik}(x_1,x_2)=\theta(1-x_1-x_2).
\ee
Such an approximation is typically justified at low $x$ values on the grounds that the population of partons is large at these values. Making use of the typically small $x_{1,2}$ and $y_{1,2}$ in hard scattering, one can drop this constraint and obtain the approximate expression Eq.~(\ref{DPSapprox}).
We name it as ``simply factorized" (SF) dPDF, which is widely used both in theoretical~\cite{Maina:2010vh, Gaunt:2010pi, Berger:2009cm, Berger:2011ep, Hussein:2007gj, Godbole:1989ti, Blok:2015afa, DelFabbro:1999tf, Bandurin:2010gn, Maina:2009sj, Kulesza:1999zh, Ceccopieri:2017oqe} and experimental studies~\cite{Akesson:1986iv, Alitti:1991rd, Abe:1993rv, ATLAS-CONF-2015-058, Abazov:2014fha, Abazov:2015nnn, Aaij:2012dz, Kumar:2016oyn, Chatrchyan:2013xxa, Aad:2013bjm, CMS-PAS-FSQ-13-001, Myska:2012dj, Myska:2013duq}. Although those experiments cover various processes such as $jj\otimes jj$~\cite{Akesson:1986iv, Alitti:1991rd, Abe:1993rv, ATLAS-CONF-2015-058, Aaboud:2016dea}, $W^\pm\otimes jj$~\cite{Kumar:2016oyn, Chatrchyan:2013xxa, Aad:2013bjm}, $W^\pm\otimes W^\pm$~\cite{CMS-PAS-FSQ-13-001, Myska:2012dj, Myska:2013duq, CMS-PAS-FSQ-16-009}, $J/\psi\otimes D$ mesons~\cite{Aaij:2012dz},  $\gamma j\otimes jj$~\cite{Abazov:2014fha} and $\gamma\gamma \otimes jj$~\cite{Abazov:2015nnn}, they all give $\sigma_{\rm eff}\sim\mathcal{O}(10)$ mb, and most of them give $\sim 15$ mb. This fact gives strong evidence to the validity of SF model and the universality of $\sigma_{\rm{eff}}$. 

The SF model, simple and supported by experimental data, ignores the longitudinal correlation between the two subprocesses. In a theoretical perspective, the SF model does not obey the dPDF sum rules and evolution equations. Ref.~\cite{Gaunt:2009re} proposes an improved dPDF named as GS09 by assuming $\mu_F=\mu'_F$ and setting
\be
\rho_{ik}(x_1,x_2)=(1-x_1-x_2)^2(1-x_1)^{-2-\alpha_i}(1-x_2)^{-2-\alpha_k},
\label{GS_model}
\ee
where $\alpha_{i}=0$ for sea partons and $0.5$ for valence partons. Nevertheless, different double parton models give nearly the same results in the small $x$  region where the parton correlation is negligible~\cite{Gaunt:2009re}.

\section{Simple Factorized dPDF versus GS09 dPDF}\label{section:parton_lumi}

In this study we use the SF model specified in Eq.~(\ref{SF_model}) to study the DPS, but before moving to the detailed phenomenological study, we compare different double parton models in the next section.

The comparison of the SF and GS09 dPDFs has been investigated in the $Z\otimes$jets channel~\cite{Maina:2010vh} and  the $W\otimes W$ channel~\cite{Gaunt:2010pi}. It was shown that both the SF and GS09 dPDFs give rise to consistent cross sections within $\sim10\%$ accuracy, and furthermore, the kinematics distributions of $p_T$, $\eta$ and invariance mass are insensitive to the choice of dPDFs. Below we examine the difference of the two dPDFs in the $W\otimes jj$, $Z\otimes jj$ and $W^\pm\otimes W^\pm$ processes.

\subsection{Parton Luminosity\label{luminosity}}

\begin{figure*}
\centering
\includegraphics[scale=0.32]{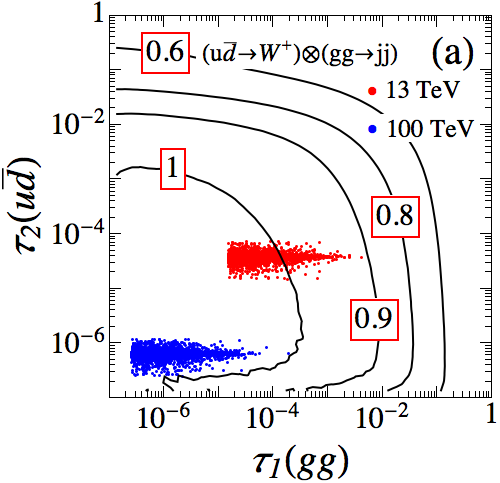}
\includegraphics[scale=0.32]{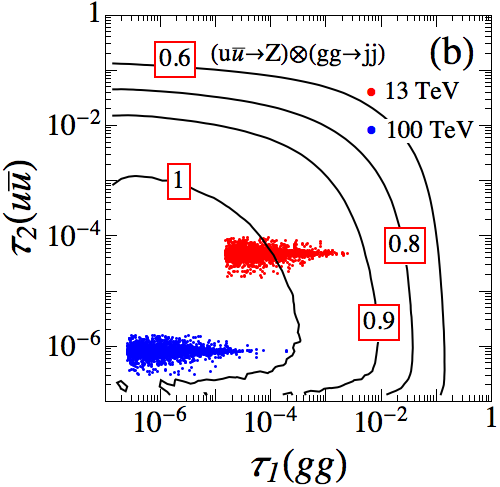}
\includegraphics[scale=0.32]{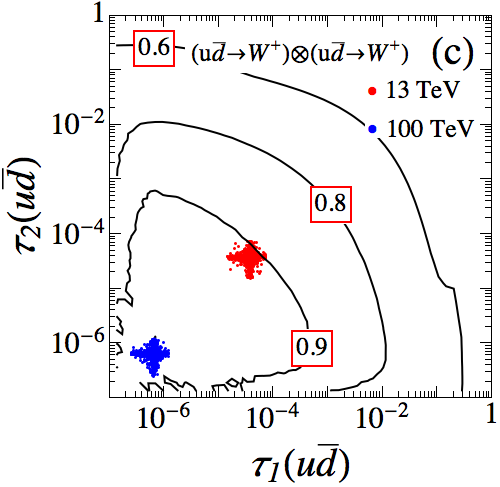}
\caption{The ratio of parton luminosities in the GS09 dPDF and SF dPDF: (a) $u\bar{d}\otimes gg\to W^+\otimes jj$; (b) $u\bar u\otimes gg\to Z\otimes jj$; (c) $u\bar d\otimes u\bar d\to W^+\otimes W^+$. The red (blue) points denote the ratio at the 13 (100) TeV colliders, respectively.}
\label{parton_luminosity} 
\end{figure*}

To compare these two kind of dPDFs, we should not only discuss the cross sections for some specific processes, but also study the parton luminosities. The parton luminosity is an important quantity to estimate the order-of-magnitude of hard process cross section in hadron collisions. In the SPS, it is defined as~\cite{Quigg:2009gg}
\bea
\frac{dL_{ij}}{d\tau}=\frac{1}{1+\delta_{ij}}\int_\tau^1\frac{dx}{x}&\Big[&f_i(x,\mu_F)f_j(\tau/x,\mu_F)\nn\\
&+&f_j(x,\mu_F)f_i(\tau/x,\mu_F)\Big],
\eea
where the indices $i$ and $j$ label the incoming partons; see Fig.~\ref{SPSandDPS}(b). The $\delta_{ij}$ symbol is used to avoid double counting. This definition is process independent and reflects the properties of PDF. In a DPS process depicted in Fig.~\ref{SPSandDPS}(a), we define double parton luminosity as
\be
\label{GS09_dPDF}
\begin{split}
\frac{dL_{ij,kl}}{d\tau_1d\tau_2}=&\frac{1}{1+\delta_{ik}\delta_{jl}}\frac{1}{1+\delta_{ij}}\frac{1}{1+\delta_{kl}}\int_{\tau_1}^1\frac{dx_1}{x_1}\int_{\tau_2}^1\frac{dx_2}{x_2}\\
&\times\Big[D_{ik}(x_1,x_2,\mu_F)D_{jl}(\tau_1/x_1,\tau_2/x_2,\mu_F)\\
&~+D_{jk}(x_1,x_2,\mu_F)D_{il}(\tau_1/x_1,\tau_2/x_2,\mu_F)\\
&~+D_{il}(x_1,x_2,\mu_F)D_{jk}(\tau_1/x_1,\tau_2/x_2,\mu_F)\\
&~+D_{jl}(x_1,x_2,\mu_F)D_{ik}(\tau_1/x_1,\tau_2/x_2,\mu_F)\Big],
\end{split}
\ee
which, in the SF dPDF model, can be simplified as 
\bea\label{simple_dPDF}
&&\left(\frac{dL_{ij,kl}}{d\tau_1d\tau_2}\right)_{\rm SF}=\frac{1}{1+\delta_{ik}\delta_{jl}}\frac{1}{1+\delta_{ij}}\frac{1}{1+\delta_{kl}}\nn\\
&&\times\int_{\tau_1}^1\frac{dx_1}{x_1}\int_{\tau_2}^1\frac{dx_2}{x_2}\theta(1-x_1-x_2)\theta\left(1-\tau_1/x_1-\tau_2/x_2\right)\nn\\
&&\qquad\times\left[f_i(x_1,\mu_F)f_j(\tau_1/x_1,\mu_F)+(i\leftrightarrow j)\right]\nn\\
&&\qquad\times\left[f_k(x_2,\mu_F)f_l(\tau_2/x_2,\mu_F)+(l\leftrightarrow k)\right].
\eea

We calculate the parton luminosities of both the GS09 and the SF dPDFs using Eqs.~(\ref{GS09_dPDF}) and~(\ref{simple_dPDF}), respectively.  In the available GS09 dPDF code, the single MSTW2008LO PDF sets~\cite{Martin:2009iq} are used to realized Eq.~(\ref{GS_model}), therefore, we use the same set of single PDF in the SF calculation. We then plot the contour of the parton luminosity ratio,  
\begin{equation}
\left(\frac{dL_{ij,kl}}{d\tau_1d\tau_2}\right)_{\rm GS09}\Big/\left(\frac{dL_{ij,kl}}{d\tau_1d\tau_2}\right)_{\rm SF},
\end{equation}
in Fig.~\ref{parton_luminosity} for the three DPS processes: (a) $(u\bar{d} \to W^+)\otimes (gg\to jj)$, (b) $(u\bar{u}\to Z)\otimes (gg\to jj)$ and (c) $(u\bar{d} \to W^+)\otimes (u\bar{d}\to W^+)$.
As shown in Sec.~\ref{sec:frac_parton} below, these parton combinations dominate in the three DPS channels.
The PDF scales are chosen as $m_W=80.4~{\rm GeV}$. Of course, the jets can also be produced from initial state quarks, but for a clear illustration, we consider only the dominant channel $gg\to jj$ in the comparison of parton luminosities. The red points denote the ratio at the 13 TeV LHC while the blue points represent the ratio at the 100 TeV  SppC/FCC-hh.

 \begin{figure*}
 \includegraphics[scale=0.3]{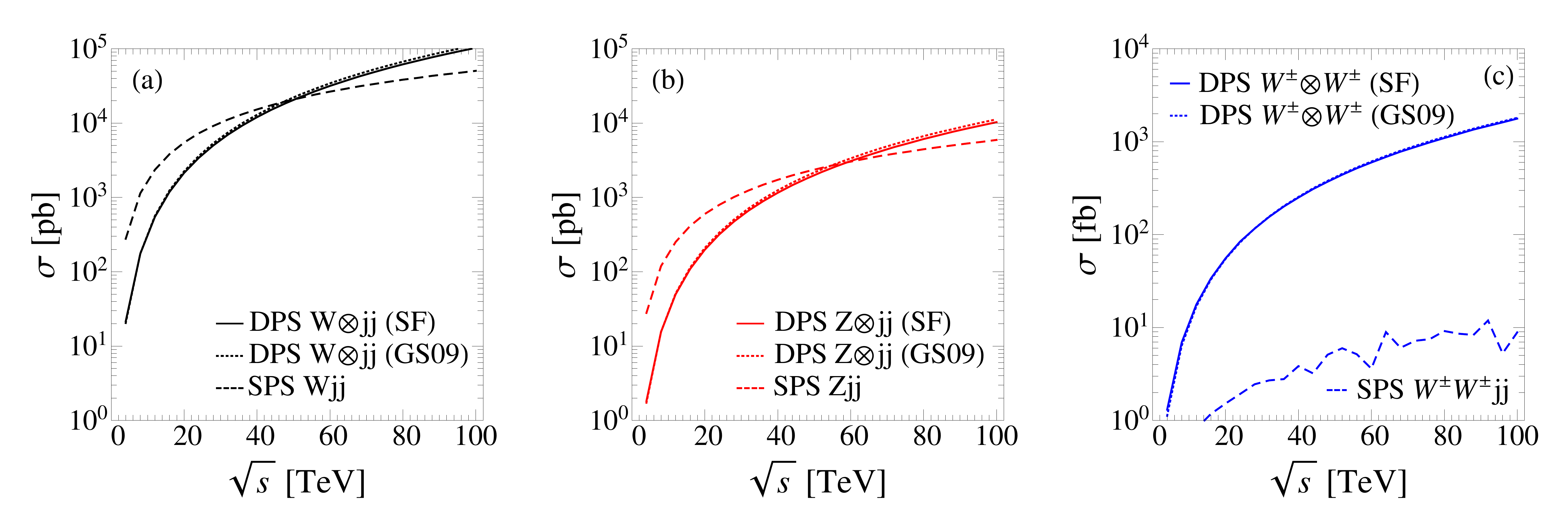}
 \caption{Cross sections of the DPS processes as a function of collider energy $\sqrt{s}$: (a) $W\otimes jj$, (b) $Z\otimes jj$ and (c) $W^\pm \otimes W^\pm$ productions.  The solid and dotted curve represents the DPS production calculated with the SF and GS09 dPDF, respectively, while the dashed curve denotes the SPS production. The jets are required to pass the selection cuts shown in Eq.~(\ref{cut0}).}
\label{tot_xsec} 
\end{figure*}

The SF and GS09 dPDFs give rise to comparable parton luminosities in the region of small $\tau$, say $\tau \sim 10^{-4}$. The difference between the two dPDFs becomes evident for $\tau \gtrsim 10^{-2}$~\cite{Gaunt:2009re}. At a collider with the fixed center of mass energy, each individual scattering channel exhibits a typical $\tau$ value. For example, the gauge bosons mass provides a natural scale in the $W$-boson or $Z$-boson production, therefore, the $\tau$ value populates mainly around $m^2_{W,\,Z}/s$. It yields $\tau\sim 4\times 10^{-5}$ at the 13 TeV LHC and $\tau\sim 10^{-6}$ at the 100 TeV  SppC/FCC-hh. For the $gg\to jj$ production the scale depends on the $p_T$ cuts imposed (which is 25 GeV in this study), making $\tau_{gg}$ distributes mostly in $(50~{\rm GeV})^2/s$. It yields a similar $\tau$ value as the $W$- or $Z$-boson production. As there is no resonance in the $jj$ production, the $\tau$ value exhibit a long tail towards larger $\tau$.

Figure~\ref{parton_luminosity} shows that the most of the luminosity ratios populates around $0.9 \sim1.1$ for both the $W\otimes jj$ and $Z\otimes jj$ channel at the 13 TeV LHC, while the luminosity ratio of $W^\pm\otimes W^\pm$ channel  is around $0.9$. It implies that the $W^\pm\otimes W^\pm$ production can be used to study the difference between GS09 and SF dPDFs at the 13 TeV LHC. For example, Ref.~\cite{Gaunt:2010pi} points out that the pseudo-rapidity asymmetry of charged leptons can be used to discriminate various dPDF sets efficiently.

\subsection{Cross Sections}

Figure~\ref{tot_xsec} displays the cross sections of the three DPS channels: (a) $W\otimes jj$ (black),  (b) $Z\otimes jj$ (red) and (c) same-sign $W^\pm \otimes W^\pm$ (blue) productions as a function of colliding energy ($\sqrt{s}$). The solid curve represents the cross sections of the DPS channel calculated with the SF dPDF while the dotted curve evaluated with the GS09 dPDF. For comparison, we also plot the SPS background processes (dashed curve). 
In order to avoid the collinear singularity, all the jets in the $W\otimes jj$ and $Z\otimes jj$ productions are required to pass the selection cuts as follows:
\be
p_T^j>25~{\rm GeV},\quad\left|\eta^j\right|<5,\quad \Delta R_{jj}\geq 0.4~.
\label{cut0}
\ee
We notice that both the SF and GS09 dPDFs generate almost identical production rates in the three DPS channels; see the solid and dotted curves. The cross sections of $W\otimes jj$ and $Z\otimes jj$ productions increase rapidly with the collider energy and exceed the SPS cross section around $\sqrt{s}=60~{\rm TeV}$; see Figs.~\ref{tot_xsec}(a) and \ref{tot_xsec}(b).

In the SPS, the same-sign $W$-boson pairs are produced in association with two extra jets. In order to mimic the DPS $W^\pm\otimes W^\pm$ production, the two additional jets in the SM SPS channel are required to escape detection, i.e. the extra jets satisfying Eq.~(\ref{cut0}) are vetoed. The jet-veto cut significantly suppresses the SPS production rate. As shown in Fig.~\ref{tot_xsec}(c), the SPS channel is about one order of magnitude smaller than the DPS channel after vetoing additional jets. 
Also, the cross section of the DPS $W^\pm\otimes W^\pm$ production increases dramatically with collider energy.

\subsection{Fraction of Double Parton Combinations}\label{sec:frac_parton}

In the study of parton luminosity ratio in Sec.~\ref{luminosity}, we only consider the parton pairs in one proton that play the leading role in the DPS channels. It is interesting to ask how often a parton pair contributes in the DPS channels of interest to us. We separate the $W^+\otimes jj$ and $W^-\otimes jj$ channels, as well as $W^+\otimes W^+$ and $W^-\otimes W^-$ channels, in order to see the difference between valence quarks and sea quarks. 

 \begin{figure}[b]
\includegraphics[scale=0.30]{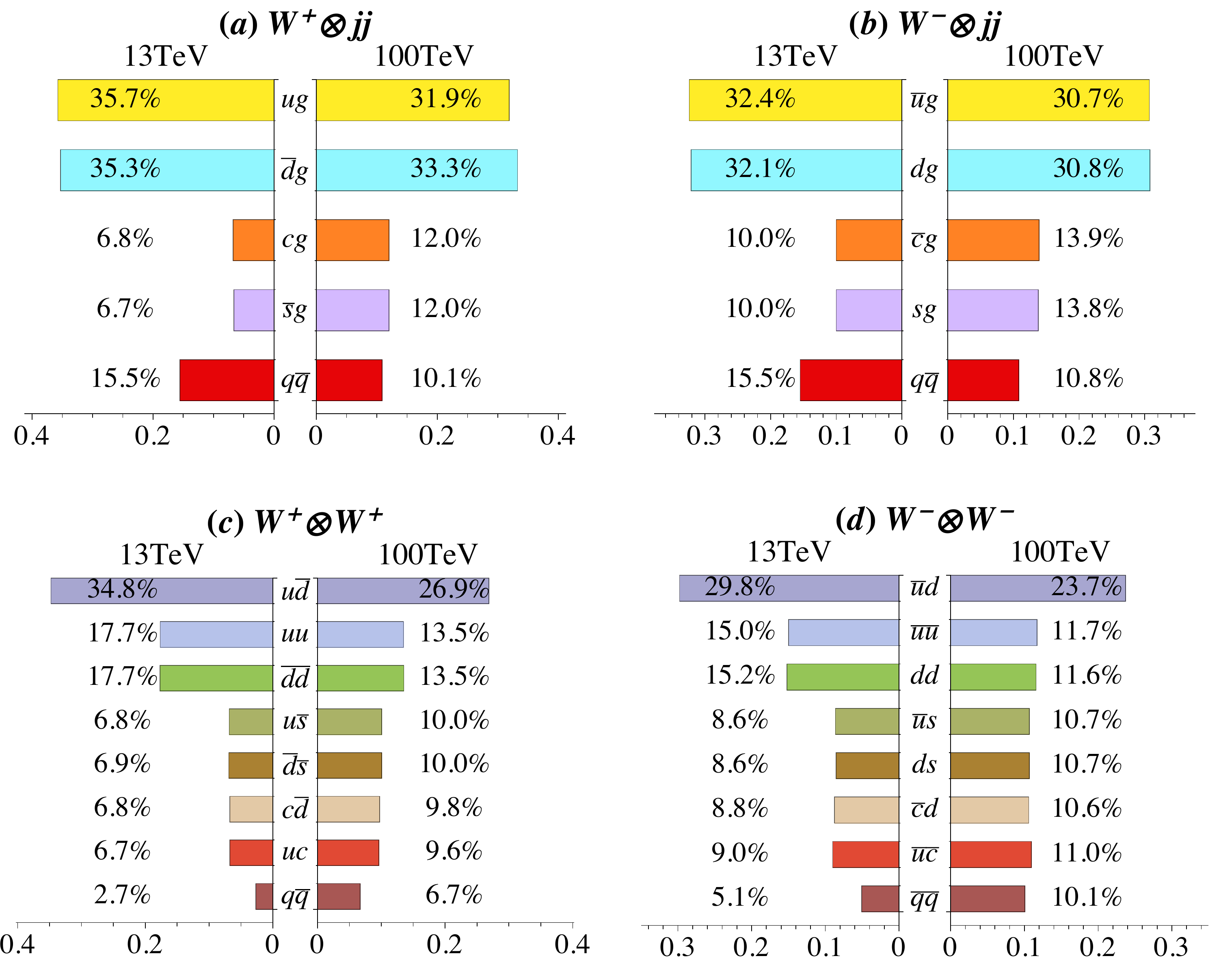}
\includegraphics[scale=0.32]{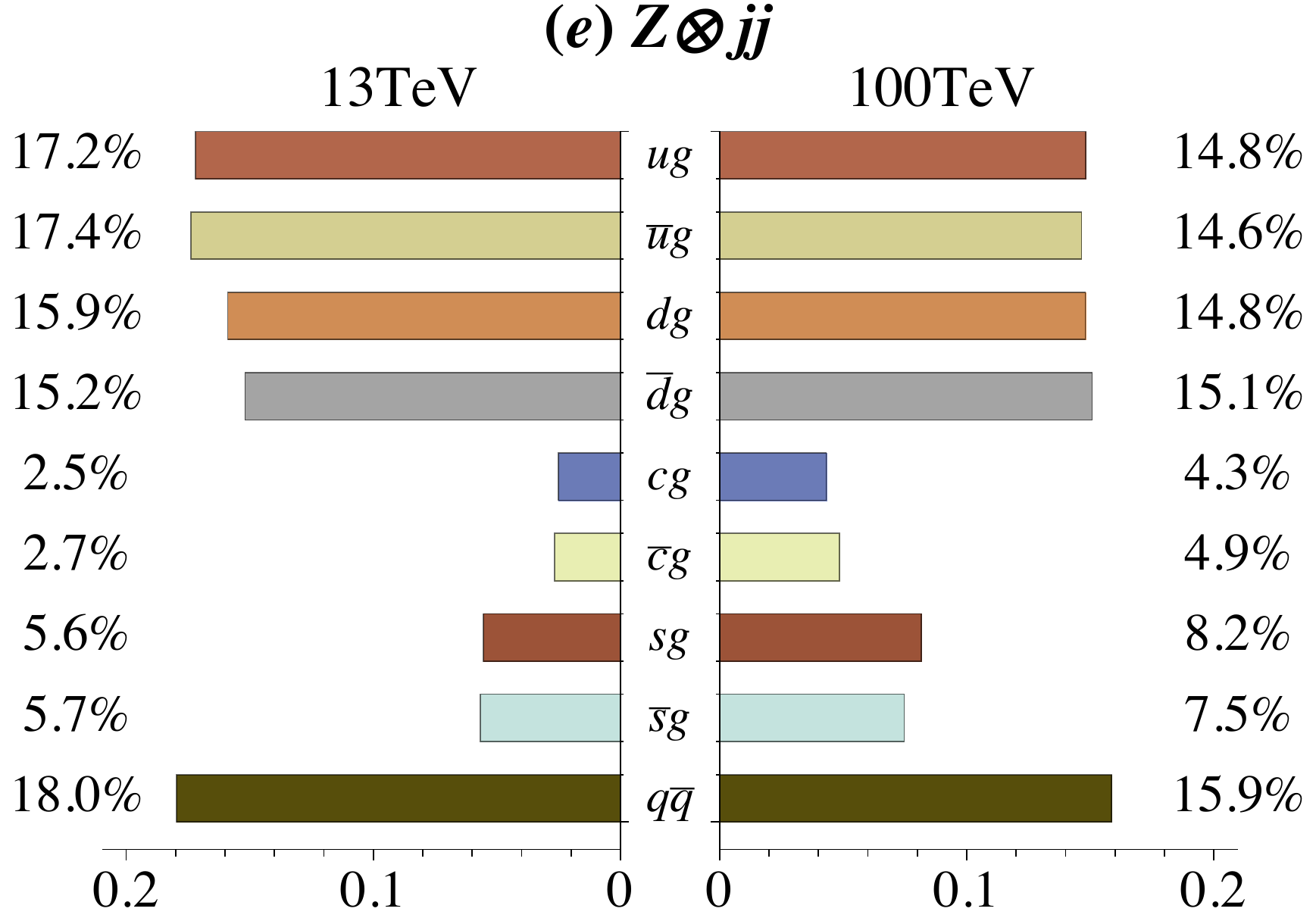}
 \caption{The fraction of double parton combinations in the DPS processes: (a) $W^+\otimes jj$, (b) $W^-\otimes jj$, (c) $W^+\otimes W^+$, (d) $W^-\otimes W^-$ and (e) $Z\otimes jj$. }
\label{partonfraction} 
\end{figure}

Electroweak gauge bosons see quarks but not gluon in the proton. To produce the $W^\pm$ or $Z$ boson, each proton need provide at least one quark. For example, the $W^+\otimes jj$ channel requires the initial state parton combinations as follows:
\bea
&& (ug)\otimes (\bar{d}g),~(uq)\otimes(\bar{d}g),~{ug}\otimes (\bar{d}q),~(uq)\otimes (\bar{d}q^\prime), \nn\\
&& (cg)\otimes (\bar{s}g),~(cq)\otimes(\bar{s}g),~{cg}\otimes (\bar{s}q),~(cq)\otimes (\bar{s}q^\prime),\cdots\nn
\eea
where $q(q^\prime)=u,d,c,s,b$. We generate ten thousand events in the $W^+\otimes jj$ channel and count the number of events with a specific parton $i$ and $j$ pair ($N_{ij}$) in one proton to obtain the fraction $\epsilon_{ij}$,
\be
\epsilon_{ij}=\frac{N_{ij}}{N_{\rm total}}.
\ee
Figure~\ref{partonfraction}(a) displays the fraction of parton pairs in one proton at the 13~TeV and 100~TeV colliders. As expected, the $ug$ pairs and $\bar{d}g$ pairs dominate the DPS process, e.g. $\epsilon_{ug}\simeq \epsilon_{\bar{d}g}\sim 35\%$ at the LHC. The subsidiary contribution is from either the $cg$ or $\bar{s}g$ pair, which yields $\epsilon_{cg}\simeq \epsilon_{\bar{s}g}\sim 6.8\%$. A pair of quarks in one proton only occurs at about 1\% of the total time, but summing over all the possible quark pairs gives rise to 15.5\%. We denote the sum of all quark pairs as $q\bar{q}$. Hence, the $W^+\otimes jj$ channel is dominated by the initial state parton configure of a pair of quark and gluon from one proton and another pair of quark and gluon from the other proton, i.e. $(qg)\otimes (\bar{q}'g)$.  The 100~TeV collider probes a much smaller $x$ at which the gluon and sea quark PDF's increase dramatically. Therefore, the fraction of $ug$ pairs decreases slightly to $\epsilon_{ug}=32\%$, but the fractions of $cg$ and $\bar{s}g$ pairs are almost doubled. A similar result is observed in the $W^-\otimes jj$ channel; see Fig.~\ref{partonfraction}(b).

The $W^+\otimes W^+$ channel has two gauge bosons and thus demand four quarks in the initial state, which are listed as follows: 
\bea
&& (u\bar{d})\otimes (u\bar{d}),~(uu)\otimes(\bar{d}\bar{d}),~{u\bar{s}}\otimes (\bar{d}c),~(uc)\otimes (\bar{d}\bar{s}), \text{ ...}\nn 
\eea
Figure~\ref{partonfraction}(c) shows the fractions of quark pairs listed above. The $u\bar{d}$ pair is the leading double partons in the $W^+\otimes W^+$ production, $\epsilon_{u\bar{d}}\simeq 35\%$. The $uu$ and $\bar{d}\bar{d}$ pairs are the second double partons, $\epsilon_{uu,\bar{d}\bar{d}}\simeq 18\%$. Other quark pairs ($u\bar{s}$, $\bar{d}\bar{s}$, $c\bar{d}$ and $uc$) contribute almost equally, $\epsilon_{u\bar{s},\bar{d}\bar{s},c\bar{d},uc}\simeq 7\%$. The rest of quark pairs not listed above only contribute 2.7\% in total. Increasing the collider energy enhances the fraction of sea quark pairs and reduces the share of $u\bar{d}$ pairs. The pattern is also applied to the $W^-\otimes W^-$ channel; see Fig.~\ref{partonfraction}(d).

The $Z\otimes jj$ channel is complicated as it involves more double parton combinations, e.g. 
\be
ug,~\bar{u}g,~dg,~\bar{d}g,~cg,~\bar{c}g,~sg,~\bar{s}g, \text{ ...}
\ee
Figure~\ref{partonfraction}(e) displays the fractions of parton pairs. Again, we use the $q\bar{q}$ to denote the sum of all quark pairs. We note that about 82\% of parton pairs are a combination of quark and gluon, which is similar to the  $W\otimes jj$ channel. 

We emphasize that the $\sigma_{\rm eff}$'s measured in the $W\otimes jj$ and $Z\otimes jj$ channels are sensitive to the double parton configuration of $(qg)\otimes (\bar{q}'g)$ while the one measured in the $W^\pm\otimes W^\pm$ channel is sensitive to the configuration of $(q\bar{q}')\otimes (q\bar{q}')$. Therefore, measuring $\sigma_{\rm eff}$ from various DPS channels involving weak bosons can check the  $\sigma_{\rm eff}$ universality. Were different $\sigma_{\rm eff}$'s reported in various DPS processes at the LHC or future colliders, the difference might shed lights on the double parton transverse correlations. 

\subsection{Rapidity difference}
 
 \begin{figure}
  \centering
    \includegraphics[scale=0.22]{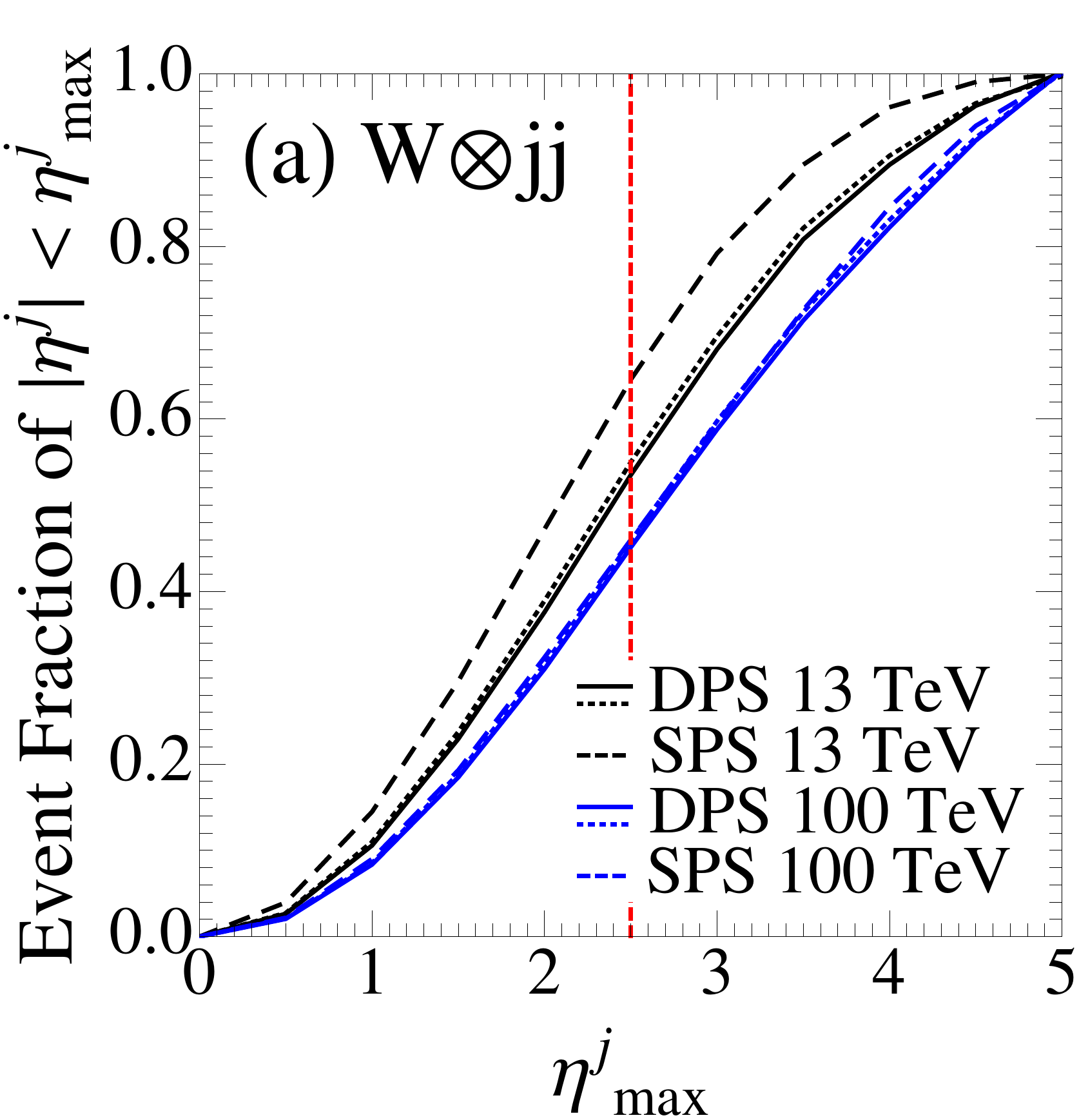}
    \includegraphics[scale=0.22]{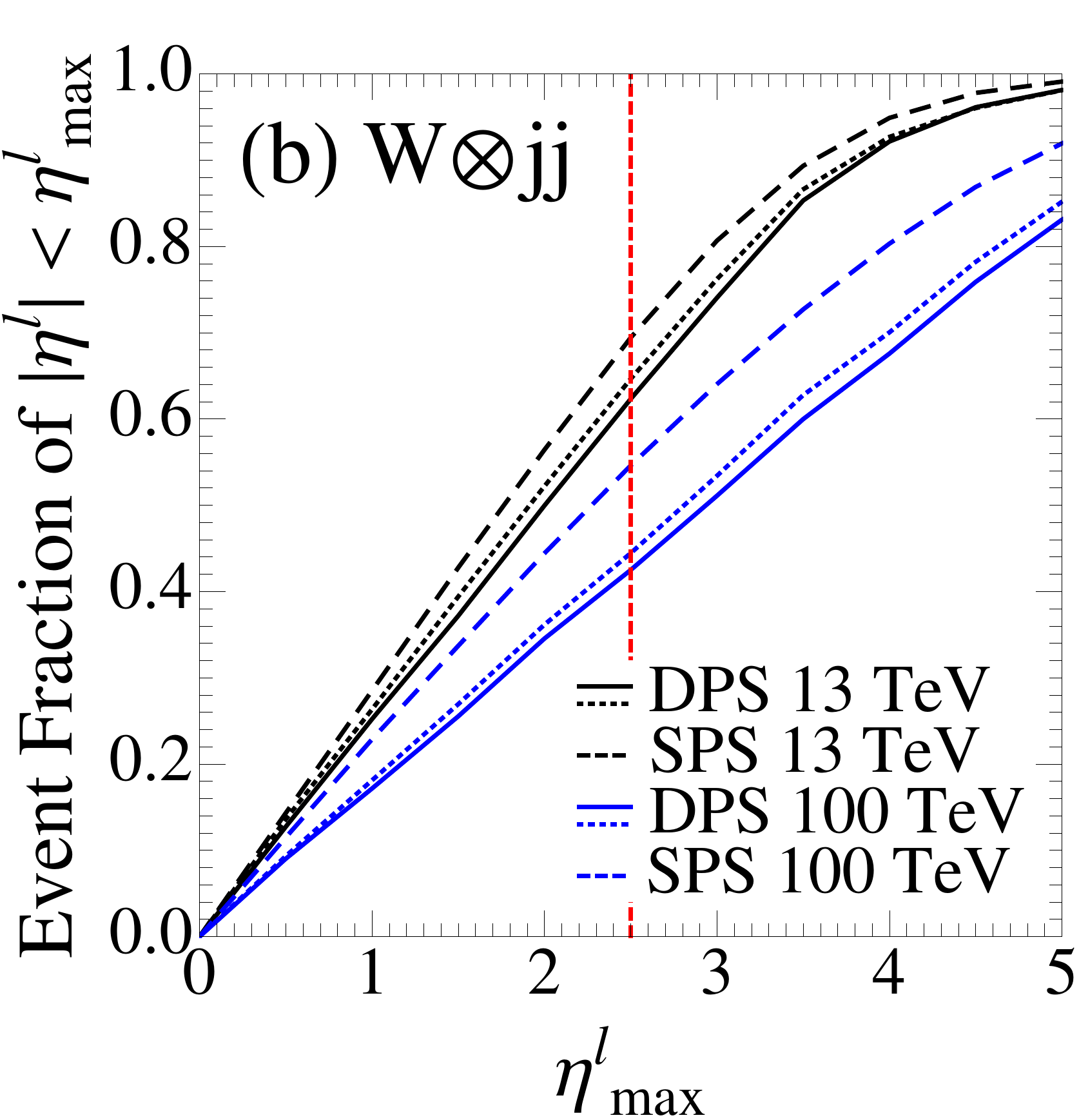}
    \includegraphics[scale=0.22]{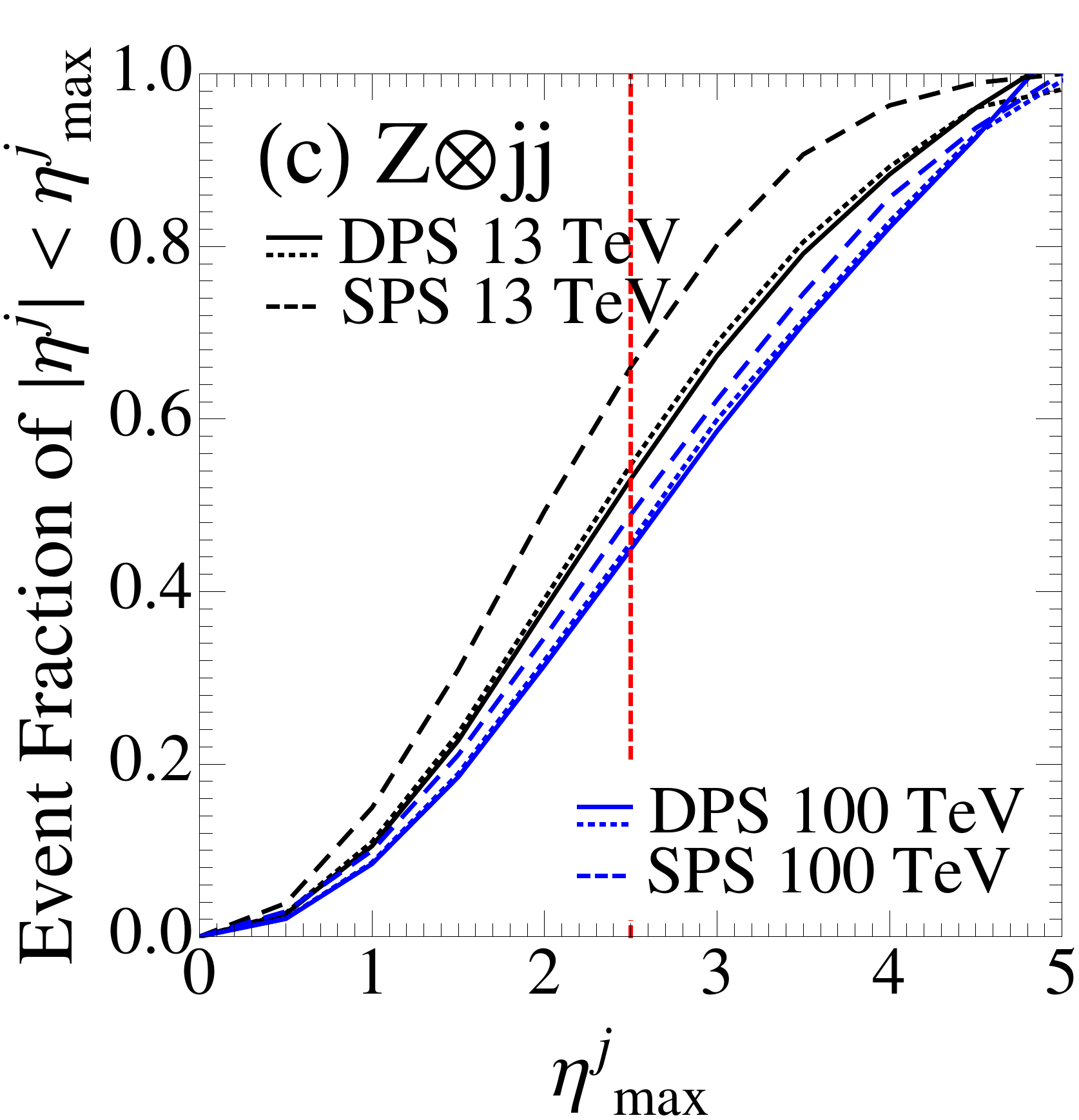}
    \includegraphics[scale=0.22]{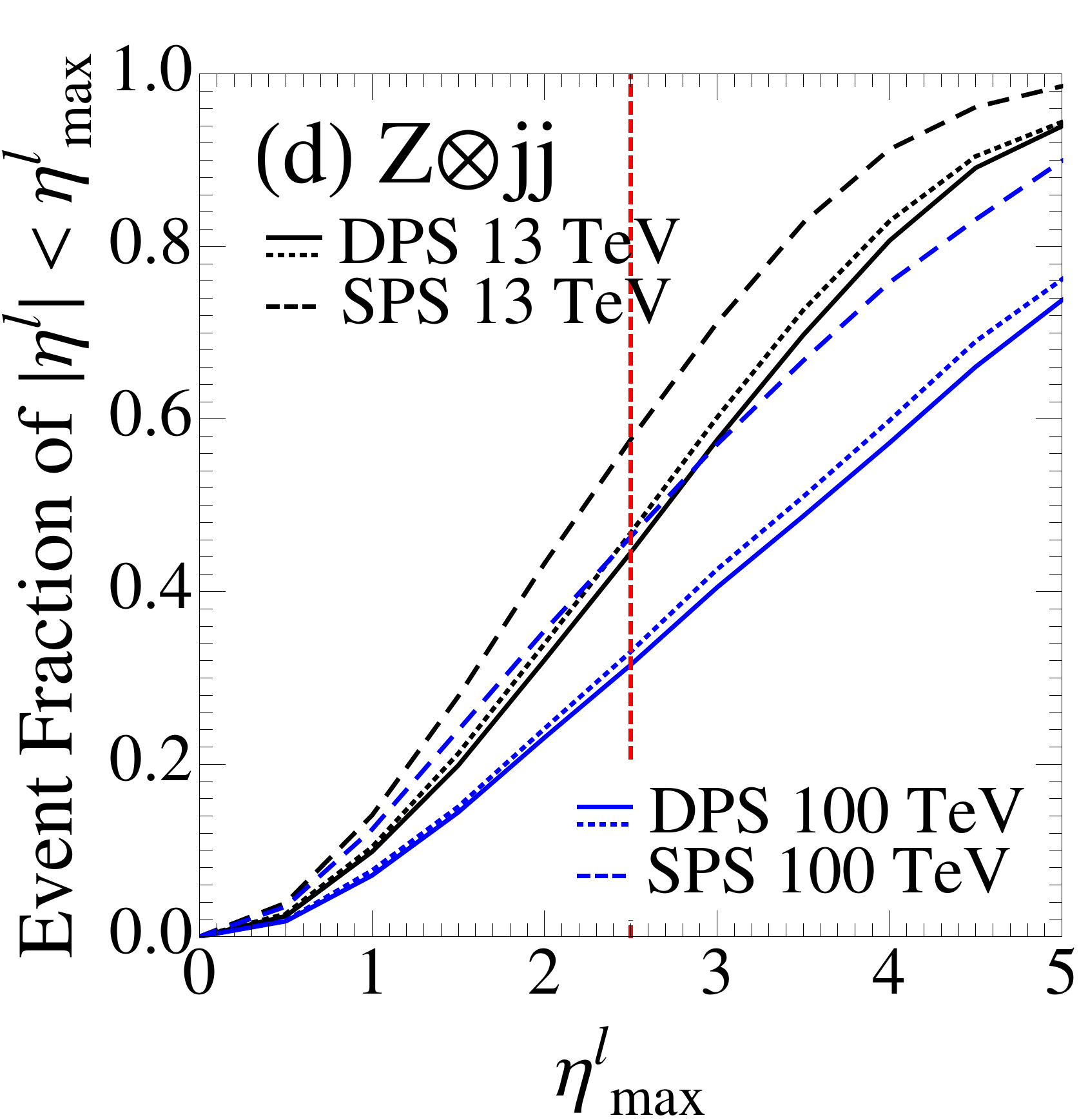}
    \includegraphics[scale=0.22]{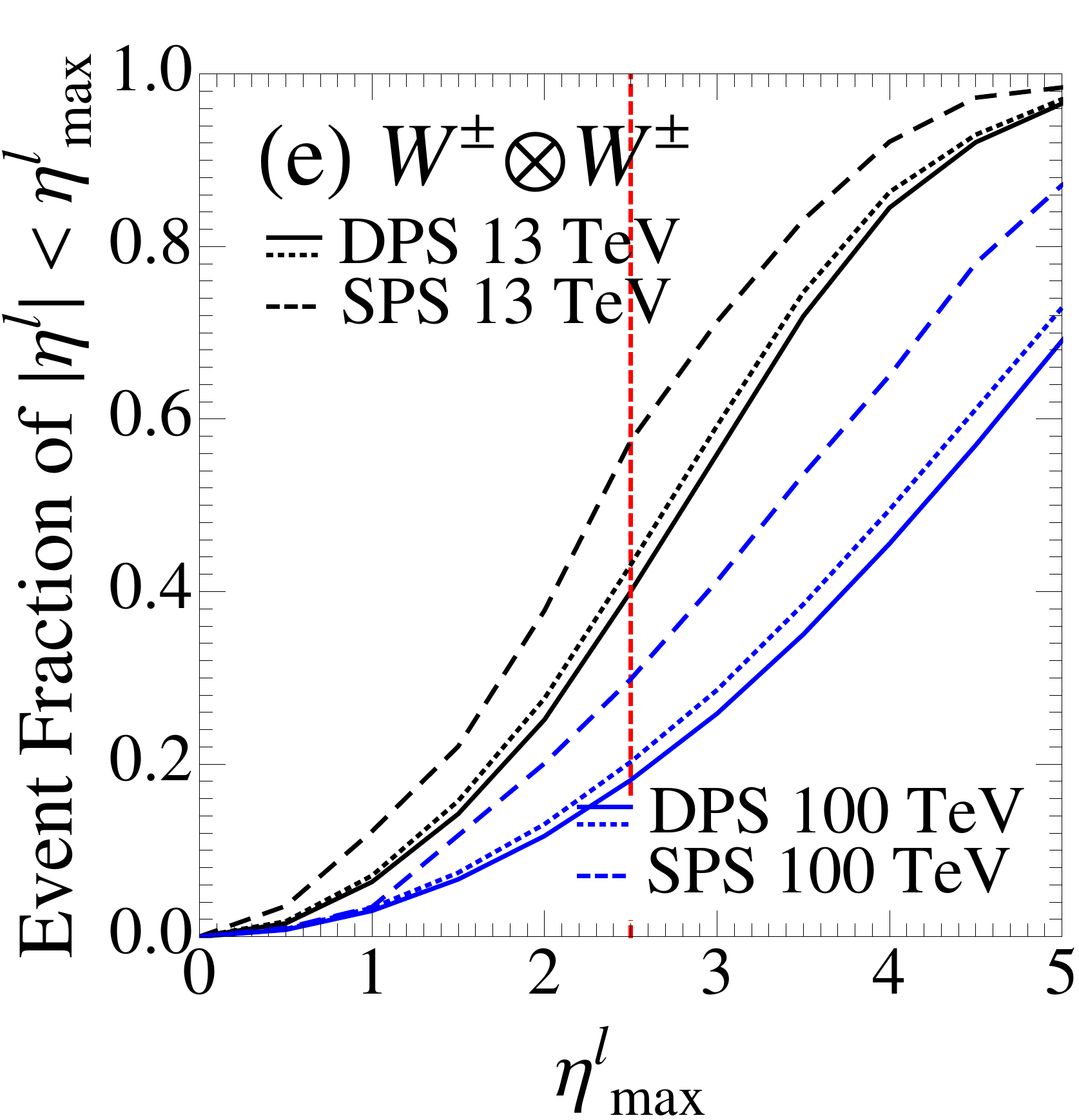}
\caption{Cross section fraction as a function of maximal pseudo-rapidity of jets and leptons: (a) $W\otimes jj$, (b) $Z\otimes jj$ and (c) same sign $W^\pm\otimes W^\pm$ process. The solid (dotted) curve represents the DPS production calculated with SF (GS09) dPDF, while the dashed curve represents the SPS channel, respectively. The cuts are listed in text.}
  \label{lepton_eta}
\end{figure}

Another important difference between the 13 TeV and the 100 TeV hadron colliders is the detector coverage of the pseudo-rapidity of final state particles. It has a significant impact on the kinematics cuts used to disentangle the signal out of the SM backgrounds. At the 13 TeV LHC, the detector can well detect jets or leptons in the centra region, say $|\eta|<2.5$ for leptons and $|\eta|<5$ for jets~\cite{Airapetian:391176, Bayatian:922757}. At the 100 TeV hadron colliders, the final state particles are often highly boosted to appear in the very forward region of the detector and thus exhibit large rapidities~\cite{Mangano:2017tke}. Below we examine the rapidity coverage of charged leptons and jets in the three DPS channels at the 13 TeV and 100 TeV hadron colliders. That guides us to decide which rapidity cut to be used at the 100 TeV colliders. Also, a comparison between the SPS and DPS processes is made.  

Figure~\ref{lepton_eta} shows the event fraction as a function of maximal pseudo-rapidity cut imposed on the charged leptons and jets in the final state of the three DPS channels and the corresponding SPS backgrounds. 
The event fraction of object $i$ is defined as 
\be
\frac{1}{\sigma_{\rm total}}\int_{-\eta^i_{\rm max}}^{\eta^i_{\rm max}}d\eta^i\dfrac{d\sigma}{d\eta^i},
\ee
where $\eta^i_{\rm max}$ is the maximal pseudo-rapidity cut imposed on the object $i$. In order to avoid the collinear divergence, we require all the jets in the $W\otimes jj$ and $Z\otimes jj$ channels to pass the selection cut as follows: 
\begin{align}
&p_T^j>25~{\rm GeV}{\rm ~at~the~13~TeV~LHC};\nn\\
&p_T^j>50~{\rm GeV}{\rm ~at~the~100~TeV~colliders}.\label{pt_thres}
\end{align}
We also veto the jets satisfying the above condition in the $W^\pm W^\pm jj$ SPS process. No $p_T$ cut is imposed on leptons.

Figures~\ref{lepton_eta}(a) and \ref{lepton_eta}(b) displays the event fraction of $\eta_{\rm max}$ of the jets and charged leptons in the $W\otimes jj$ channel, respectively. First, the SF (solid) and GS09 (dotted) dPDFs give rise to almost the same event fraction distribution. Second, at the 100 TeV  SppC/FCC-hh, both leptons and jets are distributed more in large $\eta$ ranges.  For example, there are less than $\sim 40\%$ of the jets lying in the range of $|\eta^\ell|<2.5$; see the intersection points of the red dashed vertical lines and the blue lines. In order to collect as many DPS events as possible, we have to cover a lager $\eta$ range at the 100 TeV collider. In the study we assume the lepton trigger covers the region $|\eta^\ell|<5$ at the SppC/FCC-hh~\cite{Mangano:2016jyj}.

\section{$W\otimes jj$ channel}\label{section:wjj}

\begin{figure}[h]
  \centering
    \includegraphics[scale=0.4]{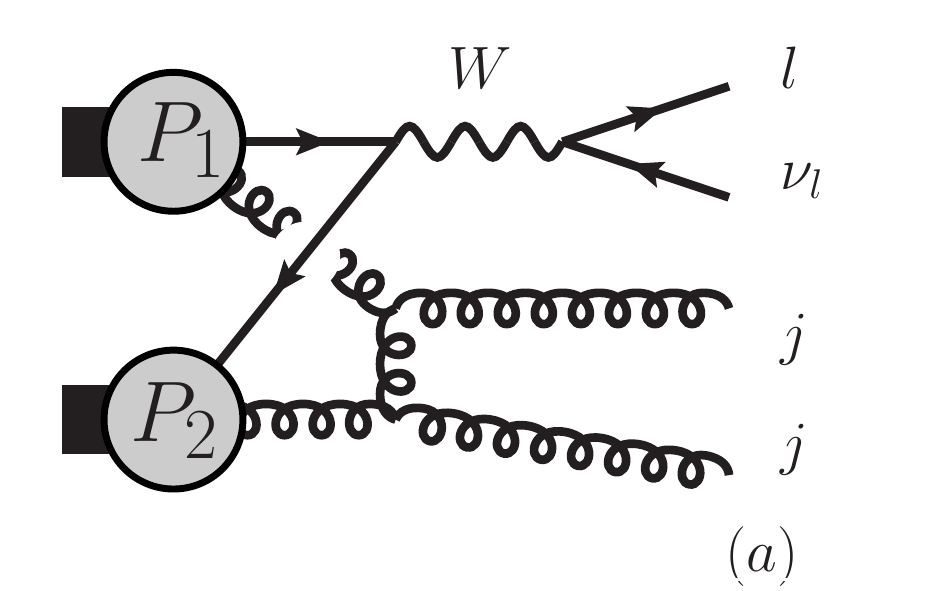}
    \includegraphics[scale=0.4]{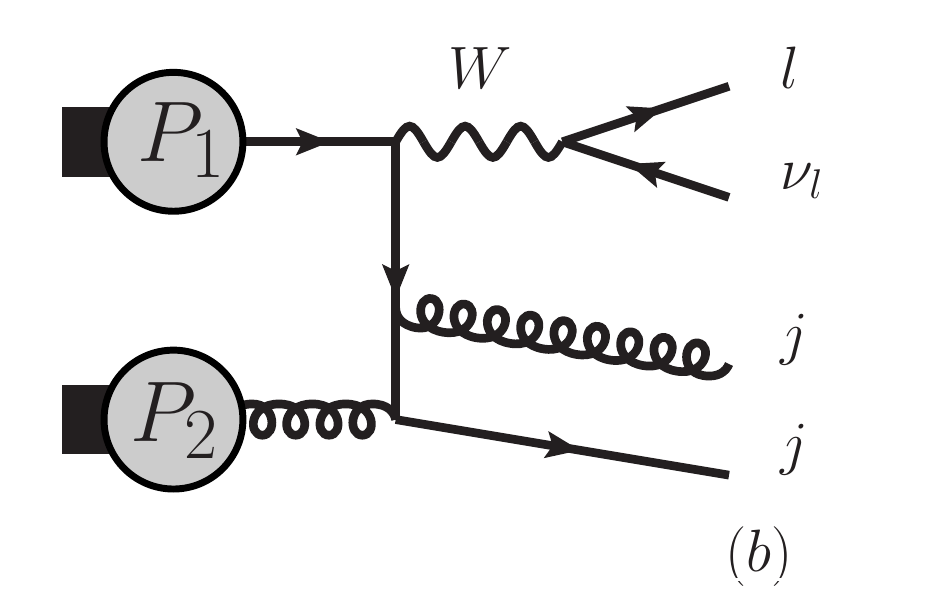}
  \caption{Typical Feynman diagrams of DPS $W\otimes jj$ (a) and SPS $Wjj$ (b).}
  \label{DPSandSPS_wjj} 
\end{figure}

Figure~\ref{DPSandSPS_wjj} shows the pictorial illustration of the DPS $W\otimes jj$ channel (a) and the SPS $Wjj$ background (b). The $W\otimes jj$ channel strikes a balance between event triggering and production rate. On one hand, the charged lepton from the $W$-boson decay provides a nice trigger of the signal events; on the other hand, the $jj$ subprocess gives rise to a large cross section. Therefore, the channel has been searched experimentally for a long time, e.g. by the CMS collaboration~\cite{Chatrchyan:2013xxa, Kumar:2016oyn} and by the ATLAS collaboration at the 7 TeV LHC~\cite{Aad:2013bjm}. Also, it has been studied theoretically both at the Tevatron~\cite{Godbole:1989ti} and the LHC~\cite{Berger:2011ep, Blok:2015afa}. In this section we first discuss various kinematic distributions and then make a hadron level simulation to explore the potential of measuring $\sigma_{\rm eff}$ at the 13~TeV LHC and 100 TeV  SppC/FCC-hh. 

\subsection{Kinematics distributions}

We generate the subprocess $pp\to W\to l\nu_l$ (merged with $pp\to Wj$) and the subprocess $pp\to jj$ (merged with $pp\to jjj$) with {\tt MadGraph 5}~\cite{Alwall:2014hca}, and then interface with {\tt Pythia 6} \cite{Sjostrand:2006za} and {\tt Delphes 3} \cite{deFavereau:2013fsa} for parton shower and detector simulations. When generating the signal in {\tt MadGraph}, we impose loose cuts on jets at the generator level as follows:
\be
p_T^j\geq 10~{\rm GeV},\quad |\eta^j|<5.
\ee
to avoid the collinear divergence in QCD radiations, while no cut is added to the leptons. We further demand a set of loose conditions on the reconstruction of jets and leptons in {\tt Delphes} package as follows:
\bea
&&p_T^\ell>10~\text{GeV},~|\eta^\ell|<2.5\;(5),\nn\\
&&p_T^j>20~\text{GeV},~|\eta^j|<5
\eea
at the 13 (100) TeV colliders, respectively.  
Next, we randomly combine the events of these two subprocesses to get the $W\otimes jj$ DPS events. At hadron colliders, massive particles are mainly produced near threshold, thus the partons participating the subprocess of $W$-boson production typically have a typical momentum fraction $\left<x\right>$ of the order of $\left<x\right>\sim m_W/\sqrt{s}\sim10^{-3}$ at the 13 TeV LHC and $\sim10^{-4}$ at the 100 TeV  SppC/FCC-hh; on the other hand, for the subprocess of $jj$ production, the momentum fraction depends on the $p^j_T$ cut, which is 10 GeV at the generator level, making $\left<x\right>\sim (10\text{ GeV})/\sqrt{s}\sim10^{-5}$ or less. As a result, the combined events can hardly break the PDF integration condition in Eq.~(\ref{DPSfactorize}). To wit, even though being combined randomly without any additional constraints, the DPS events will satisfy $x_1+x_2\le 1$ and $y_1+y_2\le 1$ automatically. The fact has been checked: we randomly combine $10^6$ events  and find that none of them breaks the above conditions. 

Additionally, we generate the DPS $W\otimes jj$ events at the parton-level with a homemade event generator which can handle two independent SPS processes simultaneously. We examine various parton-level distributions of final state particles and find good agreements with those distributions obtained by randomly combining two independent subprocesses generated by {\tt MadGraph}.

\begin{figure}
\centering
\includegraphics[scale=0.23]{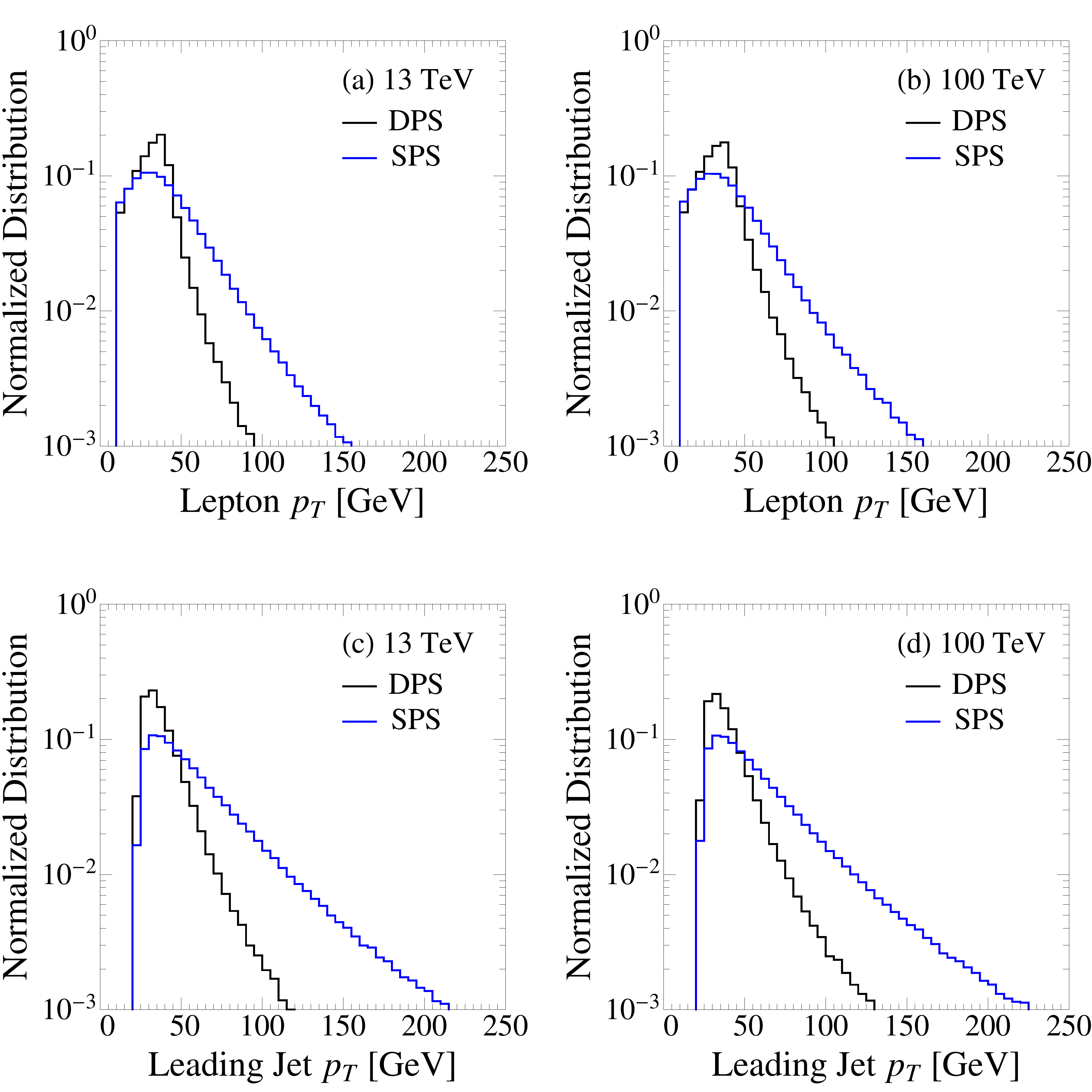}
\caption{The $p_T$ distribution of charged lepton (a, b) and leading jet (c, d) in the $W\otimes jj$ channel at the 13 TeV LHC and the 100 TeV  SppC/FCC-hh. }
\label{wjj_dist} 
\end{figure}

The DPS $W\otimes jj$ channel contains two independent hard subprocesses such that the final state particles build up two un-correlated subsystems. On the other hand, those final state particles of the dominant SM background channel, the SPS $Wjj$ production, are correlated. The difference can be used to discriminate the DPS channel from the SPS background. We plot the kinematic distributions for DPS and SPS events after {\tt Delphes} reconstruction. As shown in Figs.~\ref{wjj_dist}(a) and \ref{wjj_dist}(b), the $p_T$ distribution of the charged lepton in the DPS event (black curve) has a Jacobi peak around $m_W/2$ as the charged leptons are from an on-shell $W$-boson that exhibits small $p_T$~\cite{Balazs:1997xd,Cao:2004yy}. In the SPS background (blue curve), the two jets are produced in association with the $W$-boson. As we demand both jets carrying hard $p_T$'s, the $W$-boson exhibits a large $p_T$ to balance the two jets. It thus results in a harder $p_T$ distribution of charged leptons; see the blue curves in Figs.~\ref{wjj_dist}(a) and \ref{wjj_dist}(b).

Figures~\ref{wjj_dist}(c) and \ref{wjj_dist}(d) display the $p_T$ distributions of the leading $p_T$ jet. The jets in the SPS background are much harder than those jets in the DPS channel. The jet $p_T$ spectrum of the DPS channel peaks around the cut threshold specified in Eq.~(\ref{pt_thres}) and drops rapidly with $p_T$. On the contrary, the leading jet of the SPS channel tends to balance the $W$-boson such that it has a long tail in the large $p_T$ region. Therefore, in order to extract the DPS signal out of the SPS background, one should choose a relatively low $p_T$ cut to keep more events.

\begin{figure}
\centering
\includegraphics[scale=0.23]{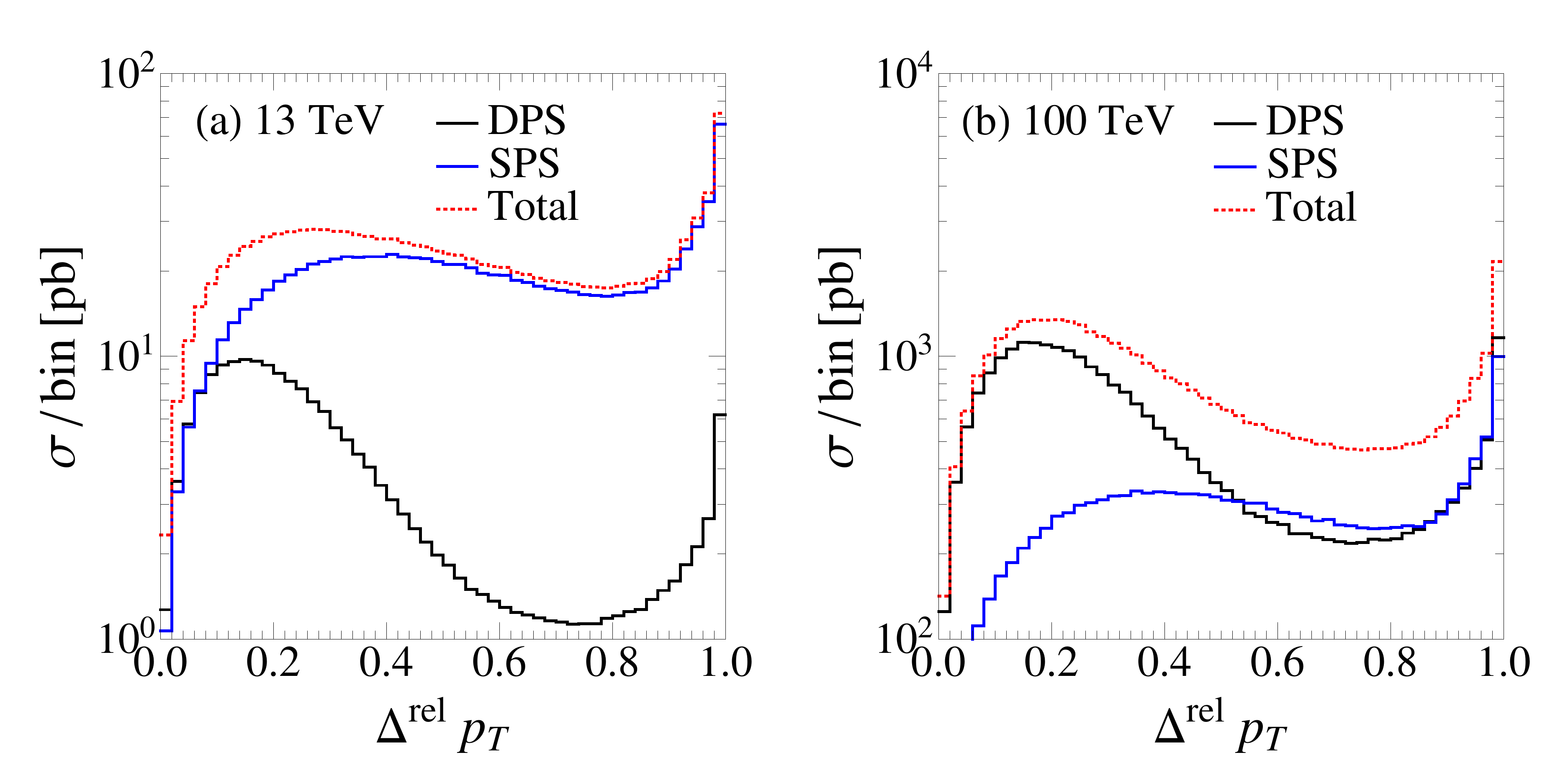}
\caption{$\Delta^{\text{rel}}p_T$ distribution for the $W\otimes jj$ channel at the 13 TeV LHC (a) and at the 100 TeV  SppC/FCC-hh (b). }
\label{wjj_DP}
\end{figure}

Besides the $p_T$ distributions, there are other optimal observables to distinguish the DPS channel form the SPS channel. The main idea is to make use of the fact that the DPS channel contains two (nearly) independent hard scatterings. For example, the $W$-boson and dijet production in the  $W\otimes jj$ DPS signal are independent, therefore, the dijet system exhibits a null transverse momentum at the leading order and develops a small transverse momentum after including the soft gluon resummation effects~\cite{Balazs:1997xd,Cao:2004yy}. The dijet system in the SPS has a large transverse momentum in order to balance the $W$-boson. The distinct difference in the $p_T$ distribution of the dijet system yields the following optimal observable~\cite{Chatrchyan:2013xxa}
\be
\Delta^{\text{rel}}p_T=\frac{|p_T(j_1,j_2)|}{|p_T(j_1)|+|p_T(j_2)|},
\ee
where $p_T(j_1,j_2)$ is the vector sum of $p_T(j_1)$ and $p_T(j_2)$. The observable denotes the relative $p_T$-balance of two tagged jets and tends to be $\sim 0$ for the DPS events. At the parton level, the $\Delta^{\text{rel}}p_T$ distribution should exhibit a sharp peak at $\Delta^{\text{rel}}p_T=0$. After parton shower and detector simulations, the sharp peak is smeared and shifted to $\Delta^{\text{rel}}p_T\sim 0.1$ due to soft/collinear radiation and acceptance cuts; see Fig.~\ref{wjj_DP}. The $\Delta^{\text{rel}}p_T$ distributions of both the DPS (black curve) and SPS (blue curve) channels have enhancements around $\Delta^{\text{rel}}p_T\sim 1$. It can be understood as follows. One factor is the collinear enhancement of QCD jets, i.e. two colored partons splitting from the same mother parton tend to have similar momentum and enhance $\Delta^{\text{rel}}p_T\sim1$. Another contribution arises from the so-called Jacobian enhancement. We define the ratio of $p_T$ magnitude of the two jets as 
\be
\lambda\equiv \frac{|p_T(j_2)|}{|p_T(j_1)|},
\ee
and obtain
\be
\Delta^{\text{rel}}p_T=\frac{\sqrt{1+\lambda^2+2\lambda\cos\Delta\phi_{jj}}}{1+\lambda},
\ee
where $\Delta\phi_{jj}$ is the azimuthal angle distance of the two jets. A simple algebra yields
\bea
&&\frac{d\sigma}{d\Delta^{\text{rel}}p_T}=\frac{d\sigma}{d\Delta\phi_{jj}}\frac{d\Delta\phi_{jj}}{d\Delta^{\text{rel}}p_T}\nn\\
&=&\frac{d\sigma}{d\Delta\phi_{jj}}\frac{2\Delta^{\text{rel}}p_T}{\sqrt{\left[(\Delta^{\text{rel}}p_T)^2-\bigg(\dfrac{1-\lambda}{1+\lambda}\bigg)^2\right]\Bigg[1-(\Delta^{\text{rel}}p_T)^2\Bigg]}}.\nn\\
\label{eq:jacobian}
\eea
The enhancement around $\Delta^{\text{rel}}p_T \sim 1$ stems from the Jacobian factor. The DPS channel is much less than the SPS channel at the 13 TeV LHC while at the 100 TeV hadron collider the DPS channel dominates.

\begin{figure}
  \centering 
  \includegraphics[scale=0.23]{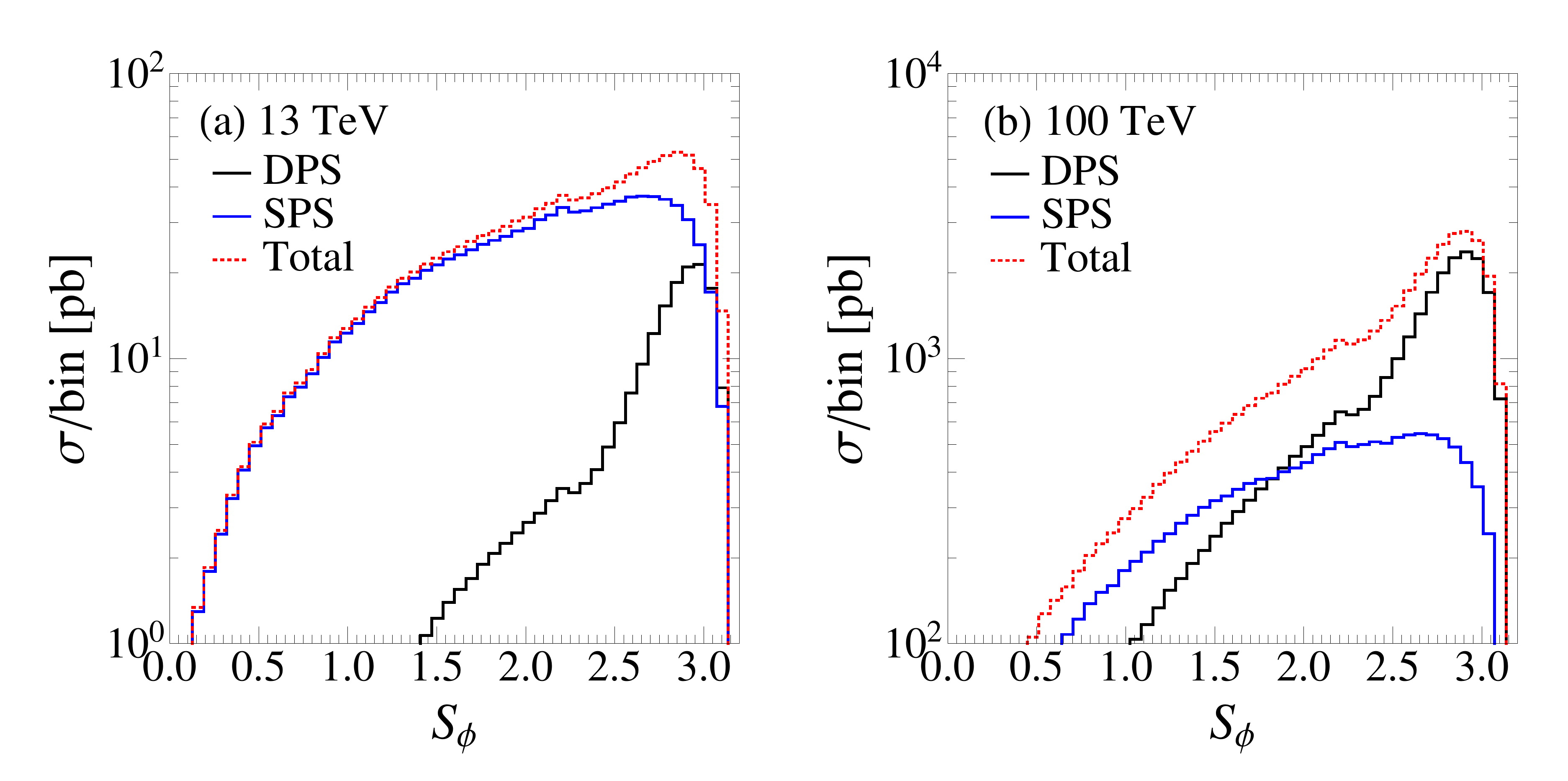} 
  \caption{$S_\phi$ distribution for the $W\otimes jj$ channel at the 13 TeV LHC (a) and at the 100 TeV  SppC/FCC-hh (b). }
  \label{wjj_SP} 
\end{figure}

Another optimal observable is the azimuthal angle correlation between the $W$ system ($\ell^\pm, \met$) and the dijet system ($j_1,j_2$), defined as
\be
S_\phi=\frac{1}{\sqrt{2}}\sqrt{\Delta\phi(\ell^\pm,\met)^2+\Delta\phi(j_1,j_2)^2},
\ee
where $\met$ denotes the missing transverse momentum generated by the invisible neutrinos from the $W$-boson decay. As $(\ell^\pm,\met)$ and $(j_1,j_2)$ are produced by two independent scatterings in the DPS channel, $S_\phi$ tends to be $\pi$. In fact, at parton level a sharp peak at $S_\phi=\pi$ will be observed, while the peak is smeared at the hadron level and the peak position is shifted to $S_\phi\sim 2.9$. See the black curves in Fig.~\ref{wjj_SP}. On the other hand, for the SPS channel, the final state particles are generally correlated and have a broader distribution; see the blue curves in Fig.~\ref{wjj_SP}. The difference in the $S_\phi$ distributions can be used to identify the DPS events.  

\subsection{Collider Simulation}

We are ready to investigate the potential of detecting the $W\otimes jj$ DPS channel in hadron collisions. 
The event topology of interest to us is one charged lepton, two hard jets and large $\met$. The major SM backgrounds are listed as follows:
i) the irreducible $Wjj$ background with a subsequent decay of $W^\pm \to \ell^\pm \nu$;
ii) the $Z/\gamma^* jj$ background wth $Z/\gamma^*\to \ell^+\ell^-$;
iii) the $t\bar t$ pair production with the top quarks decaying semi-leptonically or purely leptonically;
iv) the single top production (including $t$-channel, $s$-channel and $tW$-channel) with the top quark decaying leptonically. In the $tW$ background, we also consider the possibility of the associated $W$ boson decaying into a pair of leptons. The multijet background is shown to be less than 0.5\% at the 7~TeV LHC~\cite{Chatrchyan:2013xxa} and is ignored in our study.
The signal and backgrounds are generated and simulated using the programs mention above. For such weak gauge boson production process, the pile-up effect is expected to be small, and indeed, it has been shown to be negligible at the 7 TeV LHC for the DPS $W\otimes jj$ searches~\cite{Chatrchyan:2013xxa}. We ignore the pile-up contamination in our simulation hereafter. Following~\cite{Chatrchyan:2013xxa, Mangano:2016jyj}, we impose four {\it basic} kinematics cuts in sequence:
\begin{itemize}
\item[1.] exactly one charged lepton with $p_T^{\ell} \geq 35~{\rm GeV}$, $|\eta^{\ell}|\leq 2.5$ at the 13 TeV LHC and $|\eta^{\ell}|\le 4 $ at the 100 TeV  SppC/FCC-hh;
\item[2.] exactly two hard jets with $p_T^{j}\geq 25~{\rm GeV}$ and $|\eta^{j}|\leq 2.5 $ at the 13 TeV LHC while $p_T^{j}\geq 50~{\rm GeV}$ and $|\eta^{j}|\leq 5 $ at the 100 TeV  SppC/FCC-hh;
\item[3.] $\met \geq 30$ GeV;
\item[4.] $M_T>50~{\rm GeV}$. 
\end{itemize}
Here, $M_T$ denotes the transverse mass of the ($\ell^\pm$, $\met$) system, defined as 
\be
M_T\equiv \sqrt{2\cdot p^\ell_T\cdot \not{\!\!E}_T\cdot\Big(1-\cos\Delta\phi\Big)},
\ee
where $\phi$ denotes the open angle between the charged lepton and missing momentum in the transverse plane.

Among the four cuts listed above, the first cut (cut-1), the second cut (cut-2) and the third cut (cut-3) are meant to trigger the event. At the 100 TeV  SppC/FCC-hh, we impose a harder cut on the jet $p_T$ to suppress the QCD backgrounds. We also extend the lepton coverage to collect more signals. We adopt the same lepton $p_T$ cut and $\met$ cut at the 13 TeV and 100 TeV hadron colliders as both the lepton $p_T$ and $\met$ distributions of the DPS signal events have a unchanged Jacobian peak around $m_W/2$.

\begin{table}
\footnotesize
\caption{Cross sections (in the unit of picobarn) of the DPS $W\otimes jj$ process and the SM backgrounds at the 13 TeV LHC (top) and at the 100 TeV  SppC/FCC-hh (bottom). We choose $\sigma_{\rm eff}=15~{\rm mb}$ and impose he kinematic cuts listed in each column sequentially.
}
\label{table:wjj13}
\begin{tabular}{l|c|c|c|c|c} 
\hline
{\bf 13 TeV} & Gen. & Cut-1 & Cut-2 & Cut-3 & Cut-4 \\ \hline
DPS $W(\to l\nu_l)jj$ & 1138.94 & 248.76 & 53.85 & 36.35 & 35.07  \\ \hline
$W(\to l\nu_l)jj$ & 18591.50 & 3406.39 & 381.77 & 223.44 & 184.72 \\ \hline
$t\bar t$ (all decay modes) & 461.00 & 76.16 & 10.26 & 8.46 & 6.61 \\ \hline
$t(\to bl\nu_l)j$ & 36.58 & 13.44 & 5.78 & 4.09 & 3.36 \\ \hline
$tW$ (all decay modes) & 39.45 & 7.21 & 2.01 & 1.50 & 1.15 \\ \hline
$Z(\to ll)jj$ & 1904.81 & 513.83 & 83.60 & 8.31 & 4.72 \\ \hline
\hline
{\bf 100 TeV} \\ \hline
DPS $W(\to l\nu_l)jj$ & 128283 & 32860.7 & 1841.97 & 1259.76 & 1047.49 \\ \hline
$W(\to l\nu_l)jj$ & 189865 & 38856.1 & 2755.1 & 1869.18 & 1393.42 \\ \hline
$t\bar t$ (all decay modes) & 30675.9 & 5085.14 & 1248.42 & 1018.41 & 749.72 \\ \hline
$t(\to bl\nu_l)j$ & 915.11 & 343.69 & 111.81 & 81.42 & 64.56 \\ \hline
$tW$ (all decay modes) & 1934.79 & 364.42 & 114.52 & 89.82 & 63.21 \\ \hline
$Z(\to ll)jj$ & 14044.1 & 4756.51 & 552.36 & 117.81 & 69.98 \\ \hline
\end{tabular}
\end{table}

In the simulation, we choose $\sigma_{\rm eff}$ as the average value of current experimental results $\sigma_{\rm eff}=15~{\rm mb}$. When generating both the signal and background events in {\tt MadGraph}, we impose loose cuts on jets at the parton level as follows:
\be
p_T^j\geq 10~{\rm GeV},\quad |\eta^j|<5.
\ee
We then use {\tt Pythia} for parton shower and jet merging. 
The cross section (in the unit of picobarn) of the signal and background processes after {\tt Pythia} (denoted as ``Gen.") are summarized in the second column of Table~\ref{table:wjj13}. Next, we adapt the {\tt Delphes} for particle identification and then impose the four basic cuts. The last four columns in Table~\ref{table:wjj13} show the cross section after imposing the four selection cuts sequentially. The SPS $Wjj$ channel is the dominant background at the 13 TeV and 100 TeV colliders. It is about 5 times larger than the DPS signal at the 13 TeV LHC. The subleading background is from top quark pair production which is not important at the 13 TeV LHC. At the 100 TeV collider, owing to the dramatically enhanced productions of the dijet subprocess, the cross section of the DPS signal is comparable to the SPS $Wjj$ background. For the same reason, the top-quark pair background becomes important.   

As a matter of fact, the four kinematics cuts only select events that from $Wjj$ final state, but do not care about whether they are from DPS or SPS. So it is necessary to introduce the observables discussed last subsection to suppress SPS events and manifest DPS ones. A variable $f^{\rm DPS}$ is defined to quantitatively describe the fraction of the DPS signal event in the total event collected. It is defined as~\cite{Aad:2013bjm}
\be
f^{\rm DPS} = \frac{\sigma_{\rm DPS}}{\sigma^{\rm DPS}+\sum_i  \sigma^i_{\rm Background}},
\label{eq:fdps_def}
\ee
where summing over all the SM backgrounds are understood. As shown in Table~\ref{table:wjj13+}, $f^{\rm DPS}=15\%$ after imposing the four basic cuts at the 13 TeV LHC. The fraction increases dramatically to $f^{\rm DPS}=31\%$ at the 100 TeV collider. 

We can make use of the $\Delta^{\text{rel}}p_T$ and $S_\phi$ distributions to improve $f^{\rm DPS}$. In this study we impose a cut on either $\Delta^{\text{rel}}p_T$ or $S_\phi$ and do not require cuts on both, because cutting on one variables is good enough for identifying the DPS events. We demand either
\be
\Delta^{\text{rel}}p_T \le 0.2~,
\ee
or 
\be
S_\phi>2.5~.
\ee
The cross sections of the DPS signal and backgrounds after the optimal cut are presented in Table~\ref{table:wjj13+}; see the third row for the13 TeV LHC and the sixth row for a 100 TeV collider. It shows that either of the optimal cuts can efficiently suppress the SM backgrounds and increase the fraction $f^{\rm DPS}$ dramatically.  We notice that the $\Delta^{\text{rel}}p_T$ cut is slightly better than the $S_\phi$ cut. It yields $f^{\rm DPS}\sim 30\%$ at the 13 TeV LHC and $f^{\rm DPS}\sim 45\%$ at the 100 TeV colliders.

The numerical results of the DPS signal channel listed in Table~\ref{table:wjj13} and Table~\ref{table:wjj13+} are calculated with $\sigma_{\rm eff}=15~{\rm mb}$. Below, we study how well one can measure $\sigma_{\rm eff}$ from various distributions.

\begin{table}
\footnotesize
\caption{Cross sections (in the unit of picobarn) of the DPS signal and backgrounds after imposing optimal cuts at the 13 TeV LHC (top) and at the 100 TeV  SppC/FCC-hh (bottom). We choose $\sigma_{\rm eff}=15~{\rm mb}$.
}
\label{table:wjj13+}
\begin{tabular}{l|c|c|c|c|c|c||c} 
\hline
{\bf 13 TeV}& DPS $Wjj$ & $Wjj$ & $t\bar{t}$ & $tj$ & $tW$ & $Zjj$ & $f^{\rm DPS}$ \\ \hline
basic cuts & 35.07 & 184.72 & 6.61 & 3.36 & 1.15 & 4.72 & 0.15 \\ \hline
$S_\phi>2.5$ & 23.91 & 57.16 & 1.57 & 0.77 & 0.22 & 1.66 & 0.28 \\ 
or &&&&&&\\
$\Delta^{\text{rel}}p_T<0.2$ & 12.73 & 23.06 & 0.56 & 0.37 & 0.08 & 0.65 & 0.34 \\ \hline
\hline
{\bf 100 TeV} \\ \hline
basic cuts & 1047.49 & 1393.42 & 749.72 & 64.56 & 63.21 & 69.98 & 0.31 \\ \hline
$S_\phi>2.5$ & 479.09 & 417.96 & 173.23 & 19.99 & 11.36 & 28.68 & 0.42 \\ 
or &&&&&&\\
$\Delta^{\text{rel}}p_T<0.2$ & 312.43 & 263.56 & 92.63 & 12.84 & 6.18 & 14.33 & 0.45 \\ \hline
\end{tabular}
\end{table}

\subsection{Determining $\sigma_{\rm eff}$}

There are two methods to measure $\sigma_{\rm eff}$. One way is to extract $\sigma_{\rm eff}$ directly from the number of events collected experimentally. From the master formula given in Eq.~(\ref{DPSapprox}), one can derive $\sigma_{\rm eff}$ as following
\bea
\label{count_formula}
\sigma_{\rm eff}&=&\frac{\sigma^{\rm SPS}_{W}\times\sigma^{\rm SPS}_{jj}\times\epsilon\times\mathcal{L}}{N_{\rm DPS}}\nn\\
&=&\frac{\sigma^{\rm SPS}_{W}\times\sigma^{\rm SPS}_{jj}\times\epsilon\times\mathcal{L}}{N_{\rm OBS}-N_{\text{BKGD}}},
\eea
where $N_{\rm DPS}$ denotes the number of DPS signal events,  $N_{\rm BKGD}$ labels the number of events of the backgrounds predicted by the Monte Carlo simulation, and $N_{\rm OBS}$ denotes the total number of events which includes both signal and backgrounds events, i.e.
\be\label{pseudo_exp}
N_{\rm OBS}=N_{\text{DPS}}+N_{\text{BKGD}}.
\ee
$\mathcal{L}$ is the integrated luminosity, and $\epsilon$ represents the cut efficiency derived from theoretical simulation. In this study, we adopt the four basic cuts plus one optimal cut $\Delta^{\text{rel}}p_T<0.2$ to maximize the fraction $f^{\rm DPS}$. The cut efficiencies of the signal and background processes are derived from those numbers shown in Table~\ref{table:wjj13+}. 

\begin{figure}
  \centering
  \includegraphics[scale=0.28]{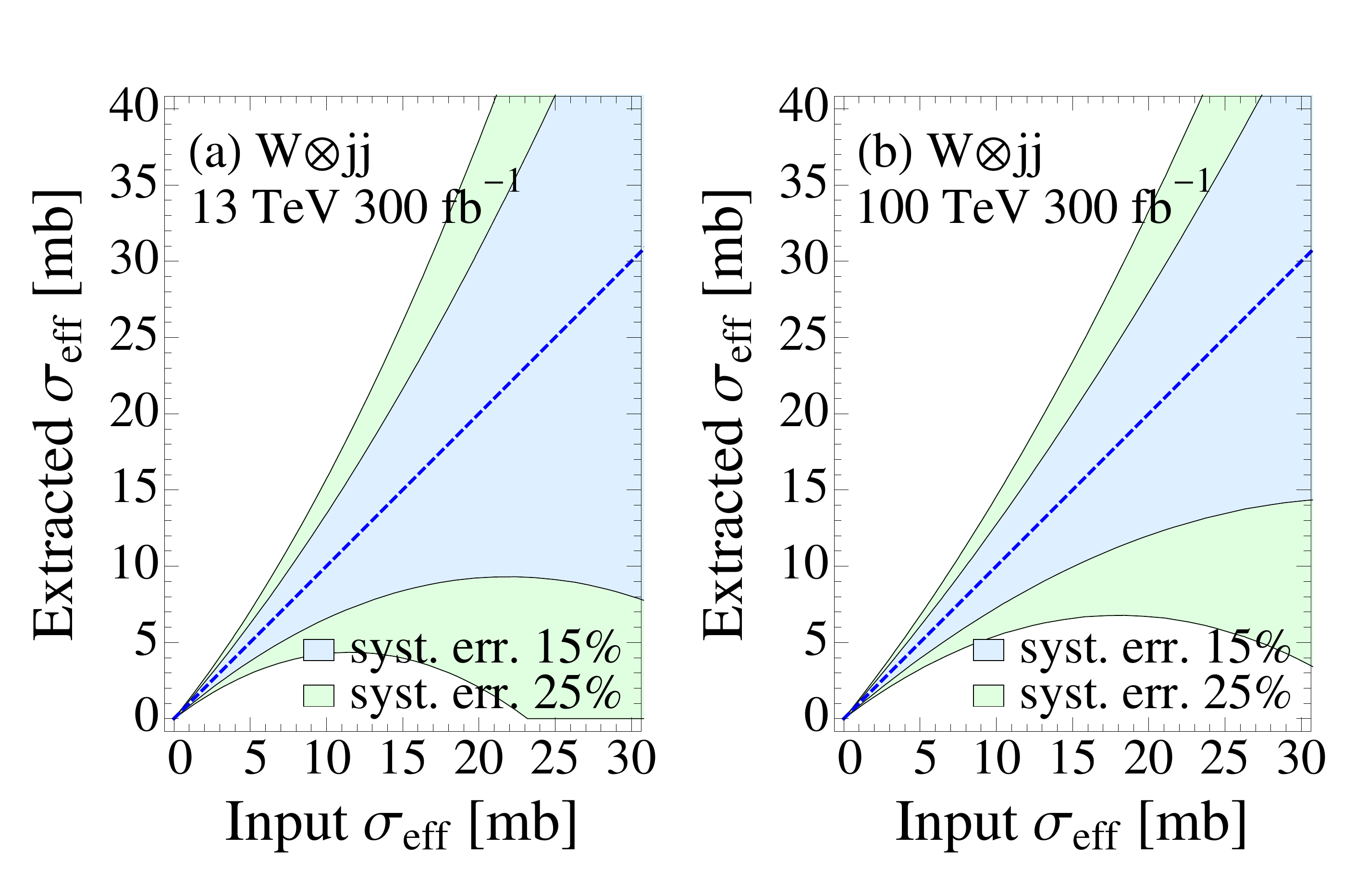}
  \caption{The projected accuracy of measuring $\sigma_{\rm eff}$ directly from event-number counting at the 13 TeV LHC (a) and 100 TeV  SppC/FCC-hh (b).}
  \label{wjj_count} 
\end{figure}

The uncertainty of measuring $\sigma_{\rm eff}$ arises from both statistical and systematics errors. In this study  the statistic error is assumed to obey a gaussian distribution, i.e. $\delta N_{\rm stat}=\sqrt{N_{\rm OBS}}$~. The systematic error can be known only after real experiments, and for a conservative estimation, we choose two benchmark uncertainties, $f_{\rm syst}=15\%$ and $25\%$, throughout this study. The total uncertainty of $N_{\rm OBS}$ is given by 
\be
\delta N=\sqrt{(\delta N_{\rm stat})^2+(\delta N_{\rm syst})^2},
\ee
with 
\be\label{error}
\delta N_{\rm stat}=\sqrt{N_{\text{OBS}}},\quad
\delta N_{\rm syst}=f_{\rm syst}\times N_{\text{OBS}}.
\ee
The accuracy of measuring $\sigma_{\rm eff}$ can be determined from Eq.~(\ref{count_formula}) for a given $\sigma_{\rm eff}$ input. 
Figure~\ref{wjj_count} displays the extracted $\sigma_{\rm eff}$ as a function of the input $\sigma_{\rm eff}$ at the 13 TeV LHC (a) and 100 TeV  SppC/FCC-hh (b) with an integrated luminosity of $300\text{ fb}^{-1}$. The blue bands denote the accuracy of $\sigma_{\rm eff}$ measurement with the choice of systematic uncertainty $f_{\rm syst}=15\%$ while the green bands label the case of $f_{\rm syst}=25\%$. 
Since the DPS rate is very large after imposing the basic and optimal cuts, the statistical uncertainty is well under control and the systematic uncertainty plays the leading role. Therefore, the uncertainty bands shown in Fig.~\ref{wjj_count} remain almost the same in the case of high luminosities.

It is obvious that the event counting method is not good for measuring $\sigma_{\rm eff}$. A better method to improve the accuracy of $\sigma_{\rm eff}$ measurement is to fit the $\Delta^{\rm rel}{p_T}$ and $S_\phi$ distributions~\cite{Chatrchyan:2013xxa, Kumar:2016oyn, Aad:2013bjm}. In the study we first generate the DPS events for a given $\sigma_{\rm eff}$ input and then combine the DPS events with the SPS backgrounds to get a pseudo-experiment data. Each bin of the distributions are allowed to exhibit a fluctuation of $\pm \delta N_i$ defined below. After that we rescale the DPS events as a function of $\sigma_{\rm eff}$ to fit the pseudo-data to obtain the accuracy of $\sigma_{\rm eff}$ measurement. In the fitting we define the $\chi^2$-function as
\be
\chi^2=\sum_{i=1}^N\frac{(N_i^{\rm th}-N_i^{\rm exp})^2}{(\delta N_i)^2},
\ee
where $N_i^{\rm exp}$ and $\delta N_i$ denotes the numbers of events and uncertainty in the $i$-th bin of the psesudo-data distribution, respectively, and $N_i^{\rm th}$ denotes the number of events in the $i$-th bin of the rescaled DPS distribution. The $\delta N_i$ contains both statistical and systematic uncertainties, defined similarly to Eq.~(\ref{error}) as   
\be
\delta N_i=\sqrt{N_i+f^2_{\rm syst} N_i^2}.
\ee
In the $W\otimes jj$ channel we divide the $\Delta^{\rm rel}{p_T}$ and $S_\phi$ distributions into 50 bins, i..e $N=50$. From the $\chi^2$ analysis we obtain the accuracy of $\sigma_{\rm eff}$ measurement at the $1\sigma$ confidence level for the two benchmark systematic uncertainties.

\begin{figure}
  \centering
  \includegraphics[scale=0.28]{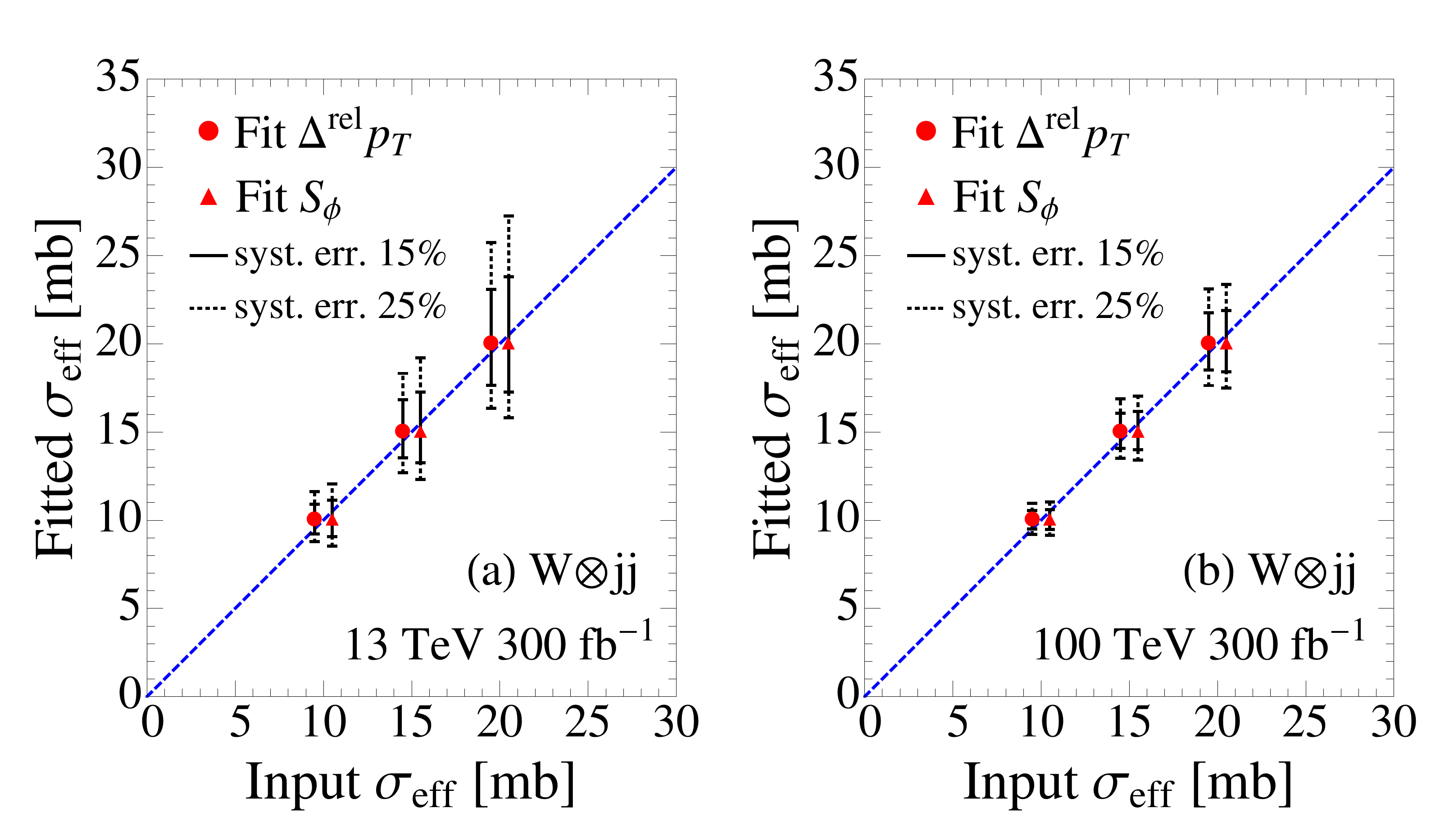}
  \caption{The fitted $\sigma_{\rm eff}$ as a function of the $\sigma_{\rm eff} $ input obtained in the $W\otimes jj$ channel at the 13 TeV LHC (a) and 100 TeV colliders (b) with an integrated luminosity of $300~{\rm fb}^{-1}$. }
  \label{wjj_fit} 
\end{figure}

We examine both the $\Delta^{\text{rel}}p_T$ and $S_\phi$ distributions at the 13~TeV LHC and 100~TeV colliders with an integrated luminosity of $300~{\rm fb}^{-1}$. Figure~\ref{wjj_fit} shows the expected $\sigma_{\rm eff}$'s versus the input values. The input values of $\sigma_{\rm eff}$ are chosen to be 10~mb, 15~mb and 20~mb. The red-circle symbol denotes the $\sigma^{\rm fit}_{\rm eff}$ obtained in fitting the $\Delta^{\text{rel}}p_T$ distribution while the red-triangle symbol labels the one obtained from the $S_\phi$ distribution. It turns out that one can get a better measurement of $\sigma_{\rm eff}$ in the $\Delta^{\text{rel}}p_T$ distribution.

The $\sigma^{\rm fit}_{\rm eff}$'s obtained from the $\Delta^{\text{rel}}p_T$ distribution are listed as follows:
\begin{itemize}[leftmargin=*]
\item $\sigma^{\rm in}_{\rm eff}=10~{\rm mb}$
\begin{align}
&\sigma^{\rm fit}_{\rm eff}=10^{+0.91~(+9.1\%)}_{-0.77~(-7.7\%)}, ~10^{+1.62~(+16.2\%)}_{-1.22~(-12.2\%)},&& {\rm ~13~TeV},\nn\\
& \sigma^{\rm fit}_{\rm eff}=10^{+0.55~(+5.5\%)}_{-0.50~(-5.5\%)},~10^{+0.95~(+9.5\%)}_{-0.80~(-8.0\%)},&& {\rm 100~TeV};\nn
\end{align}
\item $\sigma^{\rm in}_{\rm eff}=15~{\rm mb}$ 
\begin{align}
& \sigma^{\rm fit}_{\rm eff}=15^{+1.83~(+12.2\%)}_{-1.47~(~-9.8\%)}, ~15^{+3.32~(+22.1\%)}_{-2.30~(-15.3\%)}&& {\rm ~13~TeV},\nn\\
& \sigma^{\rm fit}_{\rm eff}=15^{+1.07~(+7.1\%)}_{-0.94~(-6.3\%)},~15^{+1.87~(+12.5\%)}_{-1.50~(-10.0\%)}, && {\rm 100~TeV};\nn
\end{align}
\item $\sigma^{\rm in}_{\rm eff}=20~{\rm mb}$
\begin{align}
& \sigma^{\rm fit}_{\rm eff}=20^{+3.09~(+15.5\%)}_{-2.36~(-11.8\%)}, ~20^{+5.74~(+28.7\%)}_{-3.65~(-18.3\%)},&& {\rm ~13~TeV},\nn\\
& \sigma^{\rm fit}_{\rm eff}=20^{+1.76~(+8.8\%)}_{-1.49~(-7.5\%)}, ~20^{+3.11~(+15.6\%)}_{-2.37~(-11.9\%)}, && {\rm 100~TeV},\nn
\end{align}
\end{itemize}
where the first value of $\sigma^{\rm fit}_{\rm eff}$ is for $f_{\rm syst}=15\%$ while the second value for $f_{\rm syst}=25\%$. The superscript and subscript denotes the upper and lower error at the $1\sigma$ confidential level, respectively. The percentage shown in the superscripts and subscripts denotes the percentage of the error relative to the mean fitting value of $\sigma^{\rm fit}_{\rm eff}$. 
The asymmetric errors is owing to the inverse relation between $N_i^{\rm th}$ and $\sigma_{\rm eff}$. If we fit $1/\sigma_{\rm eff}$ rather than $\sigma_{\rm eff}$, then we end up with symmetric errors.

We emphasize that, owing to the fact that the systematic errors dominate over the statistical errors, increasing luminosity does not significantly improve the accuracy of $\sigma_{\rm eff}$ measurement. Of course, accumulating more data helps with reducing the systematic errors, but on the assumption of fixed systematic uncertainty as we made, those uncertainties of $\sigma^{\rm fit}_{\rm eff}$ shown in Fig.~\ref{wjj_fit} remain almost the same for a higher luminosity. On the other hand, increasing colliding energy will greatly reduce the uncertainties of $\sigma_{\rm eff}$ measurements. The rate of the DPS channel increases dramatically with colliding energy such that the DPS channel dominates over the SM background. That enables us to reach a better precision of $\sigma^{\rm fit}_{\rm eff}$. 

Two methods of measuring $\sigma_{\rm eff}$ are presented above; one is based on event counting, the other is based on fitting the characteristic kinematics distributions of the DPS optimal observables. The fitting method works much better than the event counting method in measuring $\sigma_{\rm eff}$. Therefore, we adopt the fitting method hereafter.

\section{The $Z\otimes jj$ Channel}
\label{section:zjj}

Now consider another interesting DPS channel, the $Z\otimes jj$ process.  The channel also has advantages of clear event triggering and large production rate. A parton level analysis of the MPI contribution to the $Z\otimes$jets final states has been carried out in Ref.~\cite{Maina:2010vh} in which three colliding energies (8~TeV, 10~TeV and 14~TeV) are studied. A dynamical approach to such final state within the {\tt Pythia} event generator is studied in Ref.~\cite{Blok:2015afa}. In this work we present a hadron level study in hadron collisions.

\begin{figure}[h]
  \centering
    \includegraphics[scale=0.4]{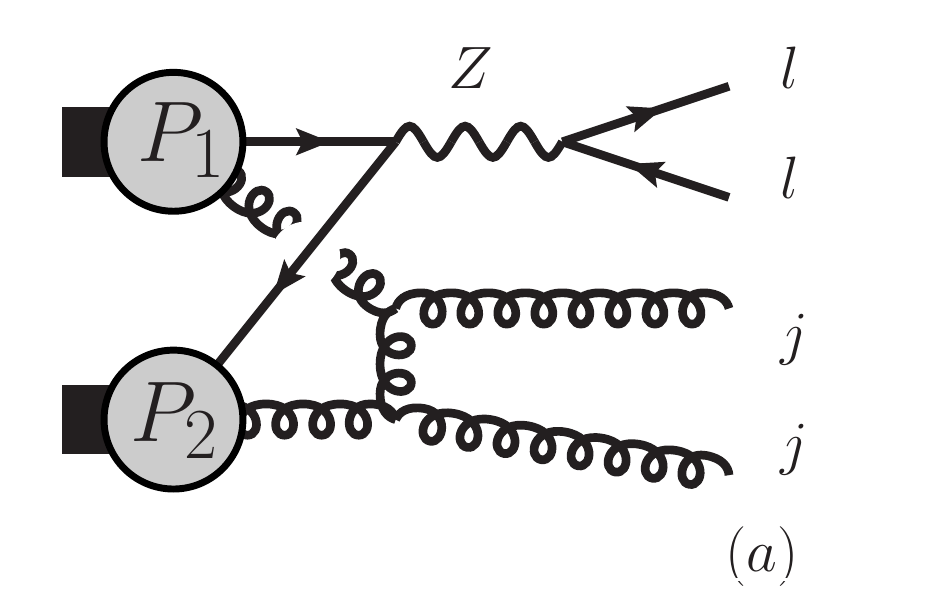}
    \includegraphics[scale=0.4]{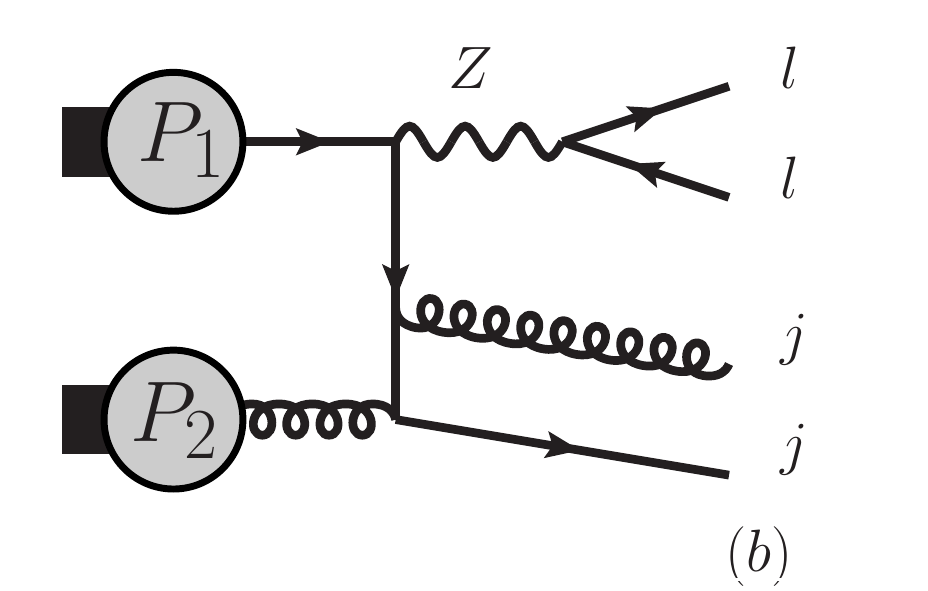}
  \caption{Representative Feynman diagrams of the $Z\otimes jj$ DPS channel (a) and the $Zjj$  SPS background (b).}
  \label{DPSandSPS_zjj} 
\end{figure}

\subsection{Kinematics distributions}

\begin{figure}
\centering
\includegraphics[scale=0.23]{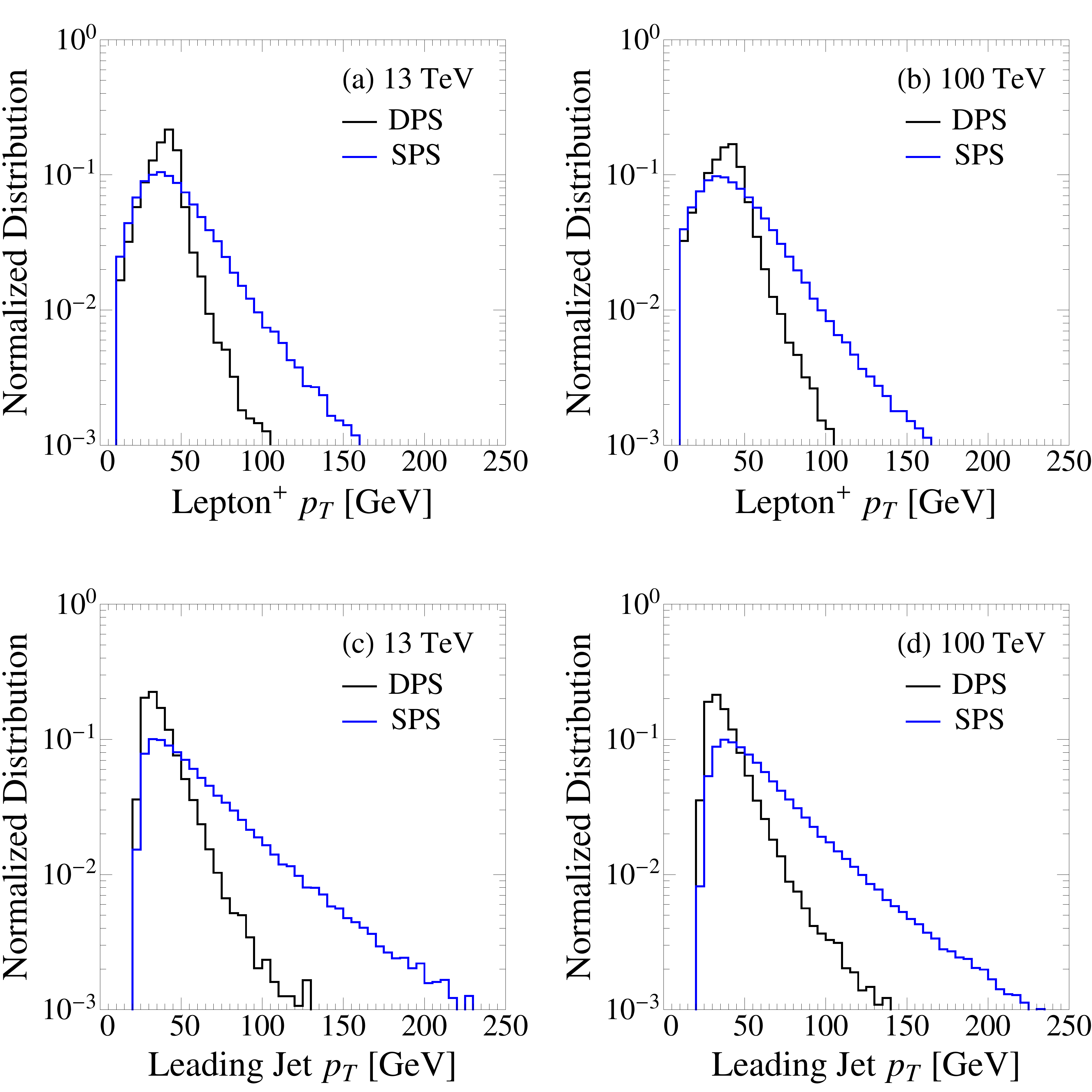}   
\caption{The $p_T$ distributions of the leading jet (a, b) and of the leading charged leptons (c, d) for the $Z\otimes jj$ channel at the 13 TeV LHC and the 100 TeV SppC/FCC-hh. }
\label{zjj_dist} 
\end{figure}

The pictorial illustration of the $Z\otimes jj$ channel and the SPS $Zjj$ background are plotted Fig.~\ref{DPSandSPS_zjj}. The $Z\otimes jj$ events are combined from two sets of hadron level event files of the SPS subprocess $Z(\to \ell^+\ell^-)$ (merged with $Zj$) and the $jj$ production  (merged with $jjj$). Similar to the $W\otimes jj$ channel, the feature of two independent subprocesses gives rise to characteristic kinematics distributions which can be used to distinguish between the DPS signal and the backgrounds.

Figure~\ref{zjj_dist} displays the transverse momentum distributions of of the leading-$p_T$ jet (a, b) and of the leading-$p_T$ charged leptons (c, d) at the 13 TeV LHC and the 100 TeV  colliders, after {\tt Delphes} reconstruction. Similar to the case of $W\otimes jj$ channel, the charged lepton $p_T$ distribution of the $Z\otimes jj$ events exhibits a Jacobi peak at $\sim m_Z/2$ while the distribution of the $Zjj$ SPS events tends to have a long tail towards the large $p_T$ region. The jet $p_T$ distribution of the $Z\otimes jj$ events peaks around the {\tt Delphes} reconstruction threshold $p_T=20~{\rm GeV}$ and drops rapidly. On the other hand, in order to balance the on-shell $Z$ boson, the $p_T$ distribution of the leading jets in the  $Zjj$ SPS events has a long tail in large $p_T$ region. Thus, we can impose a hard $p_T$ cut on the jet and a loose cut on the charged lepton to retain more DPS events.

\begin{figure}
  \centering
  \includegraphics[scale=0.23]{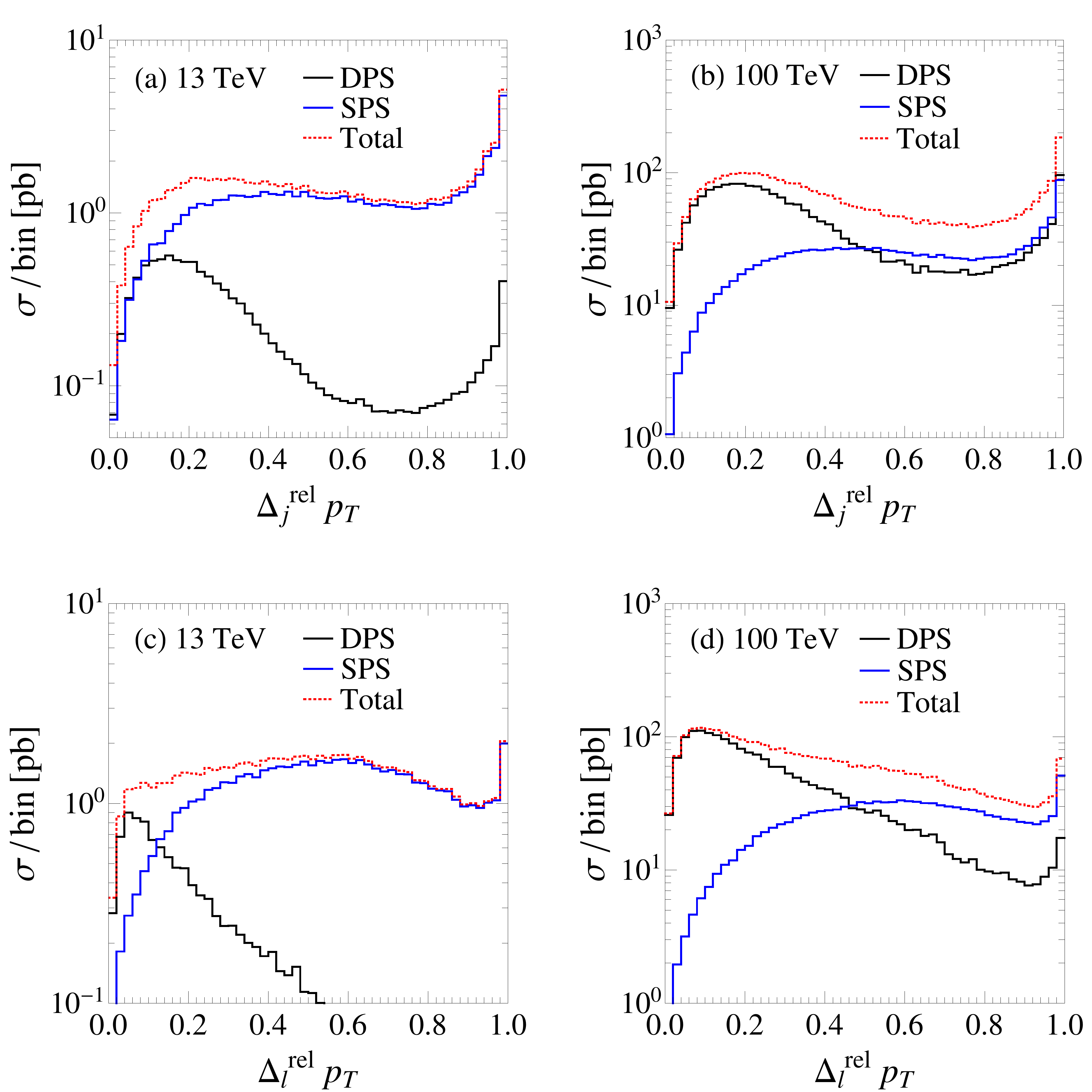}
  \caption{$\Delta^{\text{rel}}_j p_T$ and $\Delta^{\text{rel}}_\ell p_T$ distributions for the $Z\otimes jj$ channel at the 13 TeV (a, c) and 100 TeV colliders (b, d).}
  \label{zjj_DP} 
\end{figure}

Consider the optimal distributions to discriminate the DPS signal from the SPS backgrounds. Following the study of the $W\otimes jj$ channel, we define a relative $p_T$ balance of two jets as  
\be
\Delta^{\text{rel}}_jp_T=\frac{|p_T(j_1,j_2)|}{|p_T(j_1)|+|p_T(j_2)|}.
\ee
Figures~\ref{zjj_DP}(a) and \ref{zjj_DP}(b) display the $\Delta^{\text{rel}}_jp_T$ distributions of the DPS signal (black curve) and the $Zjj$ background (blue). Note that the  $\Delta^{\text{rel}}_jp_T$ distributions of the $Z\otimes jj$ channel are quite alike in shape to those distributions of the $W\otimes jj$ channel. It is no surprise as the kinematics of the two jets is identical in the both DPS channels. The peaks around $\Delta^{\text{rel}}_jp_T\sim 1$ are due to the Jacobian factor explained in Eq.~(\ref{eq:jacobian}). 

One advantage of the $Z\otimes jj$ channel is that one has full information of the two charged leptons from the $Z$ boson decay. That enables us to define a relative $p_T$ balance of two charged leptons as following:
\bea
&&\Delta^{\text{rel}}_\ell p_T=\frac{|p_T(\ell^+,\ell^-)|}{|p_T(\ell^+)|+|p_T(\ell^-)|}
\approx \frac{|p_T(Z)|}{|p_T(\ell^+)|+|p_T(\ell^-)|},\nn\\
&&
\eea
and plot the $\Delta^{\text{rel}}_\ell p_T$ distributions in Figs.~\ref{zjj_DP}(c) and \ref{zjj_DP}(d). In the both signal and background channels, most charged leptons are populated in the region of $p_T\sim m_Z/2$ such that the value of the denominator of $\Delta^{\text{rel}}_\ell p_T$ is around 90~GeV. For the DPS signal, $p_T(Z) \sim 0$, thus rendering the $\Delta^{\text{rel}}_\ell p_T$ distribution peaking around 0; see the black curves. For the $Zjj$ SPS background, the $Z$ boson, as balanced by the two hard jets, tends to have a hard $p_T$. That renders the $\Delta^{\text{rel}}_\ell p_T$ distributions of the $Zjj$ background peak around  $0.4\sim 0.6$~.

\begin{figure}
  \centering
  \includegraphics[scale=0.23]{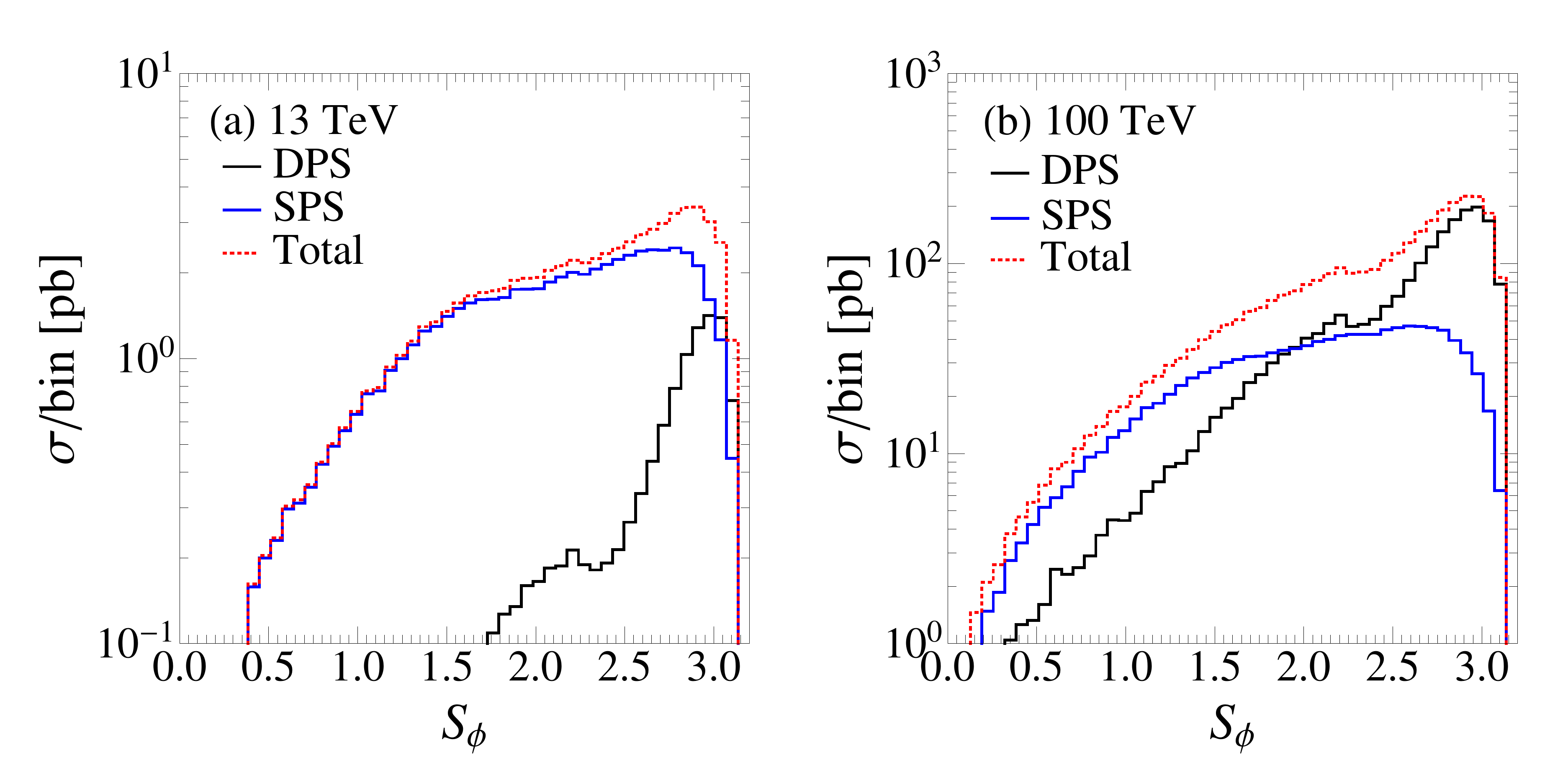}
  \caption{$S_\phi$ distribution for the $Z\otimes jj$ channel at the 13 TeV LHC (a) and 100 TeV colliders (b). }
  \label{zjj_SP} 
\end{figure}

The third optimal observable is the azimuthal angle correlation of the $Z$ system and $jj$ system, defined as
\be
S_\phi\equiv\frac{1}{\sqrt{2}}\sqrt{\Delta\phi(\ell^+,\ell^-)^2+\Delta\phi(j_1,j_2)^2}.
\ee
We plot the $S_\phi$ distributions in Fig.~\ref{zjj_SP} at the 13~TeV (a) and 100~TeV colliders (b). For the DPS channel, the two jets fly away almost back-to-back, i.e.  $\Delta \phi (j_1,j_2)\sim \pi$. Similarly, $\Delta\phi(\ell^+,\ell^-)\sim \pi$. Therefore, the $S_\phi$ distribution of the DPS signal peaks around 3; see the black curves. On the other hand, the two jets in the background events tend to move parallel such that $\Delta\phi(j_1,j_2)\sim 0$. That yields $S_\phi\sim 2.1-2.5$ in the background; see the blue curves.

\subsection{Collider Simulation}

The event topology of the $Z\otimes jj$ signal is two charged leptons with opposite charges and two hard jets. The main SPS backgrounds are listed as follows:
i)  the irreducible $Zjj$ background with $Z\to \ell^+\ell^-$;
ii) the $WZ$ pair production with $W\to q\bar{q}^\prime$ and $Z\to \ell^+\ell^-$;
iii) the $t\bar t$ pair production with the top-quark pair decaying either semi-leptonically or leptonically;
iv) the $tW$ single-top production with $t\to b\ell^\pm\nu$ and $W^\pm\to \ell^\pm\nu$.
Following Refs.~\cite{ATLAS-CONF-2016-046, Mangano:2016jyj}, we impose four basic kinematics cuts in sequence:
\begin{itemize}
\item[1.] exactly two opposite charged leptons with $p_T^{\ell}\ge25$ GeV, $|\eta^{\ell}|\le2.5$ at the 13~TeV LHC and $|\eta^{\ell}|\le 4$ at the 100~TeV  SppC/FCC-hh;
\item[2.] exactly two hard jets with $p_T^{j}\ge 30$ GeV and $|\eta^{j}|\le2.5$ at the 13 TeV LHC while $p_T^{j}\ge 50$ GeV and $|\eta^{j}|\le 5 $ at the 100~TeV  SppC/FCC-hh;
\item[3.] $\met\le 30$ GeV;
\item[4.] $M_{\ell\ell}\in[71,111]$ GeV. 
\end{itemize}
Similar to the case of $W\otimes jj$ channel, we enlarge the jet $p_T^j$ cut and the lepton $|\eta^\ell|$ cut to cover more events at the 100~TeV colliders. The third cut aims at reducing the backgrounds involving $W$ bosons, e.g. the $t\bar{t}$ and $tW$ backgrounds. The fourth cut requires that the invariant mass of the two charged leptons lies within the mass window of $Z$ boson. 

We choose the input value of $\sigma_{\rm eff}=15$ mb in our simulation. After generating both the signal and background events in {\tt MadGraph} with $p_T^j\geq 10~{\rm GeV}$ and $|\eta^j|<5$, we pass them to {\tt Pythia} for parton shower and merging. The cross section (in the unit of picobarn) of the signal and background processes after {\tt Pythia} (denoted as ``Gen.") are summarized in the second column of Table~\ref{table:zjj13}. Next, we use {\tt Delphes} for particle identifications and then impose the four kinematics cuts. The last four columns in Table~\ref{table:zjj13} show the cross section after imposing the four selection cuts sequentially.

\begin{table}
\footnotesize
\caption{Cross sections (in the unit of picobarn) of the DPS $Z\otimes jj$ process and the SM backgrounds at the 13 TeV LHC (top) and at the 100 TeV  SppC/FCC-hh (bottom). We choose $\sigma_{\rm eff}=15~{\rm mb}$ and impose the kinematic cuts listed in each column sequentially.}
\label{table:zjj13}
\begin{tabular}{l|c|c|c|c|c} 
\hline
 {\bf 13 TeV} & Gen. & Cut-1 & Cut-2 & Cut-3 & Cut-4 \\ \hline
DPS $Z(\to \ell^+\ell^-)jj$ & 108.92 & 28.10 & 3.60 & 3.46 & 3.40  \\ \hline
$Z(\to \ell^+\ell^-)jj$ & 1904.81 & 336.35 & 24.41 & 22.74 & 22.29 \\ \hline
$W(\to jj)Z(\to \ell^+\ell^-)$ & 1.623 & 0.45 & 0.14 & 0.13 & 0.12 \\ \hline
$t\bar t$ (all decay modes) & 461.00 & 6.52 & 2.64 & 0.40 & 0.12 \\ \hline
$tW$ (all decay modes) & 39.45 & 0.68 & 0.16 & 0.03 & 0.01 \\ \hline
\hline
{\bf 100 TeV} \\ \hline
DPS $Z(\to \ell^+\ell^-)jj$ & 13376.2 & 3974.08 & 175.71 & 139.70 & 137.21  \\ \hline
$Z(\to \ell^+\ell^-)jj$ & 14044.1 & 2673.22 & 240.68 & 194.4 & 190.34 \\ \hline
$W(\to jj)Z(\to \ell^+\ell^-)$ & 24.35 & 6.91 & 1.70 & 1.30 & 1.27 \\ \hline
$t\bar t$ (all decay modes) & 30675.9 & 416.44 & 132.66 & 17.63 & 4.17 \\ \hline
$tW$ (all decay modes) & 1934.79 & 33.76 & 6.53 & 0.79 & 0.13 \\ \hline
\end{tabular}
\end{table}

After the fourth cut, the intrinsic $Zjj$ SPS background still dominates over the $Z\otimes jj$ DPS signal at the 13~TeV LHC, say $\sigma(Zjj)\sim 7\times \sigma(Z\otimes jj)$. Thanks to large colliding energy of the 100~TeV colliders, the $Z\otimes jj$ DPS signal and the intrinsic $Zjj$ background are comparable. Other reducible backgrounds turn out to be negligible.

We make use of the characteristic distributions of $\Delta^{\text{rel}}_{\ell}p_T$, $\Delta^{\text{rel}}_{j}p_T$ and $S_\phi$ to further suppress the intrinsic $Zjj$ background. In this study we demand one and only one cut in the following list:
\begin{align}
&\Delta^{\text{rel}}_{\ell} p_T\le 0.2,\nn\\
&\Delta^{\text{rel}}_{j} p_T\le 0.2,\nn\\
&S_\phi>2.5~.
\end{align}
We do not require all of the three cuts simply because cutting on one variable is good enough to enhance the DPS signal. The cross sections of the DPS signal and backgrounds after the optimal cut are presented in Table~\ref{table:zjj13+}. See the third row for cross sections at the 13~TeV LHC and the sixth row for cross sections at the 100~TeV colliders. It shows that the optimal cut efficiently suppress the SM backgrounds and increase $f^{\rm DPS}$. We also notice that the $\Delta^{\text{rel}}_\ell p_T$ cut is much better than the other two cuts. It yields $f^{\rm DPS}\sim 45\%$ at the 13~TeV LHC and $f^{\rm DPS}\sim 80\%$ at the 100~TeV colliders. It is very promising to observe the DPS signal at the LHC and future hadron colliders.

\begin{table}
\footnotesize
\caption{Cross sections (in the unit of picobarn) of the DPS signal and backgrounds after imposing optimal cuts at the 13 TeV LHC (top) and at the 100 TeV  SppC/FCC-hh (bottom). We choose $\sigma_{\rm eff}=15~{\rm mb}$.}
\label{table:zjj13+}
\begin{tabular}{l|c|c|c|c|c|c} 
\hline
{\bf 13 TeV}& DPS $Zjj$ & $Zjj$ & $WZ$ & $t\bar{t}$ & $tW$ & $f^{\rm DPS}$ \\ \hline
basic cuts & 3.40 & 22.29 & 0.12 & 0.12 & 0.01 & 0.13 \\ \hline
$S_\phi>2.5$ & 2.21 & 7.47 & 0.03 & 0.04 & 0.00 & 0.23 \\ 
or $\Delta^{\text{rel}}_\ell p_T<0.2$ & 1.60 & 1.90 & 0.01 & 0.01 & 0.00 & 0.45 \\ 
or $\Delta^{\text{rel}}_jp_T<0.2$ & 1.23 & 2.77 & 0.01 & 0.02 & 0.00  & 0.31 \\ \hline
\hline
{\bf 100 TeV}  \\ \hline
basic cuts & 137.21 & 190.34 & 1.27 & 4.17 & 0.13 & 0.41 \\ \hline
$S_\phi>2.5$ & 73.19 & 62.35 & 0.27 & 1.38 & 0.03 & 0.53 \\ 
or $\Delta^{\text{rel}}_\ell p_T<0.2$ & 38.68 & 10.43 & 0.05 & 0.27 & 0.00 & 0.78 \\ 
or $\Delta^{\text{rel}}_jp_T<0.2$ & 46.13 & 35.05 & 0.17 & 0.84 & 0.01  & 0.56 \\ \hline
\end{tabular}
\end{table}

\subsection{Measuring $\sigma_{\rm eff}$}

We fit the distributions of $\Delta^{\text{rel}}_\ell p_T$, $\Delta^{\text{rel}}_j p_T$ and $S_\phi$ to measure $\sigma_{\rm eff}$. Again, we choose three benchmark inputs ($\sigma^{\rm input}_{\rm eff}=10,~15,~20~{\rm mb}$) and assume the systematic uncertainties to be 15\% and 25\% in the fitting analysis. 

\begin{figure}
  \centering
  \includegraphics[scale=0.25]{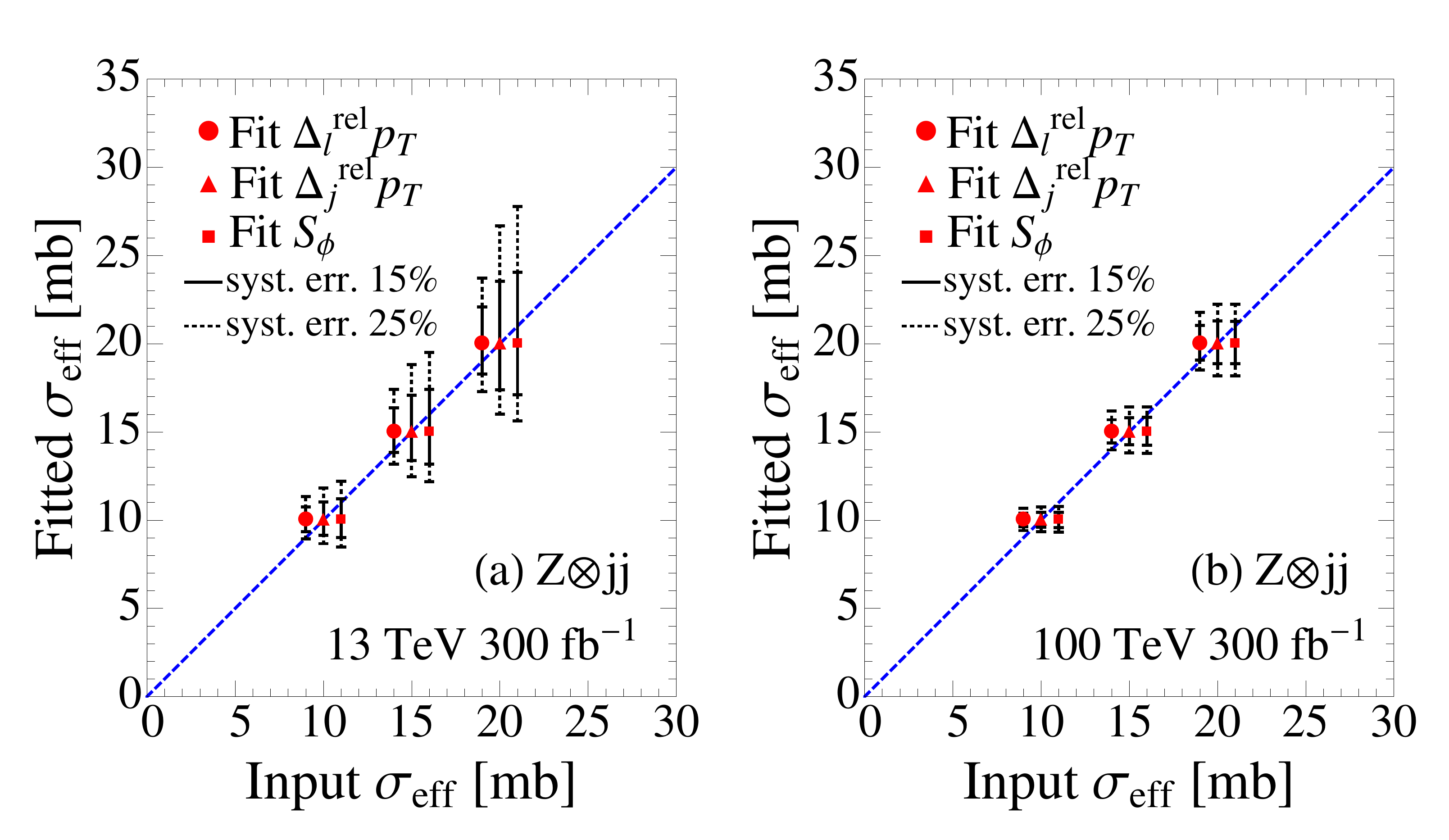}
  \caption{The fitted $\sigma_{\rm eff}$ as a function of the $\sigma_{\rm eff} $ input obtained in the $Z\otimes jj$ channel at the 13 TeV LHC (a) and 100 TeV colliders (b) with an integrated luminosity of $300~{\rm fb}^{-1}$.  }
  \label{zjj_fit} 
\end{figure}

Figure~\ref{zjj_fit} shows the fitted $\sigma^{\rm fit}_{\rm eff}$ as a function of the input $\sigma^{\rm input}_{\rm eff}$ at the 13~TeV LHC (a) and 100~TeV (b) colliders with an integrated luminosity of $300~{\rm fb}^{-1}$.
The circle (triangle, box) symbol denotes $\sigma^{\rm fit}_{\rm eff}$ obtained from fitting the $\Delta^{\text{rel}}_\ell p_T$ ($\Delta^{\text{rel}}_j p_T$, $S_\phi$) distribution, respectively. Fitting the $\Delta^{\text{rel}}_\ell p_T$ distribution gives rise to the best accuracy of $\sigma^{\rm fit}_{\rm eff}$, which are listed as follows:
\begin{itemize}[leftmargin=*]
\item $\sigma^{\rm in}_{\rm eff}=10~{\rm mb}$
\begin{align}
& \sigma^{\rm fit}_{\rm eff}=10^{+0.76~(+7.6\%)}_{-0.66~(-6.6\%)}, ~10^{+1.33~(+13.3\%)}_{-1.05~(-10.5\%)},&& {\rm ~13~TeV},\nn\\
& \sigma^{\rm fit}_{\rm eff}=10^{+0.39~(+3.9\%)}_{-0.37~(-3.7\%)},~10^{+0.67~(+6.7\%)}_{-0.59~(-5.9\%)},&& {\rm 100~TeV};\nn
\end{align}
\item $\sigma^{\rm in}_{\rm eff}=15~{\rm mb}$
\begin{align}
& \sigma^{\rm fit}_{\rm eff}=15^{+1.36~(+9.1\%)}_{-1.15~(-7.7\%)}, ~15^{+2.41~(+16.1\%)}_{-1.82~(-12.1\%)}&& {\rm ~13~TeV},\nn\\
& \sigma^{\rm fit}_{\rm eff}=15^{+0.69~(+4.6\%)}_{-0.63~(-4.2\%)},~15^{+1.18~(+7.9\%)}_{-1.02~(-6.8\%)}, && {\rm 100~TeV};\nn
\end{align}
\item $\sigma^{\rm in}_{\rm eff}=20~{\rm mb}$
\begin{align}
&\sigma^{\rm fit}_{\rm eff}=20^{+2.08~(+10.4\%)}_{-1.72~(-8.6\%)}, ~20^{+3.73~(+18.7\%)}_{-2.72~(-13.6\%)},&& {\rm ~13~TeV},\nn\\
& \sigma^{\rm fit}_{\rm eff}=20^{+1.03~(+5.2\%)}_{-0.93~(-4.7\%)}, ~~20^{+1.77~(+8.9\%)}_{-1.50~(-7.5\%)}, && {\rm 100~TeV},\nn
\end{align}
\end{itemize}
where the first value of $\sigma^{\rm fit}_{\rm eff}$ is for $f_{\rm syst}=15\%$ while the second value for $f_{\rm syst}=25\%$. The superscript and subscript denotes the upper and lower error and the percentage denotes the fraction of the error normalized to the mean value of $\sigma^{\rm fit}_{\rm eff}$. 

The systematic error also dominates over the statistical error in the $Z\otimes jj$ channel; therefore, increasing luminosity cannot significantly improve the accuracy of $\sigma^{\rm fit}_{\rm eff}$. Of course, accumulating more data helps with reducing the systematic errors, but on the assumption of fixed systematic uncertainty as we made, those uncertainties of $\sigma^{\rm fit}_{\rm eff}$ shown in Fig.~\ref{zjj_fit}(a) remain almost the same for the case of a high luminosity machine. Increasing collider energy dramatically enhance the production rate of the DPS signal such that the DPS signal dominates over the SM backgrounds after the optimal cut. That greatly improves the fitting accuracy of $\sigma^{\rm fit}_{\rm eff}$, and all the three distributions yields comparable accuracies of $\sigma^{\rm fit}_{\rm eff}$; see Fig.~\ref{zjj_fit}(b).  

We note that, in comparison with the $W\otimes jj$ channel, one can achieve a better measurement of $\sigma_{\rm eff}$ in the $Z\otimes jj$ channel. To our best knowledge, there is no experimental search for the DPS signal in the $Z\otimes jj$ channel yet. Our study shows that the relative $p_T$ balance of two charged leptons, $\Delta^{\text{rel}}_\ell p_T$, is the best variable to do the job.

\section{The $W^\pm\otimes W^\pm$ channel}\label{section:ww}

\begin{figure}[h!]
  \centering
    \includegraphics[scale=0.4]{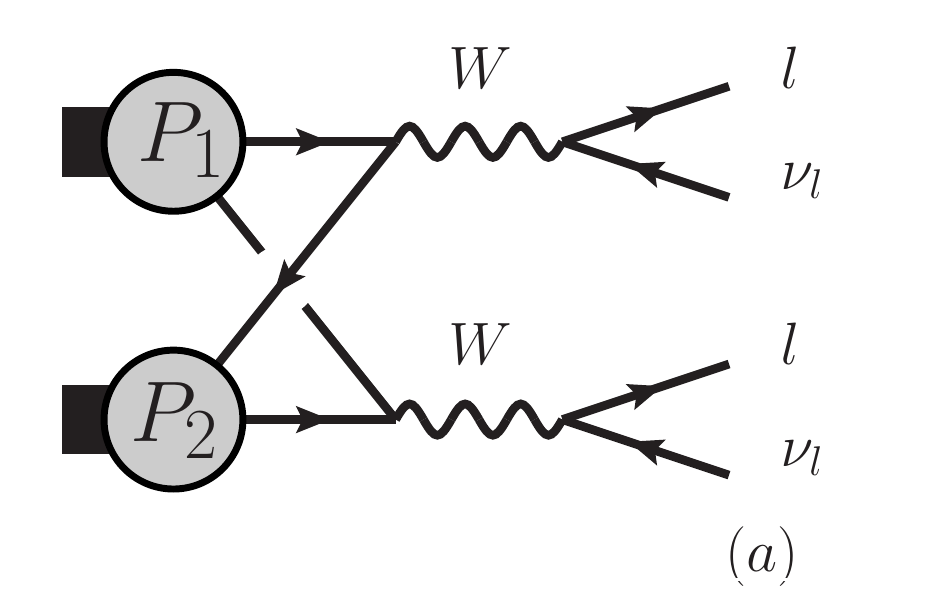}
    \includegraphics[scale=0.4]{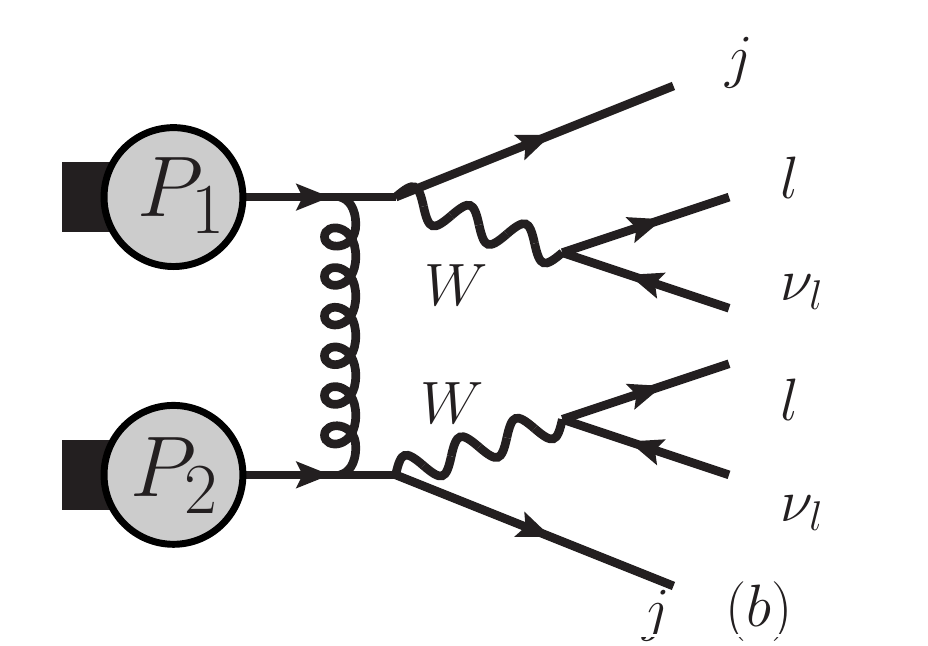}
  \caption{A pictorial illustration of  the $W^\pm\otimes W^\pm$ DPS  channel (a) and the  $W^\pm W^\pm jj$ SPS background (b).}
  \label{DPSandSPS_ww} 
\end{figure}

The $W^\pm \otimes W^\pm$ channel has a clean collider signature of two same-sign charged leptons and large missing transverse momentum induced by neutrinos. See Fig.~\ref{DPSandSPS_ww} for a pictorial illustration. The channel is often believed to offer a unambiguous measurement of $\sigma_{\rm eff}$ and has been extensively studied in the literature~\cite{CMS-PAS-FSQ-13-001,CMS-PAS-FSQ-16-009, Myska:2013duq, Myska:2012dj,Gaunt:2010pi, Maina:2009sj, Kulesza:1999zh, Ceccopieri:2017oqe}. 
Below we explore the $W^\pm\otimes W^\pm$ production at the 13~TeV LHC and future 100~TeV colliders. 

\subsection{Kinematics distributions}

\begin{figure}
  \centering
  \includegraphics[scale=0.23]{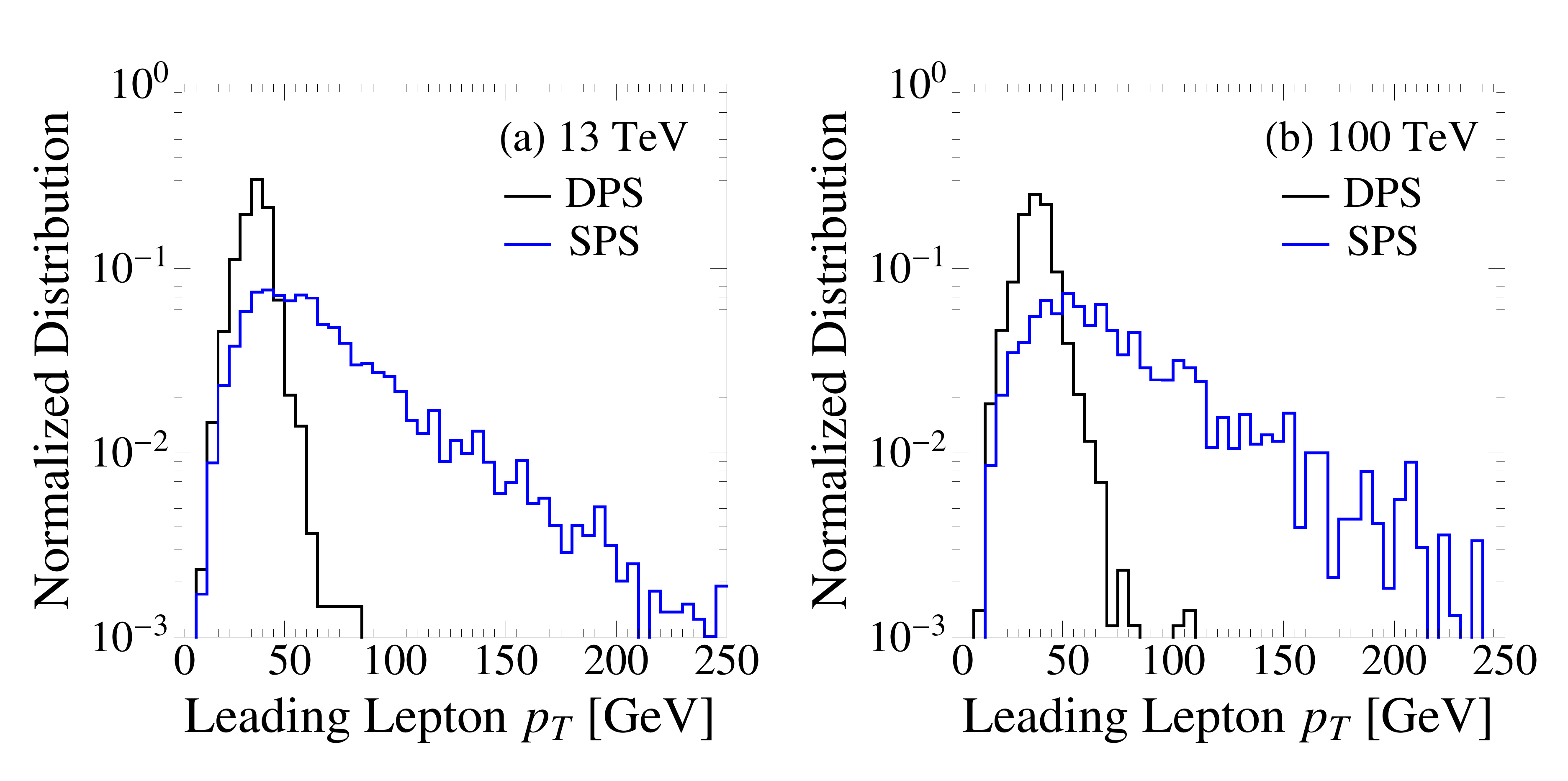}
  \caption{The $p_T$ distribution of the leading lepton in the $W^\pm\otimes W^\pm$ channel at the 13~TeV (a) and 100~TeV (b). The SPS background denotes the $W^\pm W^\pm jj$ production after vetoing additional jets as explained in text.}
  \label{ww_lepton} 
\end{figure}

The collider signature of the $W^\pm \otimes W^\pm$ DPS channel is two charged leptons plus $\met$. As shown in Fig.~\ref{DPSandSPS_ww}, the $W^\pm W^\pm jj$ SPS background has two additional jets in the final state. It can mimic the DPS signal when the two additional jets either have a small $p_T$ or appear outside of the detector coverage. We veto `` hard" jet activities in the central region of detector in the $W^\pm W^\pm jj$ SPS background, i.e. we reject any hard jet satisfying $p_T>25~{\rm GeV}$ and $|\eta|<2.5$ at the 13~TeV while $p_T>50~{\rm GeV}$ and $|\eta|<5$ GeV at the 100~TeV colliders. Figure~\ref{ww_lepton} shows the $p_T$ distribution of the leading charged lepton.  Owing to the feature of independent subprocesses of the DPS channel, the $p_T$ distribution of the leading charged lepton has a Jacobian peak around $p_T\sim m_W/2$. The sub-leading lepton also exhibits such a Jacobian peak in its $p_T$ distribution. On the contrary, the charged leptons in the SPS background are populated more around the cut threshold and have a long tail stretching far into the large $p_T$ region. 
 
The drawback of the $W^\pm\otimes W^\pm$ channel is the two invisible neutrinos, which yields a collider signature of missing transverse momentum, cannot be fully reconstructed. It is hard to determine the longitudinal component of the neutrino momentum at hadron colliders. Such a difficulty has bothered us for a long time in the single $W$-boson production through the Drell-Yan channel~\cite{Cao:2004yy} and single-top quark productions~\cite{Schwienhorst:2010je}. The situation is even worse when the final state consists of two or more invisible neutrinos. Usually, one has to use the on-shell conditions of intermediate state particles to reconstruct the neutrino kinematics~\cite{Berger:2011ua,Berger:2010fy}. However, in the $W^\pm \otimes W^\pm$ channel, we do not have enough information to determine the two neutrinos' momenta which, unfortunately, are the key of reconstructing two subsystems. Therefore, we cannot examine the independent correlations of two subsystems to probe the DPS signal as we have done in the analysis of $W\otimes jj$ and $Z\otimes jj$ channels. As only two visible charged leptons can be resolved, we need to consider their correlations to investigate the potential of measuring $\sigma_{\rm eff}$.

We first examine the azimuthal angle distance $\Delta \phi$ of the two charged leptons.
A rather flat distribution of $\Delta\phi(\ell_1,\ell_2)$ is expected for the $W^\pm\otimes W^\pm$ channel as the two charged leptons are completely independent in the DPS. Unfortunately, the SPS background also exhibits a nearly flat $\Delta\phi$ distribution such that the $\Delta \phi$ distribution is not suitable for measuring $\sigma_{\rm eff}$. 

\begin{figure}
  \centering
  \includegraphics[scale=0.23]{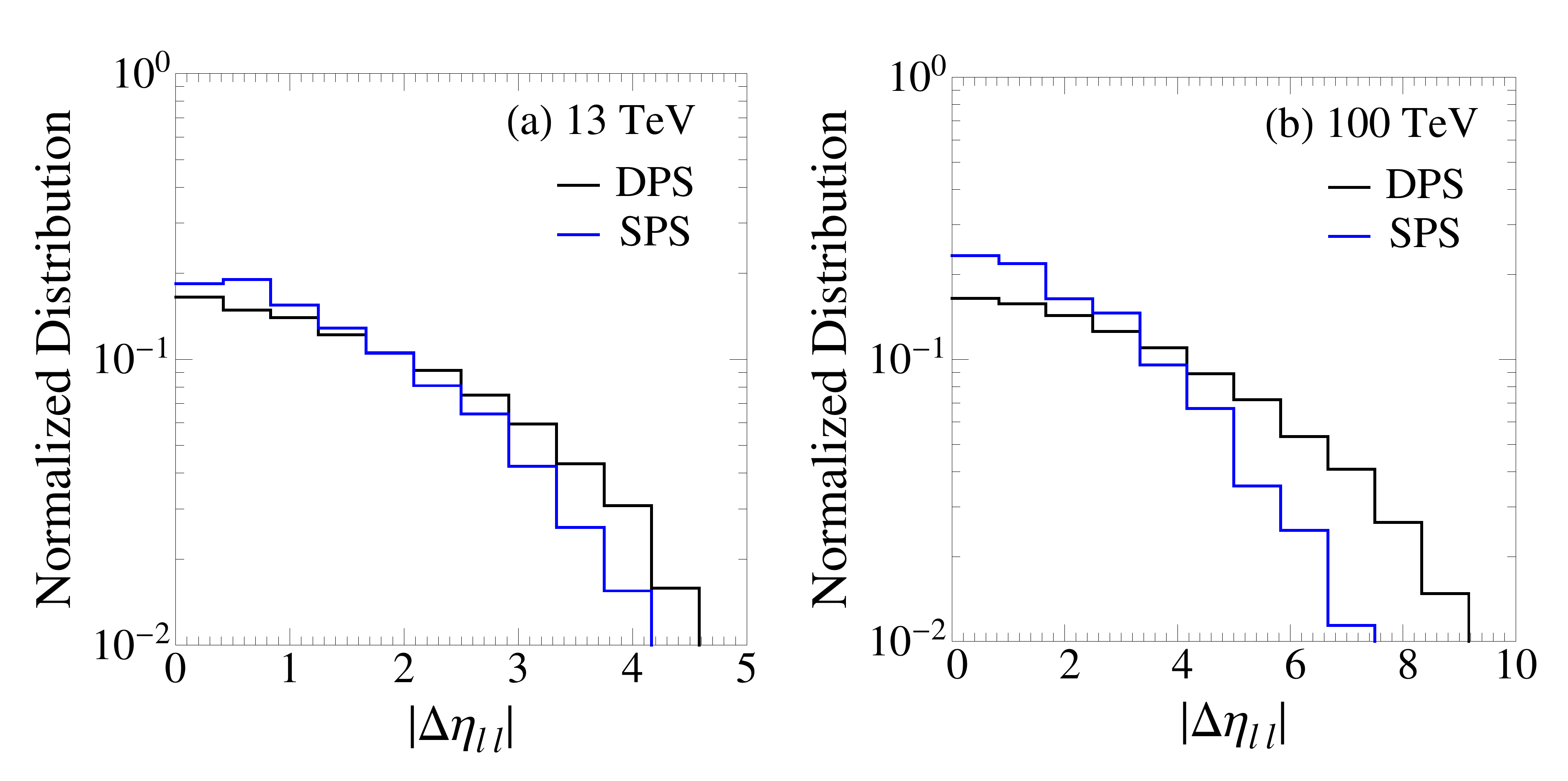}
  \caption{The $|\Delta\eta_{\ell\ell}|$ distribution in the $W^\pm\otimes W^\pm$ channel at the 13~TeV (a) and 100~TeV collider (b). The SPS background denotes the $W^\pm W^\pm jj$ production after jet-veto.}
  \label{wwDeta_gen} 
\end{figure}

Next, we consider the rapidity difference of the two charged leptons $\Delta\eta_{\ell\ell}$, defined as
\be
\Delta\eta_{\ell\ell}=\eta_{\ell^\pm_1}-\eta_{\ell^\pm_2},
\ee
where $\ell^\pm_1$ denotes the leading $p_T$ charged lepton and $\ell^\pm_2$ the subleading lepton. 
Figure~\ref{wwDeta_gen} displays the magnitude of $\Delta \eta_{\ell\ell}$ distribution of the DPS signal (black) and the SPS background (blue). The DPS signal exhibits a more flatter distribution. The difference becomes more evident at the 100~TeV collider. Hence, one can measure the DPS signal through the rapidity difference of two charged leptons. The difference can be understood as follows. In the DPS subprocess of $d\bar{u} \to W^-\to \ell^-\bar{\nu}$, the charged lepton $\ell^-$ appears predominantly along the incoming $d$-quark direction, i.e. 
\be
\mathcal{M}(d_L \bar{u}_R \to W^- \to \ell^-_L \bar{\nu}_R)\propto \frac{1+\cos\theta_{\ell d}}{2}
\ee
where the $\theta_{\ell d}$ angle denotes the open angle between the charged lepton $\ell^-$ and the moving direction of the $d$-quark in the center of mass frame, i.e. $\cos\theta_{\ell d}\equiv \vec{p}_{\ell^-}\cdot \vec{p}_{d}/|\vec{p}_{\ell^-}||\vec{p}_{d}|$ with $\vec{p}_{\ell^-,d}$ being the charged lepton ($d$-quark) three-momentum defined in the center of mass frame. As a result, it is often that one of the two charged leptons appear in the forward region and the other in the backward region, just leading to a large rapidity gap. In the SPS background, the $W$-boson pairs tend to be produced in the central regions and their decay products often appear in the central region, yielding a small rapidity gap.

\begin{figure}
  \centering
  \includegraphics[scale=0.23]{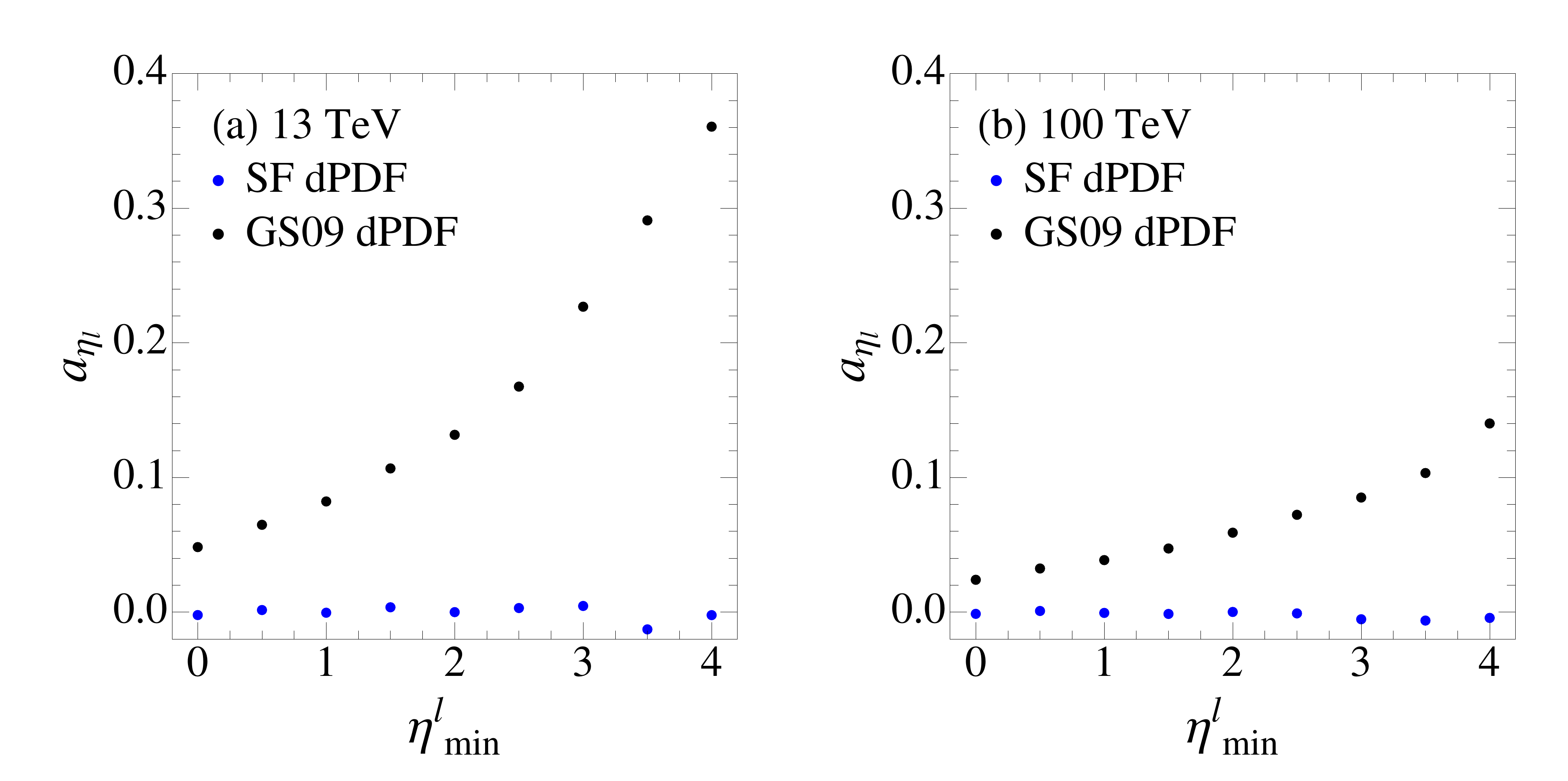}
  \caption{$\mathcal{A}_{\eta_\ell}$ versus $\eta^\ell_{\rm min}$ in SF and GS09 dPDFs at the 13~TeV LHC (a) and the 100 TeV colliders (b).}
  \label{Asym_eta} 
\end{figure}

It is interesting to ask whether the $\Delta\eta_{\ell\ell}$ distribution is sensitive to the choice of dPDF. It has been pointed out in Ref.~\cite{Gaunt:2010pi, Ceccopieri:2017oqe} that, the rapidity asymmetry of two charged leptons in the $W^\pm \otimes W^\pm$ DPS channel can manifest the difference of simple factorized dPDF and GS09 dPDF. The lepton rapidity asymmetry is defined as
\be
\mathcal{A}_{\eta_\ell}=\frac{\sigma(\eta_{\ell_1}\times\eta_{\ell_2}<0)-\sigma(\eta_{\ell_1}\times\eta_{\ell_2}>0)}{\sigma(\eta_{\ell_1}\times\eta_{\ell_2}<0)+\sigma(\eta_{\ell_1}\times\eta_{\ell_2}>0)}.
\ee
The asymmetry is sensitive to the correlations between the two partons from one proton, which is described in Eq.~(\ref{GS_model}) in the GS09 dPDF but absent in the simplified dPDF.  As the correlation effect is evident in the large $x$ region, a cut on the lepton rapidity ($|\eta_\ell|>\eta^\ell_{\rm min}$) could amplify the difference between those two dPDFs. Figure~\ref{Asym_eta} shows the $\mathcal{A}_{\eta_\ell}$ distribution as a function of $\eta^\ell_{\rm min}$ at the 13~TeV LHC (a) and 100~TeV colliders (b). The 13~TeV result agrees well with Ref.~\cite{Gaunt:2010pi}. In the SF dPDF, the two charged leptons are independent, yielding $\mathcal{A}_{\eta_\ell} \sim 0$ (see the blue points); in the GS09 dPDF, the two charged leptons tend to lie in different hemispheres with an axis defined by the beam line, giving rise to a positive $\mathcal{A_{\eta_\ell}}$ (see the black points). A much smaller $x$ is reached at the 100 TeV colliders, thus weakening the difference between dPDFs.

\begin{figure}[h]
  \centering
  \includegraphics[scale=0.23]{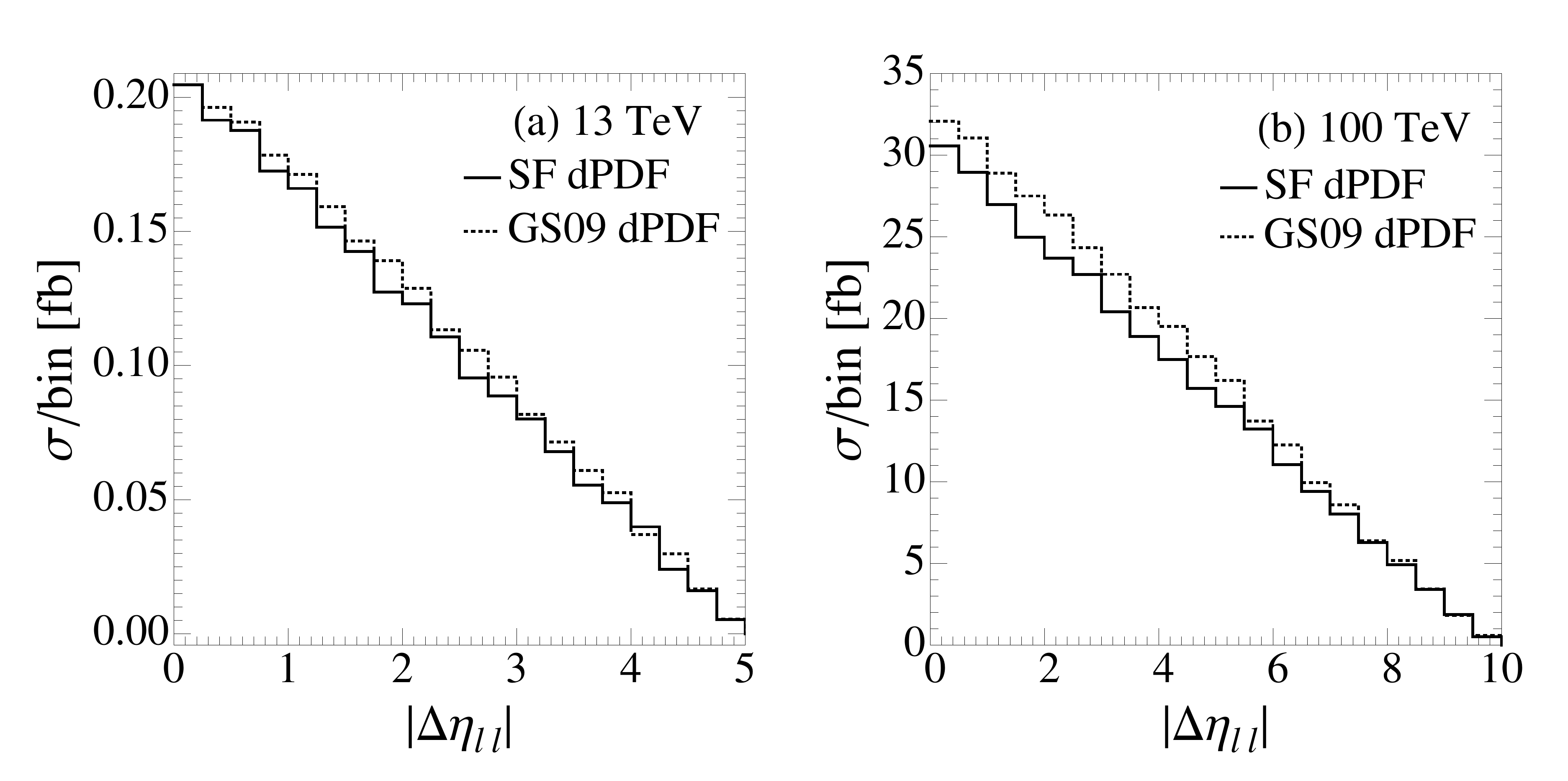}
  \caption{The $\Delta\eta_{\ell\ell}$ distribution of two same-sign charged leptons calculated with different double parton distributions at the 13~TeV LHC (a) and the 100 TeV colliders (b).}
  \label{D_eta} 
\end{figure}

In order to keep more DPS signal events, we do not impose the $\eta^\ell_{\rm min}$ cut, which is crucial to see the difference between dPDFs. Therefore, the $|\Delta \eta_{\ell\ell}|$ distribution is not sensitive to the dPDF models in our analysis. Even though the SF and GS09 dPDFs produces a mild difference in the $|\Delta \eta_{\ell\ell}|$ distribution, it does not affect our fitting results, as Fig.~\ref{D_eta} shows. 

\subsection{Collider Simulation}

The event topology of the $W^\pm\otimes W^\pm$ channel consists of two same-sign lepton and $\met$. We also demand no hard jet activity. The major SPS backgrounds are: i) the $W^\pm W^\pm jj$ production with the two additional jets being vetoed; ii)
the $t\bar t$ pair production in which a charged lepton is generated from one top quark decay while another same-sign charged lepton arises from the bottom quark emitted from the other top; 
iii) the $W^\pm Z/\gamma^*$ with $Z/\gamma^*\to \ell^+\ell^-$ and $W^\pm \to \ell^+\nu$; 
iv) the $Z(\gamma^*)Z(\gamma^*)\to \ell^+\ell^-\ell^+\ell^-$ channel which is denoted as ``$2\ell^+2\ell^-$".
In order to suppress the SPS backgrounds, we choose five basic cuts listed below:
\begin{itemize}
\item[1.] Two same-sign charged leptons with $p_T^{\ell^\pm}\ge 20~{\rm GeV}$ and $|\eta^{\ell^\pm}|\le 2.5$ at the 13 TeV LHC while $|\eta^{\ell^\pm}|\le 5$ at the 100~TeV SppC/FCC-hh;
\item[2.] No hard jet with $p_T^{j}\ge 25~{\rm GeV}$ and $|\eta^{j}|\le2.5$ at the 13~TeV LHC while $p_T^{j}\ge 50~{\rm GeV}$ and $|\eta^{j}|\le 5$ at the 100~TeV SppC/FCC-hh;
\item[3.] $\met \ge 20~{\rm GeV}$;
\item[4.] $M_{\ell_1^\pm\ell_2^\pm}\notin [75,105]~{\rm GeV}$;
\item[5.] $p_T^{\ell^\pm}<60~{\rm GeV}$.
\end{itemize}
Note that our lepton $p_T^{\ell^\pm}$ cut is slightly different from Ref.~\cite{CMS-PAS-FSQ-13-001}, which introduces asymmetric cuts on the leading lepton and trailing lepton as $p_T^{\ell_1}>20~{\rm GeV}$ and $p_T^{\ell_2}>10~{\rm GeV}$, respectively. In our analysis we demand symmetric cuts on both leptons,  $p_T^{\ell^\pm}>20~{\rm GeV}$, which can suppress the $WZ$ and $W\gamma^*$ backgrounds  efficiently, Our simulation results are consistent with Ref.~\cite{Gaunt:2010pi}. 

We choose the input value of $\sigma_{\rm eff}=15~{\rm mb}$ and generate both the signal and background events in {\tt MadGraph} with $p_T^j\geq 10~{\rm GeV}$ and $|\eta^j|<5$. We further demand the charged leptons well separated in angular distance, i.e. $\Delta R_{\ell\ell}\ge 0.4$, in order to avoid the collinear divergence in the $\gamma^*\to\ell^+\ell^-$ processes. We then pass the parton level events to {\tt Pythia} for parton shower and merging. The cross section (in the unit of picobarn) of the signal and background processes after imposing the generator-level cuts are summarized in the second column of Table~\ref{table:ww13}. Next, we adapt {\tt Delphes} for particle identifications and then impose the four kinematics cuts. The last five columns in Table~\ref{table:ww13} show the cross section after imposing the five basic cuts sequentially. 

\begin{table}
\footnotesize
\caption{Cross sections (in the unit of picobarn) of the $W^\pm\otimes W^\pm$ DPS signal process and the SM background process $W^\pm W^\pm jj$ at the 13 TeV LHC (top) and at the 100 TeV  SppC/FCC-hh (bottom). The two additional jets in the background are rejected. We choose $\sigma_{\rm eff}=15~{\rm mb}$ and impose the kinematic cuts listed in each column sequentially.
}
\label{table:ww13}
\begin{tabular}{c|c|c|c|c|c|c} 
\hline
{\bf 13 TeV} & Gen. & Cut-1 & Cut-2 & Cut-3 & Cut-4 & Cut-5 \\ \hline
$W^\pm\otimes W^\pm$ & 22.18 & 4.10 & 2.63 & 2.21 & 1.78 & 1.76 \\ \hline
$W^\pm W^\pm jj$ & 71.98 & 11.78 & 0.64 & 0.58 & 0.47 & 0.19 \\ \hline
$t\bar t$ & 461001 & 0.64 & 0.00 & 0.00 & 0.00 & 0.00 \\ \hline
$WZ/\gamma^*$ & 13749.6 & 80.75 & 33.58 & 27.88 & 20.95 & 13.55 \\ \hline
$2\ell^+2\ell^-$ & 749.94 & 10.47 & 5.25 & 0.88 & 0.69 & 0.43 \\ \hline
\hline
{\bf 100 TeV}   \\ \hline
$W^\pm \otimes W^\pm$ & 1589.33 & 512.53 & 373.72 & 336.87 & 290.81 & 275.95 \\ \hline
$W^\pm W^\pm jj$ & 1623.85 & 336.65 & 18.79 & 16.91 & 12.99 & 6.98 \\ \hline
$t\bar t$ & 30675900 & 117.69 & 5.88 & 5.88 & 5.88 & 0.00 \\ \hline
$WZ/\gamma^*$ & 85482 & 1304.95 & 581.17 & 486.63 & 390.10 & 226.03 \\ \hline
$2\ell^+2\ell^-$ & 5242.75 & 123.84 & 71.41 & 10.95 & 8.36 & 4.48 \\ \hline
\end{tabular}
\end{table}

The identification of two same-sign charged leptons in the first cut (cut-1) is the most efficient cut to suppress the SPS backgrounds; see the third column in Table~\ref{table:ww13}. While about 18\% of the DPS signal events survive the cut-1, only 0.02\% of the SPS background events remain. At the 100 TeV collider the lepton $|\eta^\ell|$ cut is extended to 5 in order to collect more signal events. We find that the lepton identification cut works better at the 100~TeV colliders; for example, about 0.006\% of the SPS background events survive while about 32\% of the DPS signal events remain. The jet-veto cut specified in the second cut (cut-2) is also very powerful in suppressing those backgrounds involving jets in the final state; see the fourth  column. We introduce the $\met$ cut (cut-3) to suppress the $2\ell^+ 2\ell^-$ backgrounds which do not have neutrinos at the parton level. Note that a potential background comes from the mis-tagging of multi-jet events. It has been shown by the CMS collaboration~\cite{CMS-PAS-FSQ-13-001} that such mis-tagged backgrounds can be efficiently suppressed by requiring the scalar sum of two charged leptons' $p_T$ larger than 45~GeV, i.e. $p_T^{\ell_1}+p_T^{\ell_2}>45~{\rm GeV}$.  Since we demand both the charged leptons exhibit $p_T^\ell>20~{\rm GeV}$ in lepton trigger, the scalar sum condition is satisfied automatically. 
It is also possible that one of the two charged leptons from the $Z$ boson decay is mis-identified as opposite charged. It then provides a faked signal of two same-sign charged leptons. In the fourth cut (cut-4), we demand the invariant mass of the same-sign lepton pair to be away from the $Z$ boson resonance so as to remove those faked events, i.e. $|M_{\ell_1^\pm\ell_2^\pm}-m_Z|>20~{\rm GeV}$. Finally, we demand an upper bound on the charged lepton's $p_T$ in the fifth cut (named as cut-5). Owing to the Jacobi peak feature of the $W$ boson decay in the $W^\pm\otimes W^\pm$ DPS channel, the charged lepton exhibits a $p_T$ mainly below  $\sim 40~{\rm GeV}$.  The cut only mildly affects the DPS signal but sizably reduce the electroweak backgrounds.

After all the five basic cuts, the $WZ/\gamma^*$ production becomes the dominant background. We end up with $f^{\rm DPS}=11\%$ at the 13~TeV LHC and $f^{\rm DPS}=54\%$ at the 100~TeV  SppC/FCC-hh. Increasing collider energy improve the fraction $f^{\rm DPS}$ significantly.

\subsection{Determining $\sigma_{\rm eff}$}

\begin{figure}
  \centering
  \includegraphics[scale=0.23]{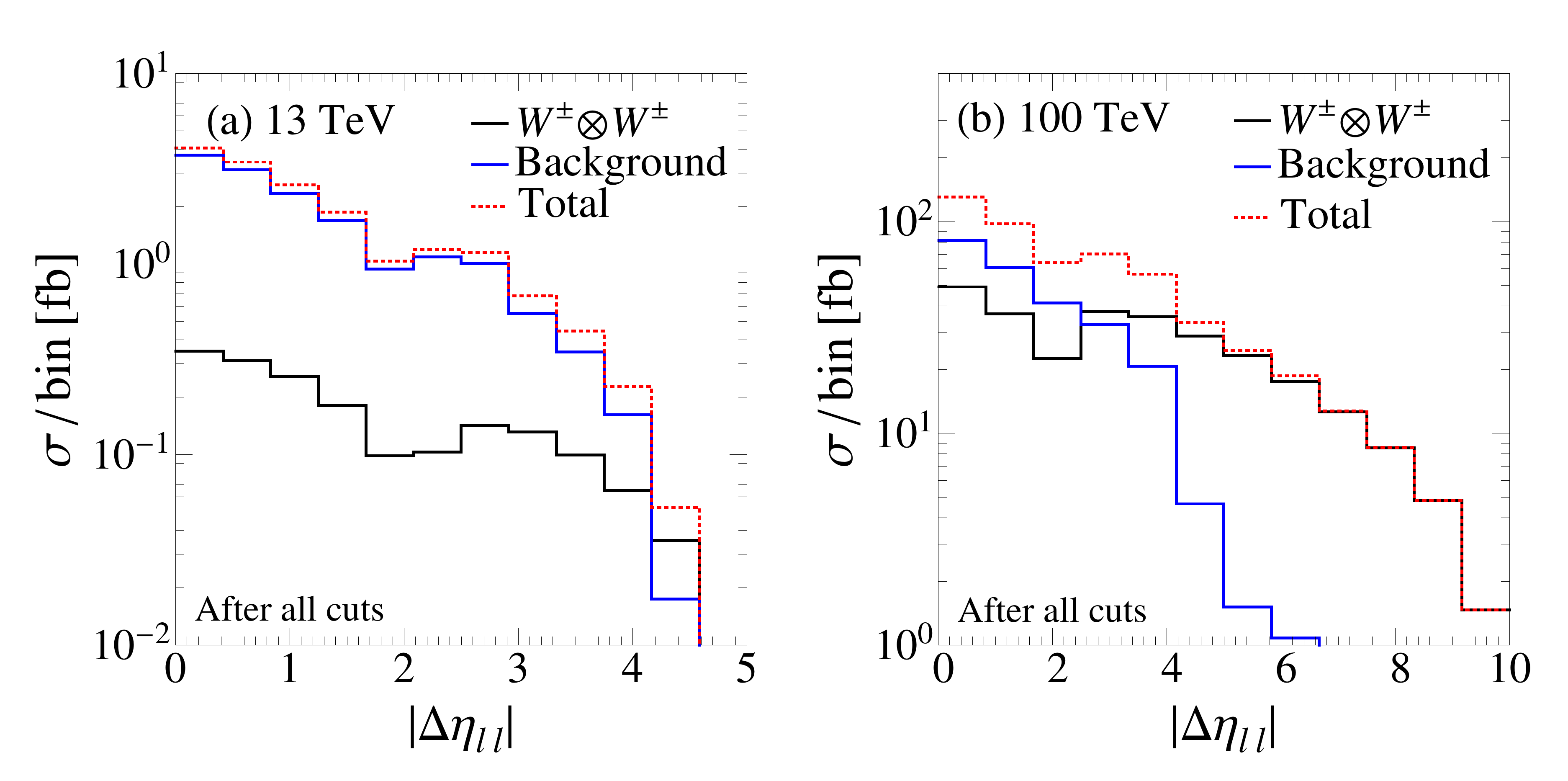}
  \caption{The $|\Delta\eta_{\ell\ell}|$ distribution in the $W^\pm\otimes W^\pm$ channel after after all of the five basic cuts at the 13~TeV (a) and 100~TeV (b). }
  \label{wwDeta} 
\end{figure}

\begin{figure}
\centering
\includegraphics[scale=0.25]{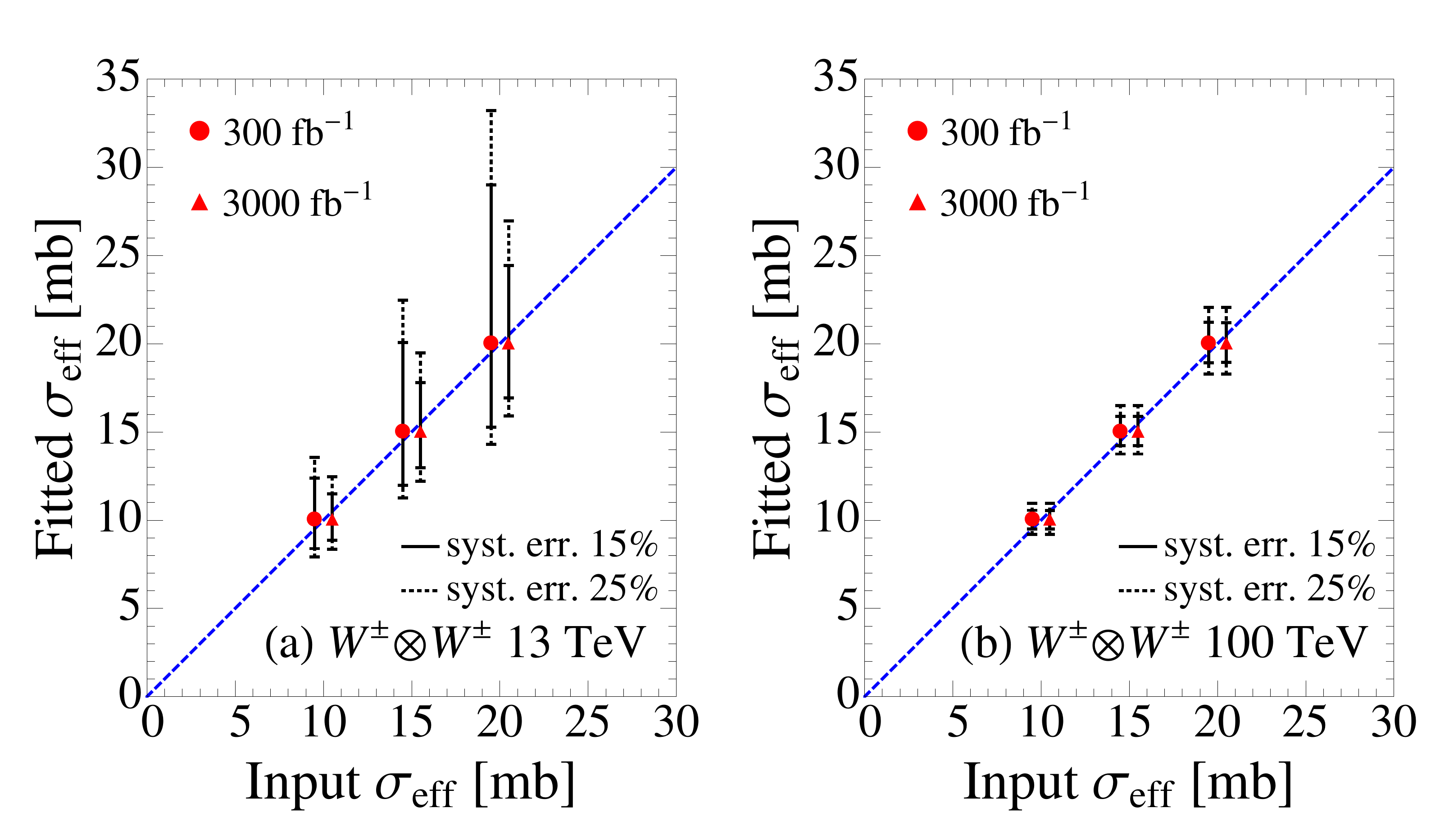}
\caption{The fitted $\sigma_{\rm eff}$ as a function of $\sigma^{\rm in}_{\rm eff}$ fitted in the $\Delta\eta_{\ell\ell}$ distribution at the 13 TeV (a) and 100 TeV colliders (b).}
\label{ww_fit} 
\end{figure}

We perform a $\chi^2$-fit of the $|\Delta\eta_{\ell\ell}|$ distribution, divided into 12 bins, to measure $\sigma_{\rm eff}$. Figure~\ref{wwDeta} displays the $|\Delta\eta_{\ell\ell}|$  distributions of the DPS signal and the sum of all the SM backgrounds (labelled as background) after the five selection cuts. Note that the $WZ/\gamma^*$ channel, not the $W^\pm W^\pm jj$ SPS background we examined, becomes the dominant background. The difference between the DPS signal and the SPS background we observed in Fig.~\ref{wwDeta_gen} still remains.  Figure~\ref{ww_fit} shows the $\sigma^{\rm fit}_{\rm eff}$ as a function of $\sigma^{\rm input}_{\rm eff}$ at the 13~TeV LHC (a) and 100~TeV colliders (b). The rate of the $W^\pm\otimes W^\pm$ DPS production is suppressed in comparison with the $W\otimes jj$ and $Z\otimes jj$ DPS channels. To study the impact of the statistical uncertainty on the fitting precision, we consider two benchmark luminosities in the fitting analysis: $300~{\rm fb}^{-1}$ (red circle) and $3000~{\rm fb}^{-1}$ (triangle). Two systematic errors, $f_{\rm syst}=15\%$ and $f_{\rm syst}=25\%$, are considered.

Due to the small rate of the $W^\pm\otimes W^\pm$ and large backgrounds ($f^{\rm DPS}=11\%$) at the 13~TeV LHC, the accuracy of $\sigma^{\rm fit}_{\rm eff}$ is  sensitive to the integrated luminosity. Upgrading the LHC to the phase of high luminosity, say $\mathcal{L}=3000~{\rm fb}^{-1}$, improves the fitting accuracy sizably; for example, see the circle and triangle points for each input $\sigma^{\rm in}_{\rm eff}$. At the 100 TeV collider, owing to the significant enhancement of the DPS production rate, the statistic uncertainty is well under control and the systematic error dominates the fitting precision. Therefore, accumulating more luminosity at the 100~TeV machine would not improve the accuracy of $\sigma^{\rm fit}_{\rm eff}$.

The fitted results for an integrated luminosity of 3000 fb$^{-1}$ are as follows:
\begin{itemize}[leftmargin=*]
\item $\sigma^{\rm in}_{\rm eff}=10~{\rm mb}$
\begin{align}
& \sigma^{\rm fit}_{\rm eff}=10^{+1.50~(+15.0\%)}_{-1.15~(-11.5\%)}, ~10^{+2.45~(+24.5\%)}_{-1.65~(-16.5\%)},&& {\rm 13~TeV},\nn\\
& \sigma^{\rm fit}_{\rm eff}=10^{+0.55~(+5.5\%)}_{-0.50~(-5.0\%)},~~10^{+0.96~(+9.6\%)}_{-0.80~(-8.0\%)},&& {\rm 100~TeV};\nn
\end{align}
\item $\sigma^{\rm in}_{\rm eff}=15~{\rm mb}$
\begin{align}
& \sigma^{\rm fit}_{\rm eff}=15^{+2.80~(+18.7\%)}_{-2.04~(-13.6\%)}, ~15^{+4.47~(+29.8\%)}_{-2.80~(-18.7\%)}&& {\rm 13~TeV},\nn\\
& \sigma^{\rm fit}_{\rm eff}=15^{+0.87~(+5.8\%)}_{-0.78~(-5.2\%)},~~15^{+1.50~(+10.0\%)}_{-1.25~(~-8.3\%)}, && {\rm 100~TeV};\nn
\end{align}
\item $\sigma^{\rm in}_{\rm eff}=20~{\rm mb}$, 
\begin{align}
& \sigma^{\rm fit}_{\rm eff}=20^{+4.44~(+22.2\%)}_{-3.07~(-15.4\%)}, ~20^{+6.96~(+34.8\%)}_{-4.10~(-20.5\%)},&& {\rm 13~TeV},\nn\\
& \sigma^{\rm fit}_{\rm eff}=20^{+1.19~(+6.0\%)}_{-1.06~(-5.3\%)}, ~~20^{+2.06~(+10.3\%)}_{-1.71~(-8.6\%)}, && {\rm 100~TeV},\nn
\end{align}
\end{itemize}
where the first value of $\sigma^{\rm fit}_{\rm eff}$ is for $f_{\rm syst}=15\%$ while the second value for $f_{\rm syst}=25\%$. The superscript and subscript denotes the upper and lower error and the percentage denotes the fraction of the error normalized to the mean value of $\sigma^{\rm fit}_{\rm eff}$.

\begin{figure*}
\centering
\includegraphics[scale=0.321]{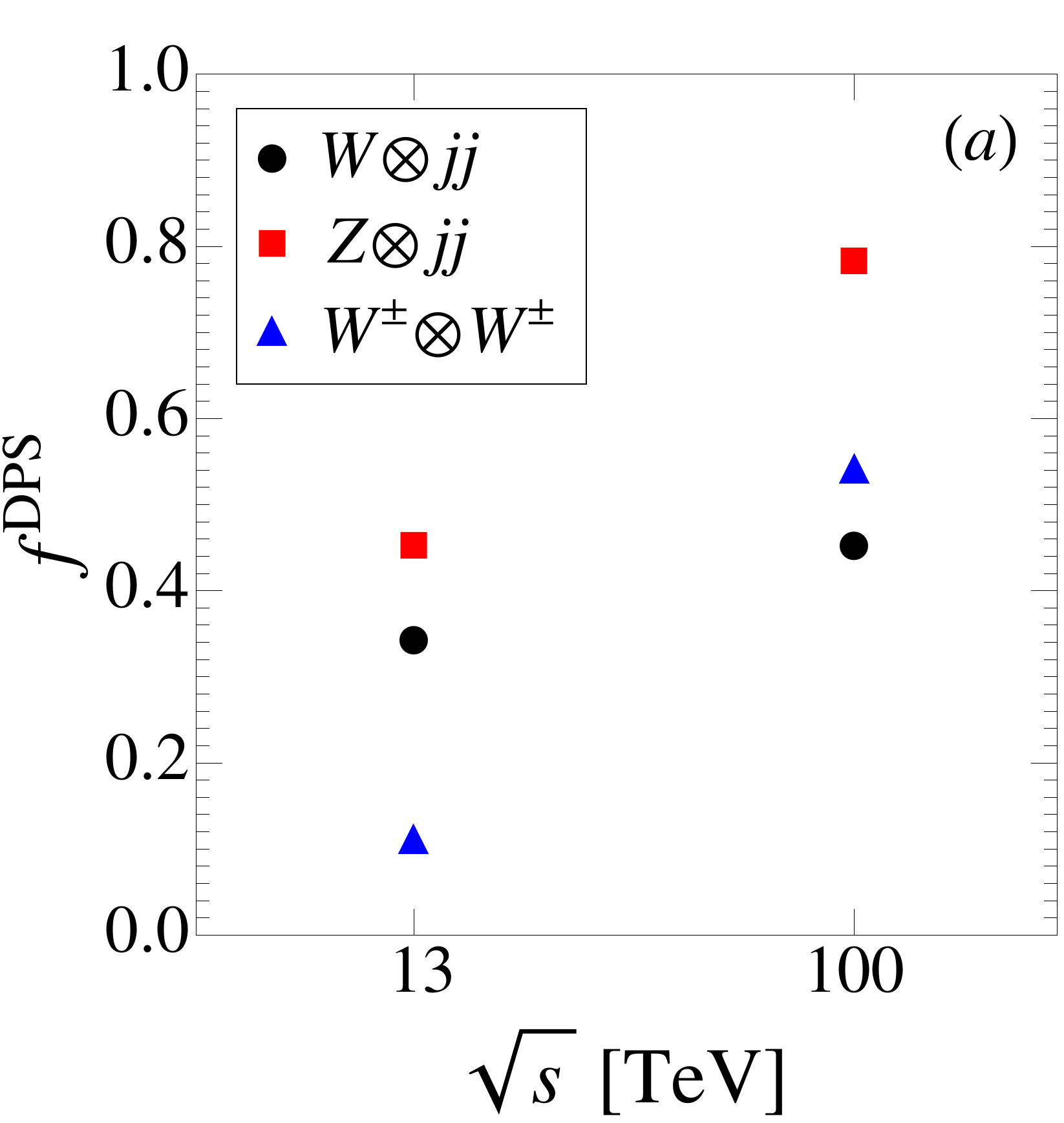}\quad
\includegraphics[scale=0.425]{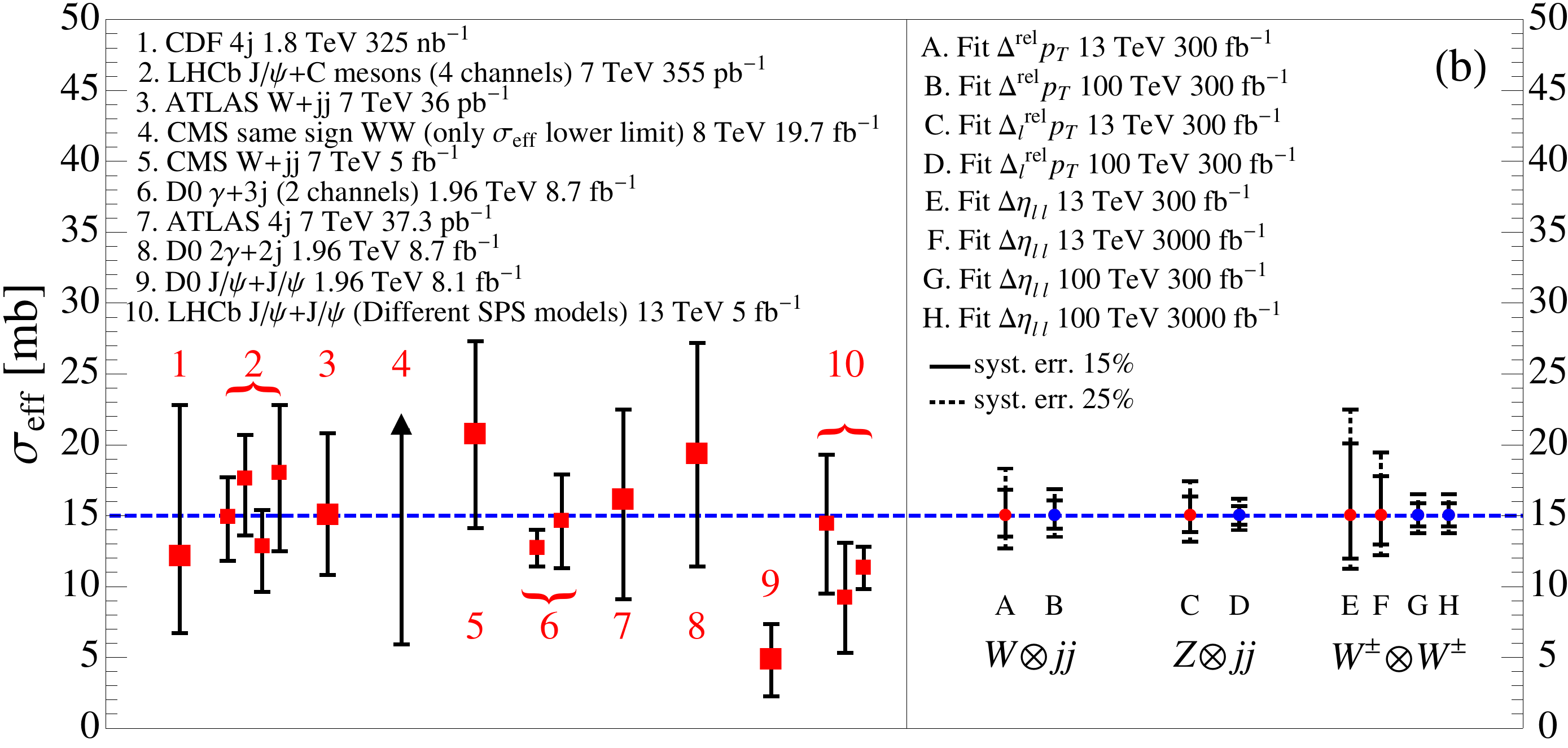}
\caption{(a) $f^{\rm DPS}$ (the fraction of the DPS signal events in the total events) in the three DPS processes after the optimal cuts at the 13~TeV and 100~TeV colliders. (b) Recent experimental data of $\sigma_{\rm eff}$ measurement (left) and the projected $\sigma^{\rm fit}_{\rm eff}$ in the $W\otimes jj$, $Z\otimes jj$ and $W^\pm \otimes W^\pm$ channels (right).}
  \label{experiments} 
\end{figure*}

\section{Discussion and Conclusions}\label{section:discussion}

In the LHC era, with much higher collision energies available, DPS has received several experimental and theoretical studies. 
Lack of theoretical ground, the phenomenological studies are based on the factorized ansatz of the double parton distribution functions, which neglect momentum correlations between partons and introduce an effective cross section $\sigma_{\rm eff}$. The latter quantity is to be extracted from experimental data and might vary for different processes. As $\sigma_{\rm eff}$ is connected with the effective size of the hard scattering core of the proton, the variation in the values of $\sigma_{\rm eff}$ may mean that $\sigma_{\rm eff}$ will have different values for $qq$, $qg$ and $q\bar{q}$ scatterings. It is desirable to establish double parton scattering in data and determine $\sigma_{\rm eff}$ in a relatively clean processes. In this work we demonstrate that the DPS production involving weak gauge bosons is important for measuring $\sigma_{\rm eff}$ because the leptons from the $W$ and $Z$ boson decays provide a nice trigger of the DPS signal. Specifically, we focus on the $W\otimes jj$, $Z\otimes jj$ and $W^\pm\otimes W^\pm$ channels and explore the potential of measuring $\sigma_{\rm eff}$ at the 13~TeV and 100~TeV proton-proton colliders.

Several observables characterizing the feature of DPS have been proposed to optimize the DPS signal in the literature~\cite{Berger:2009cm, Berger:2011ep}. Our study shows that the best observable to measure $\sigma_{\rm eff}$ is: the relative $p_T$ balance of jets ($\Delta^{\rm rel}p_T$) in $W\otimes jj$, the relative $p_T$ balance of leptons ($\Delta^{\rm rel}_\ell p_T$) in $Z\otimes jj$, and $\Delta \eta_{\ell\ell}$ in $W^\pm\otimes W^\pm$. Note that $\Delta^{\rm rel}_\ell p_T$ works better than $\Delta^{\rm rel}_j p_T$ in $Z\otimes jj$. Taking advantage of those optimal observables, we show that it is very promising to observe the DPS signal on top of the SPS backgrounds. Figure~\ref{experiments}(a) displays the fraction of the DPS signal event in the total event collected ($f^{\rm DPS}$), defined in Eq.~(\ref{eq:fdps_def}), after imposing the optimal cuts specified in main text. At the 13~TeV LHC, $f^{\rm DPS}(W\otimes jj)\sim 34\%$,  $f^{\rm DPS}(Z\otimes jj)\sim 45\%$, and $f^{\rm DPS}(W^\pm\otimes W^\pm)\sim 10\%$; at the 100~TeV colliders, $f^{\rm DPS}$ increases dramatically, say $f^{\rm DPS}(W\otimes jj)\sim 45\%$,  $f^{\rm DPS}(Z\otimes jj)\sim 78\%$, and $f^{\rm DPS}(W^\pm\otimes W^\pm)\sim 54\%$,  owing to the huge enhancement of the production rate of the DPS processes.

Once double parton scattering is established in data and $\sigma_{\rm eff}$  is determined, one can address on the three questions raised in Sec.~\ref{section:introduction}: i) how well can one measure $\sigma_{\rm eff}$? ii) does $\sigma_{\rm eff}$ vary with colliding energies? iii) is $\sigma_{\rm eff}$ universal for various DPS processes?  

Figure~\ref{experiments}(b) displays the recent experimental data (left panel) and the projected accuracies of $\sigma_{\rm eff}$ measurement obtained from our collider simulations with the choice of $\sigma^{\rm in}_{\rm eff}=15~{\rm mb}$ (right panel). The recent experimental data are summarized in Table~\ref{table:alldata}, which suggests an average value of $\sigma_{\rm eff}\simeq 15~{\rm mb}$. For the three DPS channels of interest to us, the red (blue) points denote the $\sigma^{\rm fit}_{\rm eff}$ obtained at the 13 (100)~TeV colliders, respectively. The $W\otimes jj$ and $Z\otimes jj$ channels are able to measure the $\sigma_{\rm eff}$ with errors less than the current data, assuming the systematic error is 15\% or 25\%. Since the uncertainties of the two DPS channels are dominated by the systematic errors, we present the fitting results with an integrated luminosity of $300~{\rm fb}^{-1}$. Note that accumulating more luminosity cannot improve the accuracy. At the 100~TeV colliders, the DPS production rate increases dramatically such that the $\sigma_{\rm eff}$ measurement can be sizably improved. 

The $Z\otimes jj$ channel gives a better precision than the $W\otimes jj$ channel; for example, assuming a 15\% systematic error, one can measure the $\sigma_{\rm eff}$ through the $\Delta^{\text{rel}}_\ell p_T$ distribution with a precision as 
\begin{align}
\sigma^{\rm fit}_{\rm eff}(W\otimes jj)&=15^{+1.83~(+12.2\%)}_{-1.47~(~-9.8\%)},\nn\\
\sigma^{\rm fit}_{\rm eff}(Z\otimes jj)&=15^{+1.36~(+9.1\%)}_{-1.15~(-7.7\%)},\nn
\end{align}
at the 13~TeV LHC, and \begin{align}
\sigma^{\rm fit}_{\rm eff}(W\otimes jj)&=15^{+1.07~(+7.1\%)}_{-0.94~(-6.3\%)}, \nn\\
\sigma^{\rm fit}_{\rm eff}(Z\otimes jj)&=15^{+0.69~(+4.6\%)}_{-0.63~(-4.2\%)}, \nn
\end{align}
at the 100~TeV colliders with an integrated luminosity of $300~{\rm fb}^{-1}$. See the points A, B, C and D in Fig.~\ref{experiments}(b).
Therefore, we argue that one should explore the $Z\otimes jj$ channel to measure $\sigma_{\rm eff}$. 

The $W^\pm\otimes W^\pm$ channel has been studied extensively in the literature for the reason that it has a clean signature of two charged leptons and large missing transverse momentum. However, the channel suffers from small production rate and lack of distinctive observables discriminating the DPS signal from the SPS backgrounds. Therefore, the uncertainty of $\sigma_{\rm eff}$ measurements is worse than that of the $W\otimes jj$ and $Z\otimes jj$ channels. For the same reason, the recent CMS measurement provides only a lower limit of $\sigma_{\rm eff}$; see the fourth data in the left panel of Fig.~\ref{experiments}(b). Though suffering from large uncertainties, the $W^\pm\otimes W^\pm$ signal can be measured in the $|\Delta \eta_{\ell\ell}|$ distribution at the 13~TeV LHC, and the accuracy can be further improved at the high luminosity phase. For example, choosing an input $\sigma_{\rm eff}^{\rm in}=15~{\rm mb}$ and assuming a 15\% systematic error, one can measure $\sigma_{\rm eff}$ through the $\Delta\eta_{\ell\ell}$ distribution with a precision as $15^{+5.08~(+33.9\%)}_{-3.03~(-20.2\%)}$ and $15^{+2.80~(+18.7\%)}_{-2.04~(-13.6\%)}$ with an integrated luminosity of $300~{\rm fb}^{-1}$ and $3000~{\rm fb}^{-1}$, respectively; see the points E and F in Fig.~\ref{experiments}(b). At the 100~TeV colliders, the DPS signal rate dominates over the SPS background, thus leading to a much better precision $15^{+0.87~(+5.8\%)}_{-0.78~(-5.2\%)}$ with an integrated luminosity of $\ge300~{\rm fb}^{-1}$; see the points G and H in Fig.~\ref{experiments}(b). With the projected accuracy, one might be able to check whether $\sigma_{\rm eff}$ varies with the colliding energy.

\begin{figure}
\centering
\includegraphics[scale=0.23]{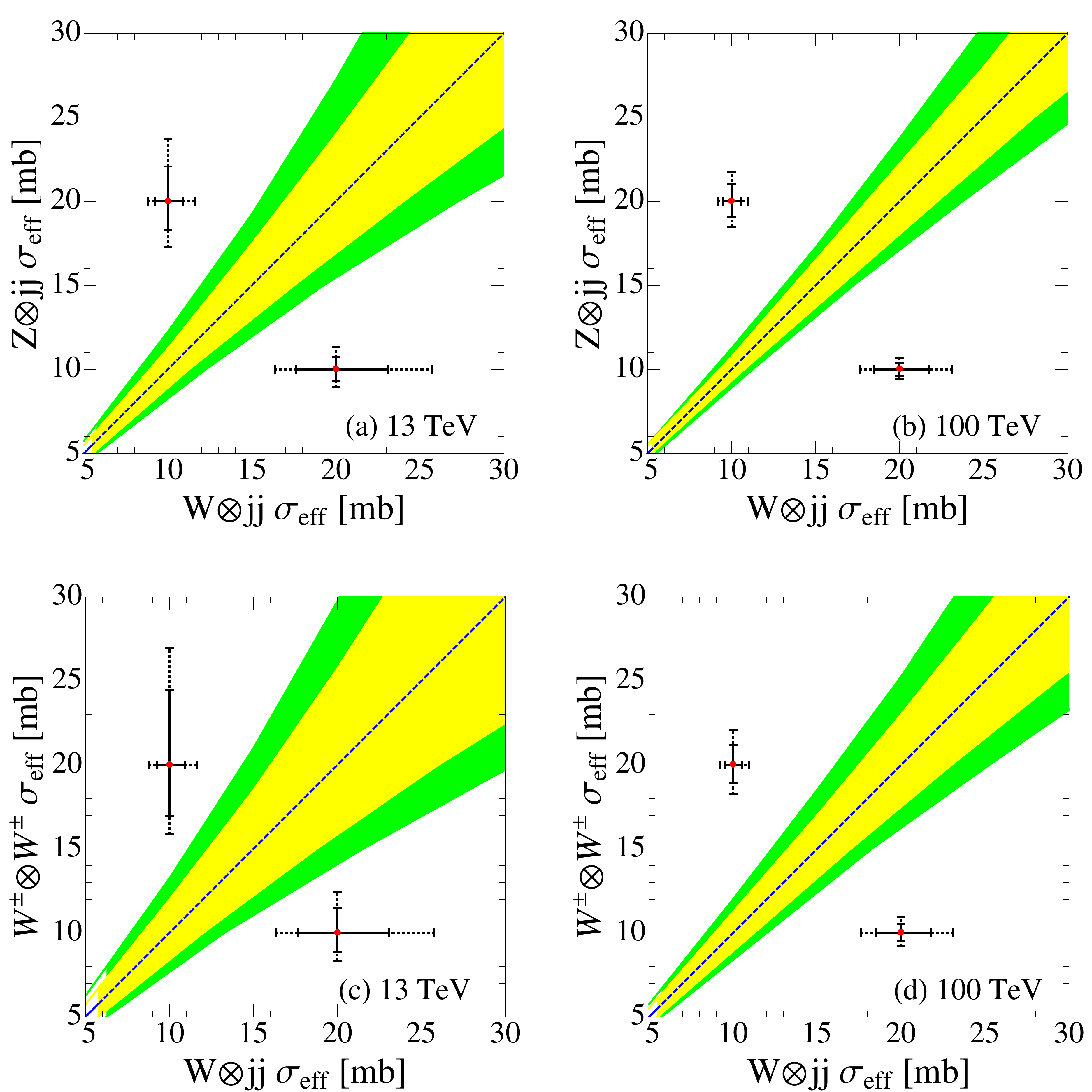}
\caption{Universality check of $\sigma^{\rm fit}_{\rm eff}$: (a, b) $Z\otimes jj$ versus $W\otimes jj$ with $\mathcal{L}\geq 300~{\rm fb}^{-1}$, (c, d) $W^\pm \otimes W^\pm$ versus $W\otimes jj$ with $\mathcal{L}= 3000~{\rm fb}^{-1}$. The yellow (green) band denotes the region of a universal $\sigma_{\rm eff}$ at the $1\sigma$ confidence level with a systematic error of 15\% (25\%), respectively. }
\label{correlation} 
\end{figure}

Now check the universality of $\sigma_{\rm eff}$ for different DPS processes. The current data implies $\sigma_{\rm eff}\sim 15~{\rm mb}$ but with large uncertainties. Figure~\ref{correlation} shows the correlations among the $\sigma^{\rm fit}_{\rm eff}$'s measured in the three DPS channels: (a, b) $Z\otimes jj$ versus $W\otimes jj$, (c, d) $W^\pm \otimes W^\pm$ versus $W\otimes jj$. The yellow and green bands represent the region of a universal $\sigma_{\rm eff}$ at the  $1\sigma$ level with a systematic error of 15\% and 25\%, respectively. Any data falling outside the band indicate that $\sigma_{\rm eff}$ is process dependent. The weak boson productions are sensitive to the flavor of quarks inside proton. As shown in Sec.~\ref{sec:frac_parton}, the DPS channels we considered depend mainly on parton combinations listed as follows: 
\bea
W\otimes jj &:& \sigma_{\rm eff}(qg\otimes \bar{q}' g), \nn\\
Z\otimes jj &:&  \sigma_{\rm eff}(qg\otimes \bar{q}g),\nn\\
W^\pm\otimes W^\pm &:&  \sigma_{\rm eff}(u\bar{d}\otimes u \bar{d}),~\sigma_{\rm eff}(qq\otimes \bar{q}' \bar{q}').\nn
\eea
A process-dependent $\sigma_{\rm eff}$ means that the $\sigma_{\rm eff}$ will have different values for $qq$, $qg$ and $q\bar{q}$ scatterings. Note that our theory calculation is based on the assumption that the $F(b)$ function is universal for any two partons in one proton. Deviation from the yellow or green band might indicates that the assumption of a universal function $F(b)$ is not valid. 

Note that the uncertainty bands in the correlation between $W\otimes jj$ and $Z\otimes jj$, shown in Figs.~\ref{correlation}(a) and \ref{correlation}(b), are dominated by the systematic errors. We plot the $1\sigma$ bands with an integrated luminosity of $300~{\rm fb}^{-1}$, and increasing luminosity will not alter the band width. The $W^\pm\otimes W^\pm$ channel has a large statistical uncertainty in the $\sigma_{\rm eff}$ measurement, therefore, it requires a high luminosity to make $W^\pm\otimes W^\pm$ usable. We obtain the $1\sigma$ bands in Figs.~\ref{correlation}(c) and \ref{correlation}(d) using an integrated luminosity of $3000~{\rm fb}^{-1}$. One is able to test the $\sigma_{\rm eff}$ universality if the $\sigma_{\rm eff}$'s of two different  DPS processes are not too close. For example, $\sigma_{\rm eff}(W\otimes jj)=10~{\rm mb}$ and $\sigma_{\rm eff}(Z\otimes jj)=20~{\rm mb}$ can be well distinguished at the 13~TeV LHC. A better test of $\sigma_{\rm eff}$ universality is expected at the 100 TeV colliders. It is worth mentioning that one should also take the $\gamma j\otimes jj$ channel into account for a better and comprehensive comparison.

In short, we affirm that the Double Parton Scattering processes involving weak bosons ($W\otimes jj$, $Z\otimes jj$ and $W^\pm\otimes W^\pm$) are promising at the LHC and future hadron colliders. The $Z\otimes jj$ channel is the best in measuring the effective cross section $\sigma_{\rm eff}$. Once DPS is established in data and $\sigma_{\rm eff}$ is determined, one can test the universality of $\sigma_{\rm eff}$  in the three channels.

\begin{acknowledgements}
The work is supported in part by the National Science Foundation of China under Grand No. 11175069, No. 11275009 and No. 11422545.
\end{acknowledgements}

\bibliographystyle{apsrev}
\bibliography{reference}

\end{document}